\documentclass[letter,11pt]{article}
\usepackage[top=1in,bottom=1in,left=1in,right=1in]{geometry}
\usepackage{comment}

\usepackage{slashed}
\usepackage{epsfig}
\usepackage{comment}
\usepackage{cancel}
\usepackage{bbm}
\usepackage{array}
\usepackage{bigints}
\usepackage{booktabs}
\usepackage{color}
\usepackage{dsfont}
\usepackage{float}
\usepackage{framed}
\usepackage{graphicx}
\usepackage{indentfirst}
\usepackage{mathrsfs}
\usepackage{multirow}
\usepackage{simpler-wick}
\usepackage{subdepth}
\usepackage{titlesec}
\usepackage[dotinlabels]{titletoc}
\usepackage{wrapfig}
\usepackage[all]{xy}
\usepackage[vcentermath]{youngtab}
\usepackage{relsize}
\usepackage{hyperref}
\usepackage{xcolor}
\usepackage{soul}
\numberwithin{equation}{section}

\usepackage[utf8]{inputenc}
\usepackage{slashed}
\usepackage{amsmath}
\usepackage{amsfonts}
\usepackage{amssymb}
\usepackage{cite}
\usepackage{mathtools}

\usepackage{cleveref}
\crefname{section}{§}{§§}
\crefname{section}{§}{§§}

   \makeatletter
  \let\over=\@@over \let\overwithdelims=\@@overwithdelims
  \let\atop=\@@atop \let\atopwithdelims=\@@atopwithdelims
  \let\above=\@@above \let\abovewithdelims=\@@abovewithdelims
\renewcommand\section{\@startsection {section}{1}{\z@}%
{-3.5ex \@plus -1ex \@minus -.2ex}%nn
{2.3ex \@plus.2ex}%
{\normalfont\large\bfseries}}

\renewcommand\subsection{\@startsection{subsection}{2}{\z@}%
{-3.25ex\@plus -1ex \@minus -.2ex}%
{1.5ex \@plus .2ex}%
{\normalfont\bfseries}}

\makeatletter
\DeclareFontFamily{OMX}{MnSymbolE}{}
\DeclareSymbolFont{MnLargeSymbols}{OMX}{MnSymbolE}{m}{n}
\SetSymbolFont{MnLargeSymbols}{bold}{OMX}{MnSymbolE}{b}{n}
\DeclareFontShape{OMX}{MnSymbolE}{m}{n}{
    <-6>  MnSymbolE5
   <6-7>  MnSymbolE6
   <7-8>  MnSymbolE7
   <8-9>  MnSymbolE8
   <9-10> MnSymbolE9
  <10-12> MnSymbolE10
  <12->   MnSymbolE12
}{}
\DeclareFontShape{OMX}{MnSymbolE}{b}{n}{
    <-6>  MnSymbolE-Bold5
   <6-7>  MnSymbolE-Bold6
   <7-8>  MnSymbolE-Bold7
   <8-9>  MnSymbolE-Bold8
   <9-10> MnSymbolE-Bold9
  <10-12> MnSymbolE-Bold10
  <12->   MnSymbolE-Bold12
}{}

\let\llangle\@undefined
\let\rrangle\@undefined
\DeclareMathDelimiter{\llangle}{\mathopen}%
                     {MnLargeSymbols}{'164}{MnLargeSymbols}{'164}
\DeclareMathDelimiter{\rrangle}{\mathclose}%
                     {MnLargeSymbols}{'171}{MnLargeSymbols}{'171}
\makeatother

\begin{document}
\begin{titlepage}
\unitlength = 1mm
\ \\
\vskip 3cm
\begin{center}

{\LARGE{\textsc{Infinite Symmetry Algebras in Four-Dimensional\\[1ex] Conformal Field Theories}}}

\vspace{1.25cm}
Elizabeth Himwich$^{*\star}$ and Monica Pate$^{ \dagger}$

\vspace{.5cm}

$^*${\it  Princeton Center for Theoretical Science, Princeton University, Princeton, NJ 08544}\\ 
$^\star${\it  Princeton Gravity Initiative, Princeton University, Princeton, NJ 08544}\\ 
$^\dagger${\it  The Center for Cosmology and Particle Physics, New York University, New York, NY 10003}\\ 

\vspace{0.8cm}

\begin{abstract} 
In generic interacting four-dimensional Lorentzian conformal field theories, an infinite set of universal light-ray operators constructed from the stress tensor is shown to generate the wedge subalgebra of the loop algebra of ${\rm w}_{1+\infty}$. This algebra was recently identified among the asymptotic symmetries of asymptotically flat spacetimes. The one-point functions of the ${\rm w}_{1+\infty}$ generators in scalar states (also known as one-point event shapes) are explicitly demonstrated to be finite and are precisely related to universal soft factors in the infinite tower of soft graviton theorems.  A second universal class of light-ray operators that generates the ``$S$ algebra,''  the gauge-theoretic analog of ${\rm w}_{1+\infty}$, is also constructed and shown to have finite one-point functions in four-dimensional conformal field theories with a spin-one conserved current.  Along with the details of these results, this paper presents a general classification of stress-tensor and conserved current light-ray operators by scaling dimension and Lorentz ${\rm SL}(2,\mathbb{C})$ weights, a general technique for computing commutators by Poincar\'e recursion, results for other light-ray operator algebras including a local version of the four-dimensional conformal symmetry algebra, and examples for free scalar fields.  

\end{abstract}

\vspace{1.0cm}
\end{center}

\end{titlepage}

\pagestyle{empty}
\pagestyle{plain}

\def\vx{{\vec x}}
\def\p{\partial}
\def\po{$\cal P_O$}
\def\i{{\rm initial}}
\def\f{{\rm final}}

\pagenumbering{arabic}
 
%%%%%%%%%%%%%%%%---------------------END OF TITLE PAGE AND ABSTRACT---------------------%%%%%%%%%%%%%%%%%%%%%%

\tableofcontents

\section{Introduction and Summary of Results}

Symmetries are among the most simple yet broadly consequential tools in physics. Principles of symmetry rigorously constrain quantitative descriptions of the natural world and have thus profoundly shaped the development of modern physical theory. The scope of a given symmetry is determined by its universality across distinct physical systems as well as the relative size of the symmetry algebra. For example, Poincar\'e symmetry, a defining feature of relativistic quantum field theories, underpins our notion of a particle and imposes further constraints on allowed particle interactions. However, absent additional restrictions, generic relativistic quantum field theories are not expected to admit larger universal symmetry algebras.  Conversely, certain conditions on particle content or interactions famously lead to large classes of quantum field theories with enhanced finite-dimensional symmetries, including conformal field theories (in $d>2$) and supersymmetric quantum field theories.  Exquisite analytic control is further afforded to theories with \emph{infinite-dimensional} symmetries, and while more rare, there still exist broad classes of such theories.  

The most notable instance of a universal infinite-dimensional symmetry is the Virasoro symmetry arising in two-dimensional conformal field theories (CFTs). More recently, universal infinite-dimensional symmetry enhancements have been established in all quantum field theories with propagating gauge bosons or gravitons \cite{Strominger:2017zoo}.  These generalize a familiar form of symmetry in general relativity known as an ``asymptotic symmetry.''\footnote{Despite the terminology, asymptotic symmetries are exact symmetries, not approximate ones.}  In asymptotically flat theories of gravity, the symmetries organize into the wedge subalgebra of the loop algebra of ${\rm w}_{1+\infty}$.  In non-abelian gauge theory, the corresponding symmetry is known as the ``$S$ algebra" \cite{Guevara:2021abz, Strominger:2021lvk}. Although large symmetry algebras are generally quite constraining, several aspects of asymptotic symmetry analyses --- including asymptotic expansions about null infinity, soft expansions in a low-energy limit, and restrictions to self-dual sectors --- obscure the dependence of these infinite-dimensional symmetry enhancements on the perturbative approximations about free theories that naturally arise in their various derivations. 

Inspired by these recent developments, in this work we present a precise avatar of the same infinite-dimensional symmetry enhancements in genuinely interacting theories that do not necessarily contain propagating gauge bosons or gravitons. Specifically, in generic interacting four-dimensional conformal field theories, we construct a universal class of light-ray operators formed from null integrals of the stress tensor which, upon smearing in the transverse directions, generate the wedge subalgebra of the loop algebra of ${\rm w}_{1+\infty}$.  We carry out an analogous analysis in theories with an additional non-abelian conserved current and establish an $S$ algebra.  In some respects, these algebras can be viewed as higher-dimensional generalizations of, respectively, the infinite-dimensional Virasoro symmetry enhancement and Kac-Moody current algebra in conformal field theories in two dimensions. Our (smeared) light-ray operators are not topological, meaning that they do not imply a set of conservation laws, but may nonetheless constrain observables.

Light-ray operators more broadly are novel intrinsically Lorentzian features of quantum field theories.  They have a long history in QCD and collider physics \cite{Balitsky:1987bk,Braun:2003rp} and more recently, have been developed substantially in the context of general conformal field theories \cite{Hofman:2008ar,Caron-Huot:2017vep,Simmons-Duffin:2017nub,Kravchuk:2018htv,Cordova:2018ygx,Kologlu:2019bco,Kologlu:2019mfz,Caron-Huot:2020adz,Chang:2020qpj,Belin:2020lsr,Besken:2020snx,Kravchuk:2021kwe,Korchemsky:2021htm,Caron-Huot:2022ugt,Chang:2022ryc,Caron-Huot:2022eqs,Hartman:2023qdn,Hartman:2023ccw,Chang:2023szz,Henriksson:2023cnh,Hartman:2024xkw,Homrich:2024nwc,Li:2025knf,Chang:2025zib,Erramilli:2025pfh,Mecaj:2025ecl}.   The connection between light-ray operators, asymptotic symmetries, soft theorems, and celestial holography has also attracted recent attention \cite{Cordova:2018ygx,He:2019ywq,Donnay:2020fof,Gonzo:2020xza,Freidel:2021ytz,Hu:2022txx,Hu:2023geb,Chang:2025zib,Gonzalez:2025ene,Moult:2025njc,Oertel:2026wsm}, starting with the work of \cite{Cordova:2018ygx} on a collection of light-ray operators that generate the BMS algebra in generic interacting conformal field theories.  This paper establishes an extension of the results in \cite{Cordova:2018ygx} and subsequent work of \cite{Besken:2020snx} to the full ${\rm w}_{1+\infty}$ symmetry algebra. In the context of a free (conformal) scalar field in four dimensions, a ${\rm w}_{1+\infty}$ and $S$ algebra were previously constructed by \cite{Hu:2022txx,Hu:2023geb}, building directly from the covariant phase space analysis of \cite{Freidel:2021ytz}. Our work extends those results to \emph{generic interacting} conformal field theories, and thereby establishes a broad class of examples in which the full ${\rm w}_{1+\infty}$ or $S$ algebra is realized in a genuinely interacting context. Our result thus indirectly supports the possibility that these symmetries might non-trivially constrain interacting sectors of asymptotically flat gravitational systems.  Finally, there have been several other recent constructions of ${\rm w}$ algebras generated by light-ray operators in conformal field theories in dimensions other than four, including \cite{Korchemsky:2021htm, Sheta:2025oep,Strominger:2026yrh}, but these  realizations are not the result of a direct application of our analysis to other dimensions. 

The central results of this paper are the ${\rm w}_{1+\infty}$ and $S$ algebras in four-dimensional conformal field theories and the explicit all-orders expressions for their generators. We also include a variety of additional applications of the general techniques that are employed to establish these main results. In particular, the organization of light-ray operators into highest-weight representations of the four-dimensional Lorentz group ${\rm SL}(2, \mathbb{C})$ plays a key role in determining explicit formulas for the ${\rm w}_{1+\infty}$ and $S$ generators as null integrals of the stress tensor or conserved current.  We also provide a more complete classification of light-ray operators in highest-weight representations of ${\rm SL}(2, \mathbb{C})$ and identify, for example, local versions of all of the generators of global conformal transformations in $d=4$.  We  recover other broad classes of light-ray operators, including generalizations of the average null energy condition operator that have been considered in the literature \cite{Casini:2017roe,Besken:2020snx,Belin:2020lsr,Huang:2020ycs,Huang:2021hye}.  Some of the light-ray operator expressions have previously appeared in the context of asymptotic symmetries, and in particular can be identified with the ``hard'' parts of the associated conserved charges. It would be interesting if some of our new expressions for highest-${\rm SL}(2, \mathbb{C})$-weight light-ray operators could be used to discover new classes of asymptotic symmetries. 

A second central technique in this paper is the recursive use of Poincar\'e constraints to fix the algebra of light-ray operators constructed from the stress tensor or a conserved current.  The methods previously employed to determine such commutators either involve  direct calculation from the operator product expansion \cite{Besken:2020snx} or rely on the existence of relations between light-ray operators and conserved charges of global symmetries \cite{Cordova:2018ygx}.  For the generators of ${\rm w}_{1+\infty}$ and other more general light-ray operators, the former becomes impractical and  the latter is not applicable.  In this work, we explain how the commutators are organized by scaling dimension into a hierarchy that can then be determined order-by-order with Poincar\'e constraints.  We establish the ${\rm w}_{1+\infty}$ and $S$ algebras via an inductive proof and additionally determine commutators involving the local versions of the global conformal generators.  

Finally, we initiate a study of observables involving our new class of light-ray operators.  Specifically, we calculate one-point functions of the ${\rm w}_{1+\infty}$ and $S$ generators evaluated in a state created by a single scalar field.  Interestingly, we find that the result precisely matches (${\rm SL}(2, \mathbb{C})$ primary descendants of) universal soft factors identified in \cite{Himwich:2023njb} that arise in the expansion of a scattering amplitude in the energy of an external graviton or gauge boson, respectively.  Notably, the result is manifestly \emph{finite}, which was not the case for other previously-considered candidates for generators of infinite-dimensional symmetry algebras in conformal field theories in $d> 2$ \cite{Casini:2017roe,Besken:2020snx,Belin:2020lsr,Huang:2020ycs,Huang:2021hye}.  The calculation relies entirely on three-point functions between two scalars and one stress tensor or conserved current, which are completely fixed by conformal symmetry.  As a result, the values of the one-point functions are non-perturbatively exact and thus provide an alternate, entirely non-perturbative meaning to the infinite tower of soft factors, which have previously only been calculated in perturbation theory. 

\subsection*{Summary of Main Results}

A ${\rm w}_{1+\infty}$ algebra is generated by light-ray operators $\mathcal{W}^p$ in  highest-weight representations of ${\rm SL}(2, \mathbb{C})$\footnote{An operator in a highest-weight representation of ${\rm SL}(2, \mathbb{C})$ transforms under $4d$ Lorentz transformations as 
\begin{equation}
    \begin{split}
        - \delta_{Y} \mathcal{O}(z, \bar{z}) &= \left(Y^z \partial_z + h \partial_z Y^z\right)\mathcal{O}(z, \bar{z}), \\
        - \delta_{\bar{Y}} \mathcal{O}(z, \bar{z}) &= \left(\bar{Y}^{\bar{z}} \partial_{\bar{z}} + \bar{h} \partial_{\bar{z}} \bar{Y}^{\bar{z}}\right)\mathcal{O}(z, \bar{z}), \\
    \end{split}
\end{equation}
where here we have parametrized the six independent Lorentz transformations by globally-defined conformal Killing vectors in two dimensions, $Y^z$ and $\bar{Y}^{\bar{z}}$, as in \eqref{ckv-retarded}.} with $(h, \bar{h}) = (p, 3-p)$ constructed from the following modes of the stress tensor $T_{\mu \nu}$:
\begin{equation} \label{def-w-intro} 
    \begin{split}
        \mathcal{W}^{\frac{m+4}{2}}(z, \bar{z}) 
             \equiv \frac{1}{2^{m+1}(m+1)!} \Bigg[&\int du~ (u\partial_z)^{m-1} 
            \left(u^{ 2}\partial_z^{2} T_{uu}^{(2)} 
                - 2(m+1)u  \partial_z  T_{uz}^{(2)}
                +3m(m+1) T_{zz}^{(2)} \right)
            \\&   + \sum_{n=3}^{m+1} \frac{(m+1)!(n-3)!}{(m-n+1)!}
                \int du~ (u\partial_z )^{m-n+1}\partial_{\bar{z}}^{2-n}\\& \quad \quad \quad \quad \times
        \left ( \partial_z^{2} T_{rr}^{(n+2)} 
            +2(n-1)\partial_z T_{rz}^{(n+1)}
            +(n+1)(n-2)   T_{zz}^{(n)} \right)\Bigg]. 
    \end{split} 
\end{equation}
Here $m\geq -1$ is the (integer-valued) length scaling dimension of the operator.  The superscript of $\mathcal{W}^p$ labels the chiral ${\rm SL}(2, \mathbb{C})$ weight $h =p= \frac{m+4}{2}$. The above expression is written in coordinates \eqref{flat-retarded-Bondi-coord} and evaluated at null infinity ($r \to \infty$).  The superscript on components of the stress tensor $T^{(n)}_{\mu \nu}$ denotes the coefficient of the $r^{-n}$ term in a large-$r$  expansion about null infinity.  Note that $\mathcal{W}^{\frac{3}{2}}$, corresponding to $m = -1$, is the average null energy condition (ANEC) operator, which has been studied extensively in the literature. Finally, these expressions are non-local in the transverse $(z, \bar{z})$ plane when $m \geq 3$.  Interestingly, in free scalar field theory, the non-locality in the $(z, \bar{z})$ plane can be traded for additional non-locality along the null direction $u$ when re-expressed in terms of the fundamental field $\phi$. 

In conformal field theories in four spacetime dimensions with no light scalars with dimensions $1 \leq \Delta \leq 2$, we prove by induction that the $\mathcal{W}$ operators satisfy an algebra of the form 
\begin{equation} \label{eq:InductiveAssumption-intro}
  \begin{aligned}
\left[\mathcal{W}^p(z,\bar{z}), \mathcal{W}^q(z',\bar{z}')\right] &= i \left[(p-1) \partial_{z'} - (q-1)\partial_z\right] \delta^{(2)}(z-z') \mathcal{W}^{p+q-2}(z',\bar{z}') \\
  &\ + \sum_{\ell = 2p-1} a_{\ell}^{i}\partial_z^{\ell}\left(\delta^{(2)}(z-z') O_i(z',\bar{z}') \right) + \sum_{k = 2q-1} b_k^j\partial_{z'}^k\left(\delta^{(2)}(z-z') O_j(z',\bar{z}') \right) ,
  \end{aligned}
\end{equation}
where $O_i$, $O_j$ denote other light-ray operators formed from null integrals of the stress tensor and contributions from the identity  are not included.  Modes of the ``wedge'' are defined by the following integration
\begin{equation} \label{eq:w-wedge-intro}
{\rm w}^{p}_{m,\bar{m}} \equiv \int  d^2z~ z^{p+ m - 1} \bar{z}^{2-p+\bar{m}} \mathcal{W}^{p}(z,\bar{z}), \qquad  1-p \leq m \leq p-1,
\end{equation}
and are labeled by the chiral ${\rm SL}(2, \mathbb{C})$ weight $p$ and chiral and anti-chiral ${\rm SL}(2, \mathbb{C})$ modes numbers $m$ and $\bar{m}$, respectively. Integrating the commutation relation \eqref{eq:InductiveAssumption-intro} to form generators in the wedge projects out the contributions from $O_i$, $O_j$ and thus projects onto the wedge subalgebra of the loop algebra of ${\rm w}_{1+\infty}$: 
\begin{equation} \label{eq:wloop-intro}
\left[{\rm w}^{p}_{m,\bar{m}}, {\rm w}^{q}_{n,\bar{n}} \right] = i \left[m(q-1) - n(p-1)\right] {\rm w}^{p+q-2}_{m+n,\bar{m}+\bar{n}}. 
\end{equation}
This symmetry algebra has recently appeared in the context of celestial holography as the algebra of asymptotic symmetries associated to the infinite tower of soft graviton theorems \cite{Guevara:2019ypd,Guevara:2021abz,Strominger:2021lvk,Himwich:2021dau,Freidel:2021ytz,Adamo:2021lrv,Adamo:2021zpw,Himwich:2023njb,Kmec:2024nmu}. The generators with $p = \frac{3}{2},2$ generate a closed subalgebra, which in the context of celestial holography can be identified as a chiral half of the bulk Poincar\'e symmetry.  Generators with $p \geq \frac{5}{2}$ increase the chiral ${\rm SL}(2, \mathbb{C})$ weight and the full algebra is finitely generated by the modes with $p = \frac{3}{2}, 2, \frac{5}{2}$.  In the context of celestial holography, there is also a central term at $p=1$, which plays a non-trivial role in the generalization of the symmetry algebra to spacetime geometries with non-vanishing cosmological constant  \cite{Taylor:2023ajd,Bittleston:2024rqe}. 

The momentum-space one-point function of the $\mathcal{W}^p$ operators in scalar states takes the form
\begin{equation} \label{eq:W1pt-intro}
    \begin{split}
        \langle\mathcal{O}(p_1)| \mathcal{W}^{\frac{m+4}{2}}(z, \bar{z})|\mathcal{O}(p_3) \rangle   
        &=\frac{1}{4\pi}{m+3 \choose 2}    \frac{(-p_1^2)^{2} \left( \partial_z \hat n^\mu \hat n^\nu \mathcal{L}_{1\mu\nu} \right)^{m+1}}{\left(-   \hat n \cdot p_1\right)^{m+4}} 
         \langle \mathcal{O}(p_1)|\mathcal{O}(p_3)\rangle,
    \end{split}
\end{equation}
where $\hat n$ is a unit null vector in the coordinates \eqref{flat-retarded-Bondi-coord}, $\mathcal{L}_{\mu \nu}$ is the orbital angular momentum
\begin{equation} 
    \begin{split}
        \mathcal{L}_{k\mu\nu} = -i \left( p_{k\mu} \frac{\partial}{ \partial p^\nu_k}-p_{k\nu} \frac{\partial}{ \partial p^\mu_k} \right),
    \end{split}
\end{equation}
and the momentum-space two-point function is 
\begin{equation}
    \begin{split}
       \langle \mathcal{O}(p_1)|\mathcal{O}(p_3)\rangle
       & \equiv\int d^4 x_1 \int d^4 x_3 ~e^{-ip_1\cdot x_1+i p_3 \cdot x_3} \langle \mathcal{O}(x_1) \mathcal{O}(x_3) \rangle. 
    \end{split}
\end{equation}
The right-hand side of \eqref{eq:W1pt-intro} is manifestly finite and precisely matches an ${\rm SL}(2, \mathbb{C})$ primary descendant of the sub$^{m+1}$leading universal soft factor that arises in the expansion of a scattering amplitude in the energy of an external graviton \cite{Himwich:2023njb}. 

Similarly, the generators $\mathcal{S}^p$ of the $S$ algebra are light-ray operators in highest-weight representations of ${\rm SL}(2, \mathbb{C})$ with $(h, \bar{h}) = (p, 2-p)$ constructed from modes of a conserved current $j_\mu$: 
\begin{equation} \label{def-s-operator-intro}
    \begin{split}
          \mathcal{S}^{\frac{m+2}{2}}(z, \bar{z})
            \equiv\frac{1}{2^m m!}& \Bigg[ \int du~ (u \partial_z )^{m-1}  \left(u\partial_z  j_{u}^{(2)} -  m  j_{z}^{(2)}\right)\\
           &   - \sum_{n=3}^{m+1} \frac{m!(n-3)!}{(m-n+1)!}
            \int du~(u\partial_z  )^{m-n+1} \partial_{\bar{z}}^{2-n} \left ( \partial_zj_{r}^{(n+1)}+ (n-2) j_{z}^{(n)} \right)\Bigg].
    \end{split}
\end{equation}
Here $m \geq 0$ labels the length scaling dimension of the operator and the superscript of $\mathcal{S}^p$ labels the chiral ${\rm SL}(2, \mathbb{C})$ weight $h =p= \frac{m+2}{2}$.

The algebra of $\mathcal{S}$ operators receives contributions from light-ray operators constructed from $j_\mu$ of the form  
\begin{equation} \label{eq:localS-intro}
\begin{aligned}
\left[\mathcal{S}^{p,a}(z,\bar{z}), \mathcal{S}^{q,b}(z',\bar{z}')\right] &= i f^{abc} \delta^{(2)}(z-z') \mathcal{S}^{p+q-1,c}(z',\bar{z}')  \\ 
&\ + \sum_{\ell = 2p-1} a_{\ell}^{i}\partial_z^{\ell}\left(\delta^{(2)}(z-z') O_i(z',\bar{z}') \right) + \sum_{k = 2q-1} b_k^j\partial_{z'}^k\left(\delta^{(2)}(z-z') O_{j}(z',\bar{z}') \right), \\
\end{aligned} 
\end{equation}
where $f^{abc}$ are the structure  constants of the global symmetry generated by $j_\mu$, $O_i$, $O_j$ denote other light-ray operators formed from null integrals of the conserved current, and the color structure is implicit but suppressed in the coefficients $a^i_\ell$, $b^j_k$ and indices $i$, $j$.  ``Wedge'' modes are constructed according to 
\begin{equation} \label{eq:Smodes-intro}
{S}^{p,a}_{m,\bar{m}} \equiv \int  d^2z~ z^{p+ m - 1}\bar{z}^{1-p+\bar{m}} \mathcal{S}^{p,a}(z,\bar{z}), \qquad  1-p \leq m \leq p-1.
\end{equation}
Integrating \eqref{eq:localS-intro} to form wedge modes again projects out the contributions from $O_i$, $O_j$, leaving the $S$ algebra:
\begin{equation} \label{eq:Salg-intro}
\left[S^{p,a}_{m,\bar{m}}, S^{q,b}_{n,\bar{n}}\right] = i f^{abc} S^{p+q-1,c}_{m+n, \bar{m}+\bar{n}}. 
\end{equation}
This symmetry algebra arises in the context of celestial holography as the algebra of asymptotic symmetries associated to the infinite tower of tree-level soft gluon theorems \cite{Guevara:2021abz,Strominger:2021lvk,Costello:2022wso,Freidel:2023gue,Kmec:2025ftx}. 

The momentum-space one-point functions of the $\mathcal{S}^p$ operators constructed from global ${\rm U}(1)$ currents in scalar states are 
\begin{equation}
    \begin{split}
        \langle \mathcal{O}(p_1) |\mathcal{S}^{\frac{m+2}{2}}(z, \bar{z})  |\mathcal{O}  (p_3)\rangle
         &= (m+1) \frac{Q}{4 \pi} \frac{(- p_1^2)\left(  \partial_z \hat n^\mu  \hat n^\nu \mathcal{L}_{1\mu\nu}\right)^m}{(-\hat n \cdot p_1)^{m+2}}  
         \langle \mathcal{O}(p_1)  |\mathcal{O}  (p_3)\rangle,
    \end{split}
\end{equation}
where $Q$ is the ${\rm U}(1)$ charge of $\mathcal{O}$. They are again manifestly finite and precisely match ${\rm SL} (2, \mathbb{C})$ primary descendants of universal soft factors that arise at sub$^{m}$leading order in the expansion of a scattering amplitude in the energy of an external gauge boson.

Finally, the mixed $\mathcal{W}$, $\mathcal{S}$ commutators receive contributions from light-ray operators constructed from $j_\mu$ of the form
\begin{equation} \label{eq:InductiveAssumptionWS-intro}
  \begin{aligned}
\left[\mathcal{W}^p(z,\bar{z}), \mathcal{S}^{q,a}(z',\bar{z}')\right] &= i \left[(p-1) \partial_{z'} - (q-1)\partial_z\right] \delta^{(2)}(z-z') \mathcal{S}^{p+q-2,a}(z',\bar{z}') \\
  &\ + \sum_{\ell = 2p-1} a^i_{\ell}\partial_z^{\ell}\left(\delta^{(2)}(z-z') O_{i}(z',\bar{z}') \right) + \sum_{k = 2q-1} b_k^j\partial_{z'}^k\left(\delta^{(2)}(z-z') O_{j}(z',\bar{z}') \right).
  \end{aligned}
\end{equation}
As before, the algebra of the wedge modes does not involve the operators $O_i$ and reproduces the adjoint action of ${\rm w}_{1+\infty}$ on $S$, namely
\begin{equation}
\left[{\rm w}^{p}_{m, \bar{m}}, {S}^{q,a}_{n, \bar{n}} \right] = i \left[m(q-1) - n(p-1)\right] {S}^{p+q-2,a}_{m+n, \bar{m}+\bar{n}}. 
\end{equation}

\subsection*{Outline of Paper}

The paper is organized as follows. 

In Section \ref{sec:preliminaries}, we introduce various coordinate systems that are employed throughout the paper, review aspects of and set conventions for the Lorentzian conformal algebra in four dimensions, and present formulas for charges and symmetry transformations in our coordinates.  

In Section \ref{sec:classification}, we classify light-ray operators constructed from the stress tensor that transform in highest-weight representations of ${\rm SL}(2, \mathbb{C})$. This analysis begins in Subsection \ref{subsec:stress-tensor-falloff} with a bottom-up conformal-theoretic reformulation of fall-off conditions that are typically employed in asymptotic symmetry analyses.  Our classification then proceeds systematically by 4$d$ scaling dimension, where at the heaviest scaling dimension we recover the ANEC operator. At the first two heaviest orders, we recover the full class of light-ray operators considered in \cite{Cordova:2018ygx}, which includes local generalizations of ($d = 4$) translation, Lorentz transformation, and dilation generators.  At further lighter scaling dimensions, we find a local version of the special conformal generators, the general tower of $\mathcal{W}$ generators, and other broad classes of light-ray operators.  Since this classification method relies solely on Lorentz symmetry, we include formulas that pertain to general relativistic quantum field theories and comment on potential corrections due to non-trivial renormalization group flow.  Appendix \ref{app:primary-derivation} contains more details of this classification.

In Section \ref{sec:LRAtensor}, we restrict to conformal field theories and derive an algebra of light-ray operators constructed from the stress tensor. We discuss general constraints in Subsection \ref{subsec:general-constraints}, and in particular exploit the covariance of the commutation relations under dilation symmetry to organize our analysis by scaling dimension as in the previous section.  In Subsection \ref{subsec:sl2c-constraints}, we derive constraints from Lorentz ${\rm SL}(2, \mathbb{C})$ symmetry that apply specifically to commutation relations between a pair of operators in highest-weight representations.  Some additional relations that are implicitly used in this subsection are derived in Appendix \ref{app:inverseDer}. Following \cite{Besken:2020snx}, we assume that there are no light neutral scalars with dimensions $1\leq \Delta \leq 2$ and find in Subsection \ref{subsec:bms-commutators} that the leading (i.e.~heaviest) few commutators are entirely determined by dilation covariance or can be calculated directly from the operator product expansion as in \cite{Besken:2020snx}.  Then, we demonstrate how commutators at lighter scaling dimensions can be recursively determined from commutators at heavier scaling dimensions by enforcing covariance under Poincar\'e.  We use this method to recover the remaining commutation relations in \cite{Cordova:2018ygx} in Subsection \ref{subsec:bms-commutators}, to determine the full algebra of the local versions of the global conformal generators in Subsection \ref{sec:LocalConfAlg} and to prove by induction that the $\mathcal{W}^p$ operators respect an algebra of the form \eqref{eq:InductiveAssumption-intro} in Subsection \ref{sec:wAlg}.  Additional details and commutation relations can be found in Appendix \ref{app:Wderivation}, while supporting formulas for the proof of the ${\rm w}_{1+\infty}$ algebra are presented in Appendix \ref{app:KerP-w}.  

Section \ref{sec:currents} contains a direct extension of the analyses in Sections \ref{sec:classification} and \ref{sec:LRAtensor} to conformal field theories in four dimensions with an additional continuous global symmetry and associated conserved current.  The classification of light-ray operators constructed from the current into  highest-weight representations of ${\rm SL}(2, \mathbb{C})$ is presented in Subsection \ref{subsec:currents-classification}, with additional details appearing in Appendix \ref{app:primary-derivation-current}. Commutation relations for these operators are derived and presented in Subsection \ref{subsec:current-algebra} and mixed commutation relations between light-ray operators constructed from the stress tensor and current are treated in Subsection \ref{subsec:w-s-commutation}. Further details pertaining to specific commutation relations appear in Appendix \ref{app:Sderivation}, while supporting formulas for the proof of the $S$ algebra are presented in Appendix \ref{app:KerP-s}. 

One-point functions in a scalar background and their relations to universal soft factors is the subject of Section \ref{sec:1pt}.  One-point functions of the $\mathcal{W}^p$ operators are derived in Subsection \ref{subsec:1-pt-W} and related to soft graviton factors.  Subsection \ref{subsec:1-pt-S} contains an analogous derivation of one-point functions of the $\mathcal{S}^p$ operators, which are related to soft factors for gauge bosons. 

Section \ref{sec:freefields} explores some of our results in the context of free scalar field theory.  In particular, we present examples of formulas for generators in which the non-locality in the transverse plane is traded for (additional) non-locality along a null direction and reproduce some of the formulas appearing in \cite{Hu:2022txx, Hu:2023geb}.  Since free scalar field theory contains neutral operators with scaling dimension in the range $1\leq \Delta \leq 2$, it violates the assumption we used to derive commutation relations in Section \ref{sec:LRAtensor}.  We present several examples of commutation relations with additional contributions due to this violation, and observe that the modifications are fairly mild. In particular, as was also observed in \cite{Hu:2022txx,Hu:2023geb}, none of the violations spoil the general form of the $\mathcal{W}$ and $\mathcal{S}$ algebras, \eqref{eq:InductiveAssumption-intro} and \eqref{eq:localS-intro}.  

We conclude in Section \ref{sec:discussion} with a broader view of the implications of our results  and how they can be used in future investigations.  We discuss several potential new consequences of our results  in conformal field theories, general quantum field theories including quantum gravity, and holography.

\section{Preliminaries} \label{sec:preliminaries}
\subsection{Coordinates}

We primarily use flat retarded coordinates $(u,r,z,\bar{z})$, which are related to the usual Cartesian coordinates $x^\mu$ by
\begin{equation} \label{flat-retarded-Bondi-coord}
    \begin{split}
        x^\mu = \frac{1}{2} \left( un^\mu + r \hat q^\mu(z, \bar{z}) \right), 
    \end{split}
\end{equation}
where $n$ and $ \hat q$ are the null vectors
\begin{equation}
    \begin{split}
        \hat q^\mu (z, \bar{z}) = \left(1+ z \bar{z}, z+\bar{z}, -i(z-\bar{z}), 1-z \bar{z}\right), 
        \quad \quad n^\mu = \partial_z \partial_{\bar{z}} \hat q^\mu(z, \bar{z}). 
    \end{split}
\end{equation}
In these coordinates, the Minkowski line element takes the form
\begin{equation} \label{line-element-urz}
    \begin{split}
        ds^2 = -dudr + r^2 dz d\bar{z}. 
    \end{split}
\end{equation}
Future null infinity $\mathscr{J}^+$ is the null surface reached by taking $r \to \infty$ at fixed $(u,z,\bar{z})$, and this paper primarily focuses on light-ray operators formed from null-integrals of local operators along $\mathscr{J}^+$.  When placed at null infinity, these light-ray operators are known as ``detector'' operators and can be defined in any quantum field theory \cite{Hofman:2008ar,Kravchuk:2018htv,Caron-Huot:2022eqs,Li:2025knf}. 

In conformal field theories, conformal symmetry maps detector operators at null infinity into the bulk.  In particular, inversions exchange null infinity and the light-cone of the origin. To see this explicitly in four dimensions, consider the coordinates \cite{Hofman:2008ar}  
\begin{equation} \label{eq:ycoords}
  y^+ = - \frac{1}{r}, \ \ \ y^- = u, \ \ \ y^1 + i y^2 = z, 
\end{equation}
in which $r\to \infty$ corresponds to $y^+ = 0$. In these coordinates, the line element \eqref{line-element-urz} becomes 
\begin{equation}
  ds^2 = \frac{1}{(y^+)^2} \left( - dy^- dy^+ + (dy^1)^2  + (dy^2)^2 \right). 
\end{equation}
This metric is conformally related to standard flat metric in light-cone coordinates
\begin{equation} 
ds^2 = - dx^-dx^+ + (dx^1)^2  + (dx^2)^2,
\end{equation}
which can be reached by the inversion coordinate transformation
\begin{equation}
  x^+ = - \frac{1}{y^+}, \ \ \ x^- = y^- - \frac{y_1^2 + y_2^2}{y^+}, \ \ \ x^{A} = -\frac{y^{A}}{y^+}. 
\end{equation}
Thus null infinity at $y^+=0$ ($r\to \infty$) is conformally mapped to the lightcone at $x^+ = 0$ ($r = 0$). It may be helpful to note that 
\begin{equation}
    \begin{split}
         x^+ = x^0 + x^3 = r, \quad \quad x^- = x^0 - x^3 = u + r z \bar{z}, \quad \quad 
         x^1 + i x^2 = r z.
    \end{split}
\end{equation}

\subsection{Generators of Conformal Symmetry}

Conformal transformations are generated by charges of the form 
\begin{equation} \label{charge-general}
    \begin{split}
        Q_\zeta = \int_{\Sigma} d \Sigma^\rho ~\zeta^\mu T_{\rho \mu}, 
    \end{split}
\end{equation}
which act on the stress tensor and conserved currents according to\footnote{Our conventions throughout are that charges generate transformations $\left[Q, \Phi(x)\right] = i \delta \Phi(x)$,  where the transformation of a field $\Phi$ under an infinitesimal isometry generated by a Killing vector $\zeta$ takes the form $\delta \Phi (x)= - \mathcal{L}_\zeta \Phi(x)$.}
\begin{equation}
    \begin{split}
        \left[Q_\zeta, T_{\mu\nu}(x)\right] &= -i \mathcal{L}_\zeta T_{\mu\nu}(x) - i [(\Delta_T-2)/d] \partial_\rho \zeta^\rho T_{\mu\nu}(x),\\
        \left[Q_\zeta, J_\mu(x)\right] &= -i \mathcal{L}_\zeta J_\mu(x) - i [(\Delta_J-1)/d] \partial_\rho \zeta^\rho J_\mu(x),
    \end{split}
\end{equation}
where $d=4$ is the spacetime dimension, $\Delta_T = d$ and $\Delta_J = d-1$. Here $\zeta$ is a conformal Killing vector and $\Sigma$ is a Cauchy surface.  In Cartesian coordinates, the conformal Killing vectors associated with translations $P_\mu$, Lorentz transformations $J_{\mu\nu}$, dilations $D$, and special conformal transformations $K_\mu$ are given respectively by   
\begin{align} \label{conformal-killing-vectors}
    \begin{split}
        P_\mu:& \quad \zeta_{(\mu)} = \partial_\mu,  \quad \quad 
        J_{\mu\nu}: \quad \zeta_{(\mu \nu)} = x_\mu \partial_\nu -x_\nu \partial_\mu, \\
        D:&  \quad \zeta = x^\mu \partial_\mu,  \quad \quad 
        K_\mu:   \quad \zeta_{(\mu)} =  2 x_\mu x^\nu \partial_\nu - x^2 \partial_\mu,  
    \end{split}
\end{align}
and \eqref{charge-general} becomes
\begin{align}
    \begin{split}
        P_\mu &= \int_{\Sigma} d \Sigma^\rho ~T_{\rho \mu}, \quad \quad \quad ~~J_{\mu\nu} = 2 \int_{\Sigma} d \Sigma^\rho~ x_{[\mu} T_{\nu] \rho}, \\
        D&= \int_{\Sigma} d \Sigma^\rho ~x^\mu T_{\rho \mu}, \quad \quad \quad  K_\mu = \int_{\Sigma} d \Sigma^\rho ~\left(2x_\mu x^\nu T_{\rho \nu}-x^2 T_{\rho \mu}\right).
    \end{split}
\end{align}
These charges act on a local operator $\mathcal{O}(x)$ of scaling dimension $\Delta$ according to 
\begin{equation}
    \begin{split}
        \left[P_\mu, \mathcal{O}(x)\right] &= -i \partial_\mu \mathcal{O}(x), \\
        \left[J_{\mu\nu}, \mathcal{O}(x)\right]&= -2i x_{[\mu} \partial_{\nu]}\mathcal{O}(x) + M_{\mu\nu} \cdot \mathcal{O}(x), \\
        \left[D, \mathcal{O}(x)\right] &= -i x^\mu \partial_\mu \mathcal{O}(x) - i \Delta \mathcal{O}(x), \\
        \left[K_\mu, \mathcal{O}(x)\right]&=-2i x_\mu x^\nu \partial_\nu \mathcal{O}(x)+ i x^2 \partial_\mu \mathcal{O}(x) - 2i \Delta x_\mu \mathcal{O}(x) + 2 x^\nu M_{\mu\nu} \cdot \mathcal{O}(x), 
    \end{split}
\end{equation}
where $M_{\mu\nu}$ is a finite-dimensional matrix for fixed $\mu \nu$ in the Lorentz representation of $\mathcal{O}(x)$.  The charges satisfy the conformal algebra
\begin{equation} \label{eq:globalg}
    \begin{split}
        \left[P_\mu, P_\nu\right] &= 0,  \quad \quad \quad  
        \left[J_{\mu \nu}, P_\rho\right] = -i \left(\eta_{\mu \rho}  P_{\nu}- \eta_{\nu \rho}  P_{\mu}\right),\\
        \left[J_{\mu \nu}, J_{\rho \sigma}\right] &=-i \left( \eta_{\mu \rho} J_{\nu \sigma} - \eta_{\nu \rho}J_{\mu \sigma} 
        +\eta_{\mu \sigma} J_{ \rho\nu} - \eta_{\nu \sigma}J_{\rho\mu } 
        \right) ,\\
        \left[D, P_\mu \right] = -i P_\mu , \quad \quad \quad &
        \left[D,J_{\mu \nu} \right] = 0, \quad \quad \quad 
        \left[D,D\right] = 0,\quad \quad \quad 
        \left[D, K_\mu\right] =i K_\mu,\\
        \left[P_\mu, K_\nu\right] = 2i \left(\eta_{\mu \nu} D - J_{\mu \nu} \right),& \quad \quad \quad 
        \left[J_{\mu \nu}, K_\rho\right] = -i \left(\eta_{\mu \rho}  K_{\nu}- \eta_{\nu \rho}  K_{\mu}\right), \quad \quad  \quad
        \left[K_\mu, K_\nu\right] = 0.
    \end{split}
\end{equation}
To evaluate the charges at $\Sigma = \mathscr{J}^+$, it is convenient to express the conformal Killing vectors \eqref{conformal-killing-vectors} in  retarded coordinates \eqref{flat-retarded-Bondi-coord} and group them into the following linear combinations:
\begin{equation} \label{ckv-retarded}
    \begin{split}
        {\rm Translations}: \quad &\zeta_f \equiv \mathcal{F}^\mu\zeta_{(\mu)}
         = f \partial_u + \partial_z \partial_{\bar{z}}f \partial_r - \frac{1}{r} \left(\partial_z f \partial_{\bar{z}} +\partial_{\bar{z}}  f \partial_z \right), \\
         {\rm Lorentz~transformations}: \quad & \zeta_{Y } + \zeta_{\bar{Y}}\equiv \mathcal{Y}^{\mu \nu}\zeta_{(\mu \nu)} 
         =\frac{1}{2} \left(\partial_z Y^z+\partial_{\bar{z}}\bar{Y}^{\bar{z}} \right) \left(u\partial_u -r \partial_r\right)\\& \quad \quad \quad  \quad \quad \quad \quad \quad +    \left(  \bar{Y}^{\bar{z}}-\frac{u}{2r} \partial_z^2 Y^z  \right) \partial_{\bar{z}} +\left ( Y^{z}-\frac{u}{2r} \partial_{\bar{z}}^2 \bar{Y}^{\bar{z}}\right) \partial_{z},\\
         {\rm Dilations}: \quad &
         \zeta_D = \frac{1}{2} (u n^\mu + r \hat q^\mu) \partial_\mu = u \partial_u + r \partial_r, \\
         {\rm SCT}: \quad &
         \zeta_g \equiv \mathcal{G}^\mu \zeta_{(\mu)} = u^2 \partial_z \partial_{\bar{z}} g  \partial_u
        +  r^2 g\partial_r +u \left(\partial_z g \partial_{\bar{z}} +\partial_{\bar{z}} g \partial_{z} \right).
    \end{split}
\end{equation}
Here the $(z, \bar{z})$-dependent parameters are related to the constant uppercase scripted  parameters by:
\begin{equation}  \label{conf-param}
    \begin{split}
        f(z, \bar{z}) \equiv - \mathcal{F}^\mu \hat q_\mu,
        \quad \quad 
        Y^z(z) \equiv \mathcal{Y}^{\mu \nu}\hat q_{[\mu}\partial_{\bar{z}} \hat q_{\nu]},
        \quad \quad 
        \bar{Y}^{\bar{z}}(\bar{z}) \equiv \mathcal{Y}^{\mu \nu}\hat q_{[\mu}\partial_z \hat q_{\nu]},  \quad \quad g(z, \bar{z}) \equiv \mathcal{G}^\mu \hat q_\mu. 
    \end{split}
\end{equation}

\section{Classification of Stress-tensor Light-ray Operators} \label{sec:classification}

In this section, we present a systematic classification of light-ray operators constructed from null integrals of the stress tensor. In particular, to facilitate our derivation of commutators in the following Section \ref{sec:LRAtensor}, we find it useful to group light-ray operators by $4d$ scaling dimension $\Delta$ and further organize the operators at fixed scaling dimension into primaries with respect to the Lorentz ${\rm SL}(2, \mathbb{C})$ group.  The first two leading (i.e.~heaviest) orders in scaling dimension fully reproduce the collection of light-ray operators that were identified in \cite{Cordova:2018ygx} as the generators of the extended BMS algebra. At the next order in scaling dimension, we find new universal light-ray operators, and at every order in scaling dimension, we fully classify a subset of light-ray operators with a light, fixed value of right ${\rm SL}(2, \mathbb{C})$ weight $\bar{h}$. The majority of this section employs the less stringent fall-off conditions in quantum field theory, provided below in \eqref{qft-fall-off}. We explicitly isolate terms involving the trace so that expressions consistent with fall-off conditions in a conformal field theory are easy to extract. 
 
\subsection{Light-ray Operators and their Fall-off Behavior at Null Infinity} \label{subsec:stress-tensor-falloff}

Although our classification can be performed on any lightsheet, it is convenient to study light-ray operators at null infinity, where we can exploit fall-off conditions on the stress tensor.  Working at null infinity also provides a direct connection to the literature on detector operators and asymptotic symmetries.  For this purpose, we work in coordinates \eqref{flat-retarded-Bondi-coord}, in which $\mathscr{J}^+$ is reached by taking $r\to\infty$ at fixed $(u,z,\bar{z})$, and determine the behavior of components of the stress tensor in the large-$r$ limit. 

In quantum field theory, fall-off conditions are typically required to be consistent with configurations of finite energy and momentum and take the form 
\begin{equation} \label{qft-fall-off}
    \begin{split}
         T_{uu}, T_{uz}, T_{zz}\sim \frac{1}{r^2}, \quad \quad 
        T_{z \bar{z}}\sim \frac{1}{r},\quad \quad T_{rz} \sim \frac{1}{r^3}, \quad \quad T_{ur}, T_{rr}\sim \frac{1}{r^4}.
    \end{split}
\end{equation}
They can be derived in free field theory from a saddle point approximation at large $r$ \cite{Strominger:2017zoo}. In interacting theories, the asymptotic behavior of the field equations is often used to extend the free-field results to these cases, although there are subtleties in four spacetime dimensions due to the slow fall-off behavior of the Coulomb potential.

In conformal field theory, tracelessness of the stress tensor $T_\mu{}^\mu = 0$ leads to faster fall-off behavior. This can either be derived by modifying \eqref{qft-fall-off} subject to the vanishing of the trace or alternatively by studying the transformation of correlation functions under conformal symmetry. Specifically, under an inversion
\begin{equation}
    \begin{split}
        x^\mu \mapsto x'^\mu  = \frac{x^\mu}{x^2}, 
    \end{split}
\end{equation}
the stress tensor transforms as
\begin{equation}
    \begin{split}
         T_{\mu\nu}(x) &\mapsto T'_{\mu\nu}(x') = (x^2)^\Delta I_\mu{}^\rho (x)I_\nu{}^\sigma (x) T_{\rho \sigma}(x),
    \end{split}
\end{equation}
where $\Delta = d = 4$ and $I$ is the inversion tensor
\begin{equation}
    \begin{split}
        \frac{\partial x'^\mu}{\partial x^\nu} = \frac{I^\mu{}_\nu(x)}{x^2}, \quad \quad \quad I^\mu{}_\nu(x) = \delta^\mu_\nu - \frac{2 x^\mu x_\nu}{x^2}.
    \end{split}
\end{equation}
Under this transformation, correlation functions transform as\footnote{Technically this transformation under inversion assumes a reflection symmetry, but the same conclusion can be reached without assuming reflection symmetry by instead using a special conformal transformation.} 
\begin{equation}
    \begin{split}
        \langle T_{\mu\nu}(x') \mathcal{O}_1(x'_1) \cdots \mathcal{O}_n(x'_n)\rangle
        & = \langle T'_{\mu\nu}(x') \mathcal{O}'_1(x'_1) \cdots \mathcal{O}'_n(x'_n)\rangle\\
        & = (x^2)^4 I_\mu{}^\rho (x)I_\nu{}^\sigma (x) \langle T_{\rho \sigma}(x)\mathcal{O}'_1(x'_1) \cdots \mathcal{O}'_n(x'_n) \rangle.
    \end{split}
\end{equation}
Next, we evaluate this expression in the flat retarded coordinates \eqref{flat-retarded-Bondi-coord}, where under an inversion
\begin{equation}
    \begin{split}
        x'^\mu = \frac{1}{2} \frac{u n^\mu + r \hat q^\mu}{-ur} = - \frac{1}{2}\left(\frac{1}{r} n^\mu+ \frac{1}{u} \hat q^\mu\right),
    \end{split}
\end{equation}
the coordinates transform as
\begin{equation}
    \begin{split}
        u' = -\frac{1}{r}, \quad \quad r' = -\frac{1}{u}, \quad \quad z' = z, \quad \quad \bar{z}' = \bar{z}. 
    \end{split}
\end{equation}
Considering the $uu$ component first, we find 
\begin{equation} \label{tuu-correlation}
    \begin{split}
        \langle T_{uu}(x') \mathcal{O}_1(x'_1) \cdots \mathcal{O}_n(x'_n)\rangle
            & =  \partial_{u'} x'^\mu\partial_{u'} x'^\nu \langle T_{\mu\nu}(x') \mathcal{O}_1(x'_1) \cdots \mathcal{O}_n(x'_n)\rangle\\
            & = (x^2)^4  \partial_{u'} x'^\mu\partial_{u'} x'^\nu I_\mu{}^\rho (x)I_\nu{}^\sigma (x) \langle T_{\rho \sigma}(x)\mathcal{O}'_1(x'_1) \cdots \mathcal{O}'_n(x'_n) \rangle\\
            & = u^2 r^6 \langle T_{rr}(x)\mathcal{O}'_1(x'_1) \cdots \mathcal{O}'_n(x'_n) \rangle.
    \end{split}
\end{equation}
Taking $r \to \infty$ sends
\begin{equation}
    \begin{split}
        x'^\mu = -\frac{1}{2}\left(\frac{1}{r} n^\mu +\frac{1}{u} \hat q^\mu\right) \to -\frac{1}{2u} \hat q^\mu,
    \end{split}
\end{equation}
which is a generic, finite point (assuming $u \neq 0$).  Hence, the left-hand side of \eqref{tuu-correlation} is manifestly regular in the limit $r\to \infty$.  Therefore, in order for the right-hand side of \eqref{tuu-correlation}  to be regular in this limit, $T_{rr}$ must fall off as
\begin{equation}
    \begin{split}
        T_{rr} \sim \frac{1}{r^6}. 
    \end{split}
\end{equation}
Repeating this analysis for each  component yields the fall-off behavior
\begin{equation} \label{cft-fall-off}
    \begin{split}
        T_{uu}, T_{uz}, T_{zz}, T_{z \bar{z}}\sim \frac{1}{r^2}, \quad \quad T_{rz}, T_{ur} \sim \frac{1}{r^4}, \quad \quad T_{rr}\sim \frac{1}{r^6}. 
    \end{split}
\end{equation}
Throughout, we use the superscript notation $T_{\mu\nu}^{(n)}$ to denote the coefficient of the $\frac{1}{r^n}$ term in an expansion about $\mathscr{J}^+$. 

Using these fall-off conditions and the expressions for the conformal Killing vectors \eqref{ckv-retarded} in flat retarded coordinates, we find the following explicit expressions for the generators of conformal transformations \eqref{charge-general} evaluated at $\Sigma = \mathscr{J}^+$: 
\begin{subequations} \label{charges-null-infinity}
    \begin{align}
        {\rm Translations:} \quad &
             Q_f  = \int_{\mathscr{J}^+} du d^2 z ~ f T_{uu}^{(2)}, \label{eq:globalP} \\ 
        {\rm Lorentz~transformations:} \quad &
             Q_Y  =   \int_{\mathscr{J}^+} du d^2 z ~  Y^{z}  \left(T_{uz}^{(2)}-\frac{1}{2} u\partial_z T_{uu}^{(2)}\right)  , \label{eq:globalY} \\ 
             &Q_{\bar{Y}}  =   \int_{\mathscr{J}^+} du d^2 z ~ \bar{Y}^{\bar{z}}  \left(T_{u\bar{z}}^{(2)}-\frac{1}{2} u\partial_{\bar{z}} T_{uu}^{(2)}\right) , \label{eq:globalYbar} \\ 
        {\rm Dilations:} \quad & 
            Q_D =    \int_{\mathscr{J}^+} du d^2 z~u T_{uu}^{(2)}, \label{eq:globalD} \\ 
        {\rm SCT:} \quad &
            Q_g =  \int_{\mathscr{J}^+} du d^2 z ~ g\left(u^2   \partial_z \partial_{\bar{z}}T_{uu}^{(2)} 
        +  T_{z \bar{z}}^{(2)} -   u \partial_zT_{u\bar{z}}^{(2)}   - u \partial_{\bar{z}}T_{uz}^{(2)}  \right).  \label{eq:globalSCT} 
    \end{align}
\end{subequations}
Note that the expression for the special conformal transformation generator $Q_g$ assumes the stricter fall-off conditions in conformal field theory \eqref{cft-fall-off}, whereas the remaining generators take the same form for both sets of fall-off conditions \eqref{qft-fall-off} or \eqref{cft-fall-off}.

\subsection{C\'ordova-Shao Light-ray Operators} \label{sec:CSlro}

In this subsection, we perform a systematic study of light-ray operators at the leading two orders in scaling dimension and fully recover the stress-tensor light-ray operators of \cite{Cordova:2018ygx}. 
We begin with null integrals of the stress tensor against non-negative powers of $u$, which have the following scaling dimensions:
\begin{equation} \label{scaling-all}
    \begin{split}
        \Delta = 3-n-m: \quad &\int du~ u^{m} T_{uu}^{(n)}, \quad \int du~ u^{m} T_{ur}^{(n)}, \quad \int du~ u^{m} T_{rr}^{(n)}, \\
        \Delta = 2-n-m: \quad &\int du~ u^{m} T_{uz}^{(n)}, \quad \int du~ u^{m} T_{rz}^{(n)}, \quad 
                            \int du~ u^{m} T_{u\bar{z}}^{(n)}, \quad \int du~ u^{m} T_{r\bar{z}}^{(n)},\\
        \Delta = 1-n-m: \quad & \int du~ u^{m}T_{zz}^{(n)}, \quad \int du~ u^{m} T_{z\bar{z}}^{(n)}, \quad \int du~ u^{m} T_{\bar{z}\bar{z}}^{(n)}. 
    \end{split}
\end{equation}
The scaling dimensions of these operators are deduced from the scaling dimensions of the coordinates,
\begin{equation}
    \begin{split}
        u, r: \quad \Delta = -1, \quad \quad z, \bar{z}: \quad \Delta = 0,
    \end{split}
\end{equation}
and the scaling dimension of components of the stress tensor $T_{\mu \nu}$ in an orthonormal frame, which is $\Delta = d = 4$.   Then in the coordinates \eqref{flat-retarded-Bondi-coord}, for example, $T_{uu}$ has $\Delta = 4$, while $T_{uz} = \partial_u x^\mu \partial_z x^\nu T_{\mu \nu}$ has $\Delta = 3$.  Accounting for the powers of $r$ implied by the $(n)$ superscript, $T_{uu}^{(n)}$ has $\Delta = 4-n$ and the additional powers of $u$ in $\int du~ u^{m} T_{uu}^{(n)}$ thus produce a net scaling dimension $\Delta = 3-n-m$. In this work, we restrict to operators with non-negative powers of $u$, because well-defined operators involving negative powers of $u$ require additional information in the form of an $i \epsilon$ prescription \cite{Belin:2020lsr}.

These scaling dimensions, together with the fall-off behavior from the previous subsection,
imply that there is a unique light-ray operator of the stress tensor with the largest scaling dimension
\begin{equation} \label{anec}
    \begin{split}
        \mathcal{W}_{-1}(z, \bar{z}) \equiv \int du~  T_{uu}^{(2)}, \quad \quad \quad  \Delta =1. 
    \end{split}
\end{equation}
The subscript on the operator $\mathcal{W}_{m}$ labels the negative of the scaling dimension $m=-\Delta$. The operator \eqref{anec} is called the average null energy condition (ANEC) operator and its properties in quantum field theory have been extensively studied in the literature \cite{Hofman:2008ar,Hofman:2016awc,Faulkner:2016mzt,Hartman:2016lgu,Cordova:2017zej,Cordova:2017dhq,Hartman:2022njz,Hartman:2023qdn,Hartman:2023ccw,Hartman:2024xkw}; see \cite{Moult:2025nhu} for a recent review and more complete list of references.  Comparing with  \eqref{eq:globalP}, we observe that the ANEC operator takes the form of a local version of the translation generator. Indeed the ANEC operator plays precisely this role in the context of the asymptotic symmetries of asymptotically flat spacetimes.  Specifically, the ANEC operator is the ``hard'' part of the charge that generates supertranslations \cite{Strominger:2013jfa,He:2014laa}. Its momentum-space action on asymptotic states is the universal factor in the leading soft graviton theorem \cite{Weinberg:1965nx}.

Although there is only a single light-ray operator of this scaling dimension, we note that it transforms as an ${\rm SL}(2, \mathbb{C})$ primary under Lorentz transformations generated by \eqref{eq:globalY} and \eqref{eq:globalYbar}:
\begin{equation} \label{eq:Wm1underY}
    \begin{split}
        -\delta_Y\mathcal{W}_{-1}(z, \bar{z})& = \left(Y^z \partial_z + \frac{3}{2}\partial_z Y^z\right) \mathcal{W}_{-1}(z, \bar{z}), \\
        -\delta_{\bar{Y}}\mathcal{W}_{-1}(z, \bar{z})& = \left( \bar{Y}^{\bar{z}} \partial_{\bar{z}} + \frac{3}{2}\partial_{\bar{z}} \bar{Y}^{\bar{z}}\right) \mathcal{W}_{-1}(z, \bar{z}).
    \end{split}
\end{equation}
In particular, $\mathcal{W}_{-1}$ carries left and right ${\rm SL}(2, \mathbb{C})$ weights 
\begin{equation} \label{eq:Wm1weights}
    \mathcal{W}_{-1} : \quad \left(h, \bar{h}\right) = \left(\frac{3}{2}, \frac{3}{2}\right). 
\end{equation}
Here and throughout, we use $\left(h, \bar{h}\right)$ to label the left and right ${\rm SL}(2, \mathbb{C})$ weights under $4d$ Lorentz transformations and reserve $\Delta$ to denote the $4d$ scaling dimension under $4d$ dilations. 

Finally, the ANEC operator is annihilated by translations \eqref{eq:globalP}
\begin{equation} \label{level--1-translations}
    \begin{split}
        \delta_f \mathcal{W}_{-1} = 0. 
    \end{split}
\end{equation}
Special conformal transformations lower the bulk scaling dimension and therefore involve light-ray operators that we have not yet  discussed. 

At scaling dimension $\Delta = 0$, we find the following set of light-ray operators:  
 \begin{equation} \label{subleading-basis}
      \begin{split}
          \Delta=0: \quad  \int du~ uT_{uu}^{(2)}, \quad \int du~T_{uz}^{(2) },\quad \int du~T_{u\bar{z}}^{(2) } .
     \end{split}
  \end{equation}
Note that in principle, the operators 
\begin{equation} \label{subleading-extra}
    \begin{split}
        \int du ~T^{(3)} = 2\int du~T_{z\bar{z}}^{(1)}, \quad \quad  \int du~T_{uu}^{(3)},
    \end{split}
\end{equation}
are also of dimension $\Delta=0$, where  
\begin{equation}
T^{(n)} \equiv   -2\left(T_{ur}^{(n)}-T_{z\bar{z}}^{(n-2)}\right)
\end{equation}
is used to denote coefficients in the expansion of the trace $T_\mu{}^\mu$.  However, the conservation equation $\partial_\mu T^{\mu \nu} = 0$ implies that
\begin{equation}
    \begin{split}
        0& =   - \partial_u  T_{ur}^{(4)} 
        +  T_{uu}^{(3)} +  \partial_z  T_{u\bar{z}}^{(2)}  +  \partial_{\bar{z}}  T_{uz}^{(2)}, \\
       0& =  -\partial_u T_{r r}^{(4)} - T^{(3)} ,
    \end{split}
\end{equation}
meaning that, up to total derivatives that vanish upon integration against $u$, \eqref{subleading-extra} are not independent from the operators in \eqref{subleading-basis}.  Thus \eqref{subleading-basis} is the complete basis for null integrals of the stress tensor with scaling dimension $\Delta = 0$.  Note that since this basis is independent of the trace, it does not rely on the stricter fall-off conditions in a conformal field theory.  

Next, we consider the transformations of these basis elements under ${\rm SL}(2, \mathbb{C})$.  We find that 
\begin{equation} \label{eq:Dlocal}
    \begin{split}
        \mathcal{D}(z, \bar{z}) \equiv \int du~ uT_{uu}^{(2)}
     \end{split}
\end{equation}
transforms as a primary with ${\rm SL}(2, \mathbb{C})$ weight
\begin{equation}
    \begin{split}
         \mathcal{D}: \quad \left(h, \bar{h}\right) = \left(1, 1\right),
    \end{split}
\end{equation}
but the remaining generators do not:
\begin{equation} \label{eq:Tuz2trans}
    \begin{split}
         -\delta_Y\int du~T_{uz}^{(2) }& = \left(Y^z \partial_z + 2\partial_z Y^z\right) \int du~T_{uz}^{(2) }+ \frac{1}{2} \partial_z^2 Y^z \int du ~uT_{uu}^{(2)}, \\
        -\delta_{\bar{Y}}\int du~T_{uz}^{(2) }& = \left( \bar{Y}^{\bar{z}} \partial_{\bar{z}} + \partial_{\bar{z}} \bar{Y}^{\bar{z}}\right) 
                \int du~T_{uz}^{(2) }.
    \end{split}
\end{equation}
Instead, taking an appropriate linear combination, we find that
\begin{equation} \label{eq:W0local}
    \begin{split}
        \mathcal{W}_{0}(z, \bar{z}) &\equiv \frac{1}{2} \int du~ \left(u \partial_z T_{uu}^{(2)} - 2T_{uz}^{(2)}\right)
        ,\\
        \overline{\mathcal{W}}_{0}(z, \bar{z}) &\equiv 
            \frac{1}{2} \int du~ \left(u \partial_{\bar{z}} T_{uu}^{(2)} - 2T_{u\bar{z}}^{(2)}\right),
    \end{split}
\end{equation}
each transform as primaries of respective weights 
\begin{equation} \label{eq:W0weights}
    \begin{split}
           \mathcal{W}_{0}: \quad \left(h, \bar{h}\right) = \left(2, 1\right),
           \quad \quad \quad \quad \quad 
          \overline{\mathcal{W}}_{0}: \quad \left(h, \bar{h}\right) = \left(1, 2\right).
    \end{split}
\end{equation}
Comparing with the dilation \eqref{eq:globalD} and Lorentz charges \eqref{eq:globalY}, \eqref{eq:globalYbar} at $\mathscr{J}^+$, we note that $\mathcal{D}$ and $\mathcal{W}_{0}$, $\overline{\mathcal{W}}_{0}$ respectively take the form of generators of local dilations and Lorentz transformations.  In the context of asymptotic symmetries, $\mathcal{W}_{0}$ and $\overline{\mathcal{W}}_{0}$ can be identified with hard part of the superrotation charge \cite{Kapec:2014opa,Campiglia:2015kxa,Campiglia:2016efb,Conde:2016rom,Strominger:2017zoo} that generates the symmetry associated to the subleading soft graviton theorem \cite{Cachazo:2014fwa}, while $\mathcal{D}$ takes the form of a local ``superdilation" charge considered in \cite{Donnay:2020fof}.

Under translations, these operators transform as
\begin{equation} \label{level-0-translations}
    \begin{split}
        \delta_f \mathcal{D}& = f \mathcal{W}_{-1}
        ,\\
        \delta_f \mathcal{W}_0 &= \frac{1}{2} \partial_z \left(f \mathcal{W}_{-1}\right) + \partial_z f \mathcal{W}_{-1}.
    \end{split}
\end{equation}
Finally, these operators carry the appropriate scaling dimension to appear in the transformation of the leading operator $\mathcal{W}_{-1}$ under special conformal transformations parameterized by $g(z, \bar{z})$ as in \eqref{ckv-retarded}. We find
\begin{equation} \label{level--1-sct}
    \begin{split}
        \delta_g \mathcal{W}_{-1}&= 
         \partial_{\bar{z}} g \mathcal{W}_{0}
         +\partial_{\bar{z}} \left(g \mathcal{W}_{0}\right)
        + \partial_{z} g \overline{\mathcal{W}}_{0} 
        +\partial_{z} \left(g \overline{\mathcal{W}}_{0}\right) \\& \quad \quad 
        - \partial_z \partial_{\bar{z}}\left(g \mathcal{D}\right) 
        -\partial_z \left( \partial_{\bar{z}}g \mathcal{D}\right)
       - \partial_{\bar{z}}\left(\partial_zg \mathcal{D}\right)
       - \partial_z   \partial_{\bar{z}}g \mathcal{D} .
    \end{split}
\end{equation}
Notice that the individual \emph{modes} of $\mathcal{W}_0$, $\overline{\mathcal{W}}_0$, and $\mathcal{D}$ can be obtained by acting with special conformal transformations on the modes of $\mathcal{W}_{-1}$, similar to the analysis in \cite{Strominger:2026yrh}. However, the full operators $\mathcal{W}_0$, $\overline{\mathcal{W}}_0$, and $\mathcal{D}$ are \emph{not} conformal descendants of $\mathcal{W}_{-1}$. The modes of $\mathcal{D}$ are also related to the modes of $\mathcal{W}_{-1}$ by collinear ${\rm SL}(2)$ transformations \cite{Belin:2020lsr}. In addition, the operator $\mathcal{D}$ is related to $\mathcal{W}_{-1}$ by a weight-shifting operator in the embedding space \cite{Kologlu:2019mfz,Kologlu:2019bco}. Special conformal transformations of $\mathcal{D}$ and $\mathcal{W}_0$ involve light-ray operators of scaling dimension $\Delta=-1$ and are presented in the next subsection. 

\subsection{Light-ray Operators of Scaling Dimension $\Delta = -1$} \label{sec:m1lro}

At the next order in scaling dimension, we find universal light-ray operators that generalize beyond the light-ray operators considered in \cite{Cordova:2018ygx}.  Working modulo current conservation relations, the independent light-ray operators with scaling dimension $\Delta = -1$ are 
\begin{equation} \label{delta-1-basis}
    \begin{split}
        \Delta = -1: \quad \int du~ u^2T_{uu}^{(2)}, \quad 
            \int du~uT_{uz}^{(2) },\quad \int du~T_{zz}^{(2)}, \quad   \int du~ T_{z\bar{z}}^{(2)}, \quad \int du ~u T^{(3)},
    \end{split}
\end{equation}
plus the complex conjugate of the complex-valued generators. A systematic derivation of the basis at all orders in scaling dimension is presented in the following Subsection \ref{subsec:classification-all-orders}.

Only the first and last light-ray operators in \eqref{delta-1-basis},
\begin{equation}
    \begin{split}
        \mathcal{L}_{1}(z, \bar{z}) \equiv \int du~ u^2T_{uu}^{(2)},
        \quad \quad 
        \widehat{\mathcal{T}}_{1}(z, \bar{z}) \equiv \int du ~u T^{(3)}, 
    \end{split}
\end{equation}
transform as primaries on their own with the same left and right ${\rm SL}(2, \mathbb{C})$ weights:
\begin{equation}
    \begin{split}
         \mathcal{L}_{1}, \widehat{\mathcal{T}}_{1}: \quad \left(h, \bar{h}\right)  = \left(\frac{1}{2}, \frac{1}{2}\right). 
    \end{split}
\end{equation}
The remaining light-ray operators can be organized into the following ${\rm SL}(2, \mathbb{C})$ primaries together with their complex conjugates:
\begin{equation} \label{def-primaries-level-1}
    \begin{split}
        \mathcal{K}(z, \bar{z}) &\equiv\int du \left(u^2 \partial_z\partial_{\bar{z}}T_{uu}^{(2)} + T_{z\bar{z}}^{(2)} - u \partial_z T_{u\bar{z}}^{(2)} - u \partial_{\bar{z}}T_{uz}^{(2)}-\frac{1}{2}u \partial_z \partial_{\bar{z}}T^{(3)} \right)
        ,\\
        \mathcal{X}_{1}(z, \bar{z}) &\equiv \int du \left(u^2 \partial_z T_{uu}^{(2)} -  u T_{uz}^{(2)}- \frac{1}{2}u \partial_z T^{(3)}  \right)
        , \\
        \mathcal{W}_{1}(z, \bar{z}) &\equiv\frac{1}{8}\int du\left(  u^2 \partial_{z}^2 T_{uu}^{(2)} - 4 u \partial_z T_{uz}^{(2)} + 6 T_{zz}^{(2)}-2 u \partial_z^2 T^{(3)} \right)
        ,
    \end{split}
\end{equation} 
with ${\rm SL}(2, \mathbb{C})$ weights
\begin{equation}
    \begin{split} 
          \mathcal{K}: ~~  \left(h, \bar{h}\right)  = \left(\frac{3}{2}, \frac{3}{2}\right),  
          \quad \quad \quad 
         \mathcal{X}_{1} :~~ \left(h, \bar{h}\right)  = \left(\frac{3}{2}, \frac{1}{2}\right), 
         \quad\quad  \quad 
          \mathcal{W}_{1}:~~ \left(h, \bar{h}\right)  = \left(\frac{5}{2}, \frac{1}{2}\right). 
    \end{split}
\end{equation}
Note that appropriate expressions in conformal field theory respecting the stricter fall-off conditions \eqref{cft-fall-off} are readily obtained by setting $T^{(3)}=0$. In this case, $\mathcal{K}$ takes the form of a local version of the special conformal transformation generator \eqref{eq:globalSCT}.  At this order, we find additional light-ray operators $\mathcal{L}_1$, $\mathcal{X}_{1}$ and $\mathcal{W}_{1}$, which cannot be identified with local versions of the conformal generators. However, $\mathcal{W}_1$ can be identified as the hard part of the charge that generates the symmetry associated to the subsubleading soft graviton theorem \cite{Campiglia:2016jdj,Conde:2016rom,Campiglia:2016efb,Horn:2022acq}. In particular, we observe a pattern that the light-ray operators that are related to asymptotic symmetry generators of ${\rm w}_{1+\infty}$ are ${\rm SL}(2, \mathbb{C})$ primary light-ray operators of maximal ${\rm SL}(2, \mathbb{C})$ spin. 

Under translations parametrized by $f$ as in \eqref{ckv-retarded}, these operators transform as 
\begin{equation} \label{level-1-translations}
    \begin{split}
        \delta_f \mathcal{L}_{1}&= 2 f\mathcal{D}, \\
        \delta_f \widehat{\mathcal{T}}_{1}&=  0, \\
        \delta_f \mathcal{K}&=  \partial_z \partial_{\bar{z}} \left(f\mathcal{D}\right) 
    +  \partial_z \left(\partial_{\bar{z}}f  \mathcal{D}  + f \overline{\mathcal{W}}_0  \right)  
    +  \partial_{\bar{z}} \left(\partial_zf  \mathcal{D}  + f  \mathcal{W}_0 \right)    
    +   \partial_z\partial_{\bar{z}}f \mathcal{D}
        +\partial_{\bar{z}}f\mathcal{W}_0 + \partial_zf \overline{\mathcal{W}}_0 , \\
        \delta_f\mathcal{X}_{1}&=f\mathcal{W}_0+ \frac{3}{2}\partial_z \left(f   \mathcal{D}   \right)+   \frac{3}{2}\partial_z f  \mathcal{D} 
          ,\\
        \delta_f\mathcal{W}_{1}&=  \frac{1}{2}\partial_z \left(f \mathcal{W}_0\right)+ \frac{3}{2}\partial_z f \mathcal{W}_0 + \frac{3}{4}\partial_z^2 f \mathcal{D}
        .
    \end{split}
\end{equation}
Here we have allowed $f$ to be a general function of $z, \bar{z}$ rather than restricting to the more limited form in \eqref{conf-param}.  Interestingly, in the next section we find that the local light-ray commutation relations at these scaling dimensions are consistent with these ``supertranslation'' transformations.\footnote{At this order in scaling dimension, consistency with local light-ray commutation relations also extends to transformations under superrotations in which $Y^z$ is allowed to have arbitrary dependence on $z$.} Note that the previously-determined transformations under translations \eqref{level--1-translations} and \eqref{level-0-translations} are not modified for more general forms of $f$. 

Finally, we present the transformation of the light-ray operators with $\Delta = 0$ under special conformal transformations \eqref{ckv-retarded} parametrized by $g$, which we also allow to be a general function of $(z, \bar{z})$:
\begin{equation} \label{SCT-level-0}
    \begin{split}
        \delta_g \mathcal{D}&=
            -g \mathcal{K} 
            + 2 \partial_z \left(g \bar{\mathcal{X}}_{1}\right)
            + 2 \partial_{\bar{z}} \left(g  \mathcal{X}_{1}\right)   
            -3 \partial_z \partial_{\bar{z}} \left(g  \mathcal{L}_{1}  \right)   
        ,\\
        \delta_g \mathcal{W}_0 &= 
             \frac{1}{2}\partial_z \left(g \mathcal{K}\right)
            + \partial_z g \mathcal{K}
            + \frac{4}{3} \partial_{\bar{z}} \left(g \mathcal{W}_1\right)
            -\frac{2}{3} \partial_{z} \partial_{\bar{z}} \left(g \mathcal{X}_1\right)
            - \frac{4}{3} \partial_{\bar{z}} \left( \partial_z g \mathcal{X}_1\right) -\frac{1}{2} \partial_{\bar{z}}\left(\partial_z^2 g \mathcal{L}_1\right).
    \end{split}
\end{equation}
These expressions are obtained by working in a conformal field theory where trace terms are absent. Note that the previously-determined special conformal transformation \eqref{level--1-sct} is not modified for this more general form of $g$. As we will see in the next section, these ``super-special conformal transformations'' of $\mathcal{W}_{-1}$, $\mathcal{D}$, and $\mathcal{W}_0$ are consistent with the light-ray commutation relations in a conformal field theory. 

Finally, successive special conformal transformations of the ANEC operator give
\begin{equation}
    \begin{split}
        \delta_{g_2}& \delta_{g_1} \mathcal{W}_{-1}\\
            &= 2 \partial_z \partial_{\bar{z}} \left(g_1 g_2 \mathcal{K}\right)
                + \partial_z  \left( \partial_{\bar{z}} \left(g_1 g_2 \right)\mathcal{K}\right)
                    + \partial_{\bar{z}} \left( \partial_{z} \left(g_1 g_2 \right)\mathcal{K}\right)
                        + \left(\partial_zg_1 \partial_{\bar{z}} g_2+\partial_{\bar{z}} g_1 \partial_z g_2 \right) \mathcal{K}\\
        & \quad
            + \frac{4}{3} \partial_{\bar{z}}^2 \left(g_1 g_2 \mathcal{W}_1\right)
                -  \frac{4}{3} \partial_{\bar{z}}^2  g_1 g_2 \mathcal{W}_1
            + \frac{4}{3} \partial_{z}^2 \left(g_1 g_2 \overline{\mathcal{W}}_1\right)
                -  \frac{4}{3} \partial_{z}^2  g_1 g_2 \overline{\mathcal{W}}_1\\
        & \quad 
            -\frac{8}{3} \partial_z \partial_{\bar{z}}^2\left(g_1 g_2  \mathcal{X}_1\right)
                -\frac{4}{3} \partial_{\bar{z}}^2\left( \partial_z \left(g_1 g_2\right)  \mathcal{X}_1\right)
                    + \frac{8}{3}\partial_z \left(\partial_{\bar{z}}^2 g_1 g_2  \mathcal{X}_1\right)
                        + \frac{4}{3}  \partial_z \left(\partial_{\bar{z}}^2 g_1 g_2\right)  \mathcal{X}_1 
        \\
        & \quad 
            -\frac{8}{3} \partial_z^2 \partial_{\bar{z}}\left(g_1 g_2 \bar{\mathcal{X}}_1\right)
                -\frac{4}{3}\partial_z^2 \left( \partial_{\bar{z}}\left(g_1 g_2\right) \bar{\mathcal{X}}_1\right)
                    +\frac{8}{3}  \partial_{\bar{z}}\left(\partial_z^2g_1 g_2 \bar{\mathcal{X}}_1\right)
                        +\frac{4}{3} \partial_{\bar{z}}\left(\partial_z^2g_1 g_2\right) \bar{\mathcal{X}}_1
        \\
        & \quad
            + 3 \partial_z^2 \partial_{\bar{z}}^2 \left(g_1g_2 \mathcal{L}_1\right)
                -\frac{1}{2} \partial_z^2 \left( \left(6 \partial_{\bar{z}}^2 g_1 g_2 + g_1 \partial_{\bar{z}}^2 g_2\right)\mathcal{L}_1\right)
                -\frac{1}{2} \partial_{\bar{z}}^2 \left( \left(6 \partial_z^2 g_1 g_2 + g_1 \partial_z^2 g_2\right)\mathcal{L}_1\right)
        \\& \quad 
                +\frac{1}{2} \left( \partial_z^2 g_1 \partial_{\bar{z}}^2 g_2 +  \partial_{\bar{z}}^2g_1\partial_z^2  g_2 +6 \partial_z^2 \partial_{\bar{z}}^2g_1 g_2\right) \mathcal{L}_1.
    \end{split}
\end{equation}
As in prior discussions of individual single special conformal transformations, this transformation illustrates that the modes of $\mathcal{W}_1$ and $\overline{\mathcal{W}}_1$ can be constructed by acting with special conformal transformations on the modes of $\mathcal{W}_{-1}$, but the operator $\mathcal{W}_1$ is not simply a 4$d$ conformal descendant of $\mathcal{W}_{-1}$. 

\subsection{Classification at All Orders} \label{subsec:classification-all-orders}

In this subsection, we extend our classification of light-ray operators formed from null integrals of the stress tensor to all orders in scaling dimension.  In particular, we find analogs of the $\mathcal{L}_m$, $\widehat{\mathcal{T}}_m$, $\mathcal{X}_m$, $\mathcal{W}_m$ and $\mathcal{K}_m$ light-ray operators at all scaling dimensions $\Delta=-m$. We also find additional spin-2 ${\rm SL}(2, \mathbb{C})$ primaries $\mathcal{Y}_m$, higher-spin ${\rm SL}(2, \mathbb{C})$ primaries $\mathcal{V}_m$ and primaries constructed solely from the trace $\mathcal{T}_m$, which degenerate with the previously-found primaries or primary descendants at leading orders in scaling dimension. 

First we identify a minimal set of independent light-ray operators at fixed scaling dimension $\Delta = -m$.  Using \eqref{scaling-all}, we find an over-complete set is given by
\begin{equation}  
    \begin{split}
          \quad &\int du~ u^{m-n+3} T_{uu}^{(n)}, \quad
            \int du~ u^{m-n+2} T_{uz}^{(n)}, \quad 
                    \int du~ u^{m-n+1}T_{zz}^{(n)},  
        \\
        \quad &
            \int du~ u^{m-n+1} T_{rr}^{(n+2)}, \quad 
                \int du~ u^{m-n+1} T_{rz}^{(n+1)}, 
        \\
       & \int du~ u^{m-n+1} T_{ur}^{(n+2)}, \quad 
            \int du~ u^{m-n+2} T_{z\bar{z}}^{(n-1)},\quad + \quad \text{complex conjugates},
    \end{split}
\end{equation} 
for $n\geq2$, assuming the fall-off conditions in quantum field theory \eqref{qft-fall-off}. 

For ease of restriction to conformal field theory, we use light-ray operators involving the trace $T$ in place of light-ray operators involving $T_{z \bar{z}}$:
\begin{equation}
    \begin{split}
        \int du~ u^{m-n+1} T^{(n+2)} = -2 \int du~ u^{m-n+1}  \left(T_{ur}^{(n+2)}-T_{z\bar{z}}^{(n)} \right). 
    \end{split}
\end{equation}
Then, expanding the conservation equation   $\partial_\mu T^{\mu \nu}  = 0$, we find 
\begin{equation}  \label{current-conservation-expanded}
    \begin{split}
      - (n-2)  \int du ~u^{p} T_{uu}^{(n)} 
      & =   p\int du ~u^{p-1}   T_{ur}^{(n+1)} 
         +   \int du ~u^{p} \left( \partial_zT_{u\bar{z}}^{(n-1)}  +    \partial_{\bar{z}}T_{uz}^{(n-1)} \right)  , \\
     - (n-4)  \int du ~u^{p} T_{u r}^{(n)} & = 
            p\int du ~u^{p-1} T_{r r}^{(n+1)} 
         +    \int du ~u^{p} \left(\partial_z T_{ r\bar{z}}^{(n-1)} 
         +  \partial_{\bar{z}}T_{ rz}^{(n-1)} -   T^{(n)}  \right)  , \\
      - (n-2) \int du ~u^{p} T_{u z}^{(n)}
           &= 
            p\int du ~u^{p-1} T_{r z}^{(n+1)} 
            +  \int du ~u^{p} \left(\partial_zT_{ur}^{(n+1)} 
       +  \frac{1}{2} \partial_zT^{(n+1)} 
         +  \partial_{\bar{z}} T_{z z}^{(n-1)} \right)  ,
    \end{split}
\end{equation}
which enables us to eliminate light-ray operators involving $T_{uu}^{(n)}$, $T_{uz}^{(n)}$ and $T_{ur}^{(n+2)} $ for $n \geq3$.  We also find additional constraints from the second equation in \eqref{current-conservation-expanded} evaluated at $n=3,4$ and the third equation at $n=2$:
\begin{equation}  \label{current-conservation-qft-only}
    \begin{split}
    0 & = p\int du ~u^{p-1} T_{r r}^{(4)} -  \int du ~u^{p} T^{(3)} , \\ 
    0  & = p\int du ~u^{p-1} T_{r r}^{(5)} 
         +    \int du ~u^{p} \left(\partial_z T_{ r\bar{z}}^{(3)} 
         +  \partial_{\bar{z}}T_{ rz}^{(3)} -   T^{(4)}  \right)  , \\
      0 &= 
            p\int du ~u^{p-1} T_{r z}^{(3)} 
            + \frac{1}{2} \int du ~u^{p}  \partial_zT^{(3)}  .
    \end{split}
\end{equation}
This allows us to eliminate $T_{rr}^{(4)}$,  $T_{rr}^{(5)}$, $T_{rz}^{(3)}$, and $T_{r\bar{z}}^{(3)}$ in favor of $T^{(3)}$ and $T^{(4)}$. Thus, we find a minimal basis
\begin{equation} \label{basis}
    \begin{split}
        \quad &\int du~ u^{m+1} T_{uu}^{(2)}, \quad
            \int du~ u^{m} T_{uz}^{(2)}, \quad 
            \int du~ u^{m} T_{u\bar{z}}^{(2)}, \quad \int du~ u^{m-1} T_{z\bar{z}}^{(2)}, \\
        n\geq 2:
        \quad &\int du~ u^{m-n+1}T_{zz}^{(n)}, \quad
        \int du~ u^{m-n+1}T_{\bar{z}\bar{z}}^{(n)}, \quad
                 \int du~ u^{m-n+2} T^{(n+1 )} ,
        \\
        n\geq 3: \quad &
            \int du~ u^{m-n+1} T_{rr}^{(n+2)}, \quad 
                \int du~ u^{m-n+1} T_{rz}^{(n+1)}, \quad
                 \int du~ u^{m-n+1} T_{r\bar{z}}^{(n+1)},  
    \end{split}
\end{equation}
where the trace terms are absent in a conformal field theory. Note that we have chosen to use $T_{z\bar{z}}^{(2)}$ in place of $T_{ur}^{(4)}$.  When $p=0$, there is an additional constraint arising from the second equation in \eqref{current-conservation-qft-only}, which ultimately allows us to eliminate $T^{(4)}$ in favor of $ T^{(3)}$ at $m=1$:
\begin{equation} \label{basis-caveate}
    \begin{split}
        \int du ~ T^{(4)}&= -\partial_z\partial_{\bar{z}}\int du ~u T^{(3)}. 
    \end{split}
\end{equation}

As in the previous subsections, only two of the operators in the minimal basis \eqref{basis} transform as primaries on their own.  These generalize the operators $\mathcal{L}_1$ and $\widehat{\mathcal{T}}_1$ from the previous subsection and are accordingly labeled
\begin{equation} \label{def-l}
    \begin{split}
        \mathcal{L}_m(z, \bar{z})& = \int du~ u^{m+1} T_{uu}^{(2)}, 
        \quad\quad 
         \widehat{\mathcal{T}}_m(z, \bar{z}) = \int du~ u^{m} T^{(3)}.
    \end{split}
\end{equation}
Note that at $\Delta =0$, $\mathcal{L}_m$ coincides with the local dilation $\mathcal{D} = \mathcal{L}_{0}$.  The notation $\mathcal{L}_m$ is inspired by the existing literature on a Virasoro algebra generated by light-ray operators of that form \cite{Casini:2017roe,Besken:2020snx,Belin:2020lsr,Huang:2020ycs,Huang:2021hye}.  Note that \cite{Besken:2020snx,Belin:2020lsr} label their operators $L_{n} \equiv \int du~ u^{n+2} T_{uu}^{(2)}$ by their weight under a collinear subalgebra of the full $4d$ conformal symmetry so $\mathcal{L}_m = L_{m-1}$.  

The operators in \eqref{def-l} transform with left and right ${\rm SL}(2, \mathbb{C})$ weights
\begin{equation}
    \begin{split}
        \mathcal{L}_{m}, \widehat{\mathcal{T}}_{m}: \quad \left(h, \bar{h}\right)  = \left(\frac{2-m}{2}, \frac{2-m}{2}\right),
    \end{split}
\end{equation}
transform under translations as 
\begin{equation}
    \begin{split}
        \delta_f \mathcal{L}_m &=(m+1)f \mathcal{L}_{m-1},  \quad \quad \quad
        \delta_f \widehat{\mathcal{T}}_m = m f \widehat{\mathcal{T}}_{m-1},
    \end{split}
\end{equation} 
and transform under special conformal transformations as 
\begin{equation} \label{l-SCT}
    \begin{split}
        -\delta_g \mathcal{L}_m &=\frac{1}{(m-1)^2}  \Bigg[(m+1) g \mathcal{K}_{m+1}  -2 \partial_z  \left(g \bar{\mathcal{X}}_{m+1}\right)
        -2\partial_{\bar{z}} \left( g  \mathcal{X}_{m+1}\right)
        + 2 m\partial_z g \bar{\mathcal{X}}_{m+1}
        + 2 m\partial_{\bar{z}} g  \mathcal{X}_{m+1} \\& \quad \quad \quad  
           +(m-3) \left(  -\partial_z \partial_{\bar{z}} 
                 \left(g   \mathcal{L}_{m+1}  \right) 
              + m  \partial_{\bar{z}} \left(\partial_z  g  \mathcal{L}_{m+1}  \right)   
             +m  \partial_z \left(\partial_{\bar{z}} g   \mathcal{L}_{m+1} \right)  
          - m^2 \partial_z \partial_{\bar{z}}g   \mathcal{L}_{m+1}   \right) 
        \Bigg].
    \end{split}
\end{equation}
Note that \eqref{l-SCT} pertains specifically to light-ray operators in a conformal field theory where trace terms are absent. Here $\mathcal{X}_{m}$, $\bar{\mathcal{X}}_{m}$ and $\mathcal{K}_{m}$ generalize the light-ray operators from the previous subsections to lower scaling dimension and are defined later in the subsection in \eqref{def-x-and-y} and \eqref{def-K}. Since $\widehat{\mathcal{T}}_{m+1}$ is built entirely from the trace, we do not include its transformation under special conformal transformations.  

Next, we observe that $\mathcal{L}_m$ and $\widehat{\mathcal{T}}_{m}$ transform with the lightest ${\rm SL}(2, \mathbb{C})$ weights of any operator in the minimal basis \eqref{basis}. Following the pattern observed above, for the purpose of finding a ${\rm w}_{1+\infty}$ symmetry algebra, we are interested light-ray operators of higher ${\rm SL}(2, \mathbb{C})$-spin and with a light value of anti-chiral ${\rm SL}(2, \mathbb{C})$ weight $\bar{h} = \frac{2-m}{2}$. We will see in Section \ref{sec:wAlg} that such ${\rm SL}(2, \mathbb{C})$ primaries of maximal spin generate the wedge algebra of ${\rm w}_{1+\infty}$, while other primaries with this light value of $\bar{h}$ can appear in the commutator of two ${\rm w}_{1+\infty}$ generators (although only \emph{outside} of the wedge \eqref{eq:w-wedge-intro}). 

To classify the primaries with $\bar{h} = \frac{2-m}{2}$, we first note that there are four elements of the basis \eqref{basis} with ${\rm SL}(2, \mathbb{C})$ scaling dimension $\bar{h} = \frac{2-m}{2}$:
\begin{equation}
    \begin{split}
        \quad &\int du~ u^{m+1} T_{uu}^{(2)}, \quad
            \int du~ u^{m} T_{uz}^{(2)}, \quad 
           \int du~ u^{m-1}T_{zz}^{(2)}, \quad   \int du~ u^{m} T^{(3)}. 
    \end{split}
\end{equation}
Considering a general linear combination of these operators at the same ${\rm SL}(2, \mathbb{C})$ spin $h-\bar{h}=\ell$ and canceling as many terms proportional to $\partial_z^2Y^z$ in its transformation under $\delta_Y$ as possible uniquely specifies the coefficients of the linear combination to the following: 
\begin{equation}
    \begin{split}
        \mathcal{Q}_{m, \ell}(z, \bar{z}) &\equiv
             \int du~ u^{m-1} 
                \left(u^{ 2} \partial_z^{\ell} T_{uu}^{(2)} + \ell(m{-}\ell{-}1)u   \partial_z^{\ell-1} T_{uz}^{(2)}
         + \frac{1}{2}\ell(\ell{-}1)(m{-}\ell{-}1)(m{-}\ell{-}2)\partial_z^{\ell-2}T_{zz}^{(2)}
         \right.  \\& \quad  \quad \quad \quad \quad\quad \quad \quad
         \quad  \quad \quad \quad \quad\quad \quad \quad
         \quad  \quad \quad \quad \quad\quad \quad \quad\left.
         +  \frac{(\ell{-}1)(m{-}\ell{-}2)-m}{2m} u  \partial_z^{\ell} T^{(3)}\right),
    \end{split}
\end{equation}
which transforms under ${\rm SL}(2, \mathbb{C})$ as
\begin{equation}
    \begin{split}
         - \delta_Y\mathcal{Q}_{m, \ell} &= \left(Y^z \partial_z + \frac{2-m+2\ell}{2}\partial_zY^z \right)  \mathcal{Q}_{m, \ell}\\& \quad ~- \frac{\partial_z^2Y^z}{4}\ell(\ell-1)(\ell-2)(m-\ell-1)(m-\ell-2)(m-\ell-3)\int du~u^{m-1} \partial_z^{\ell-3}T_{zz}^{(2)}
        ,\\
        - \delta_{\bar{Y}}\mathcal{Q}_{m, \ell} &= \left(\bar{Y}^{\bar{z}} \partial_{\bar{z}} + \frac{2-m}{2}\partial_{\bar{z}}\bar{Y}^{\bar{z}} \right)\mathcal{Q}_{m, \ell}.
    \end{split}
\end{equation}
Here we have labeled $\mathcal{Q}_{m,\ell}$ by the scaling dimension $\Delta =-m$ and the ${\rm SL}(2, \mathbb{C})$ spin $h -\bar{h} = \ell$. Notice that $\mathcal{Q}_{m, \ell}$ is a primary when $\ell = 0,1,2$.  The $\ell=0$ primary is $\mathcal{L}_m =\mathcal{Q}_{m, 0}$.  The $\ell = 1$ and $\ell =2$ primaries generalize $\mathcal{X}_1$ and $\mathcal{W}_1$ to higher values of $m$:
\begin{equation} \label{def-x-and-y}
    \begin{split}
        \mathcal{X}_m (z, \bar{z}) & \equiv
             \int du~ u^{m-1} 
                \left(u^{ 2} \partial_z  T_{uu}^{(2)} +  (m-2)u T_{uz}^{(2)}
                -\frac{1}{2} u  \partial_z  T^{(3)}\right)
        ,\\
        \mathcal{Y}_m (z, \bar{z})& \equiv
         \int du~ u^{m-1} 
                \left(u^{ 2} \partial_z^{2} T_{uu}^{(2)} + 2(m-3)u   \partial_z  T_{uz}^{(2)}
         + (m-3)(m-4) T_{zz}^{(2)} 
         -  \frac{2}{m} u  \partial_z^{2} T^{(3)}\right).
    \end{split}
\end{equation}
Note that $\mathcal{X}_m$ is defined for $m \geq 0$ with $\mathcal{X}_0 \propto \mathcal{W}_0$, and  $\mathcal{Y}_m$ is defined for $m \geq 1$ with $\mathcal{Y}_1 \propto \mathcal{W}_1$. They transform with  ${\rm SL}(2, \mathbb{C})$ weights
\begin{equation}
    \begin{split}
        \mathcal{X}_m: \quad \left(h, \bar{h}\right) = \left(\frac{4-m}{2}, \frac{2-m}{2}\right), 
        \quad \quad \quad 
         \mathcal{Y}_m: \quad \left(h, \bar{h}\right) = \left(\frac{6-m}{2}, \frac{2-m}{2}\right),
    \end{split}
\end{equation}
and transform under translations as
\begin{equation}
    \begin{split}
         \delta_f  \mathcal{X}_m &=\frac{1}{m-3} \left( m(m-2) f \mathcal{X}_{m-1} + (m-2) \partial_z f \left(3 \mathcal{L}_{m-1}- \frac{m}{2}\widehat{\mathcal{T}}_{m-1}\right)- \partial_z \left(f \left(3 \mathcal{L}_{m-1}- \frac{m}{2} \widehat{\mathcal{T}}_{m-1}\right) \right) \right)
        ,\\
         \delta_f  \mathcal{Y}_m &=
        \frac{1}{m-5} \bigg( (m-1)(m-3)f \mathcal{Y}_{m-1} - 8 \partial_z \left(f \mathcal{X}_{m-1}\right)+ 8(m-4)  \partial_z f \mathcal{X}_{m-1}  \\& \quad \quad \quad \quad  \quad  - (m-5) \partial_z^2 f \left( (m-7) \mathcal{L}_{m-1}  -(m-2)(m-5)\widehat{\mathcal{T}}_{m-1}\right)\bigg).
    \end{split}
\end{equation}  
Due to the heavier right ${\rm SL}(2, \mathbb{C})$ weight $\bar{h}$ of $\mathcal{K}$, these operators mix under the action of special conformal transformations with operators with heavier $\bar{h}$ weight that fall outside of our classification scheme. Therefore we do not include general expressions for these transformations.\footnote{At $m=2$, $\mathcal{X}_m$ degenerates with the chiral primary descendant of $\mathcal{L}_m$, as does $\mathcal{Y}_m$ at $m=3$. The transformations $\delta_g \mathcal{X}_2$ and $\delta_g \mathcal{Y}_3$ simplify accordingly and can be deduced from \eqref{l-SCT}.}

The generators of ${\rm w}_{1+\infty}$ carry increasing ${\rm SL}(2, \mathbb{C})$ spin and are thus not contained in this class except for at low values of $m$, where for example $\mathcal{W}_{-1} = \mathcal{L}_{-1}$,  $\mathcal{W}_0 \propto \mathcal{X}_0$, and $\mathcal{W}_1  \propto \mathcal{Y}_1$.  Moreover, none of the other elements of the minimal basis \eqref{basis} carry low-enough $\bar{h}$ to be added to these operators to form additional primaries.  However, notice that for large enough $m \geq 2$, $\bar{h}$ becomes non-positive, meaning that these operators admit anti-chiral ${\rm SL}(2,\mathbb{C})$ primary descendants with $\bar{h} = \frac{m}{2}$.\footnote{We always use the term ``primary descendant" to refer to an ${\rm SL}(2,\mathbb{C})$ primary descendant, unlike the $4d$ conformal primary descendants discussed e.g. in \cite{Chang:2020qpj}.} Specifically, for $m \geq 2$,  $\partial_{\bar{z}}^{m-1}$ derivatives of a ``pre"-primary of weight $\bar{h}=\frac{2-m}{2}$ generates a primary descendant of weight $\bar{h}=\frac{m}{2}$, and this procedure can also be inverted.  At $\bar{h}=\frac{m}{2}$, there are additional available elements of the basis \eqref{basis} that can be added to form these primary descendants, and the associated ``pre"-primaries are thus candidates for the desired maximal-spin primaries.   We therefore now classify these primary descendants with $\bar{h}=\frac{m}{2}$. 

If the primary descendants involve $T_{uu}^{(2)}$, $T_{uz}^{(2)}$, and $T_{zz}^{(2)}$, then for the same reason as above, they must come in the combination
\begin{equation}
    \begin{split}
        \mathcal{O}_{m, \ell} &\equiv \partial_{\bar{z}}^{m-1} \mathcal{Q}_{m, \ell+m-1}\\
        &=\int du~ (u\partial_z \partial_{\bar{z}})^{m-1}\left(u^{ 2} \partial_z^{\ell} T_{uu}^{(2)} - \ell(m{+}\ell{-}1)u   \partial_z^{\ell-1} T_{uz}^{(2)}
         + \frac{1}{2}\ell(\ell{+}1)(m{+}\ell{-}1)(m{+}\ell{-}2)\partial_z^{\ell-2}T_{zz}^{(2)}
         \right.  \\& \quad  \quad \quad \quad \quad\quad \quad \quad
         \quad  \quad \quad \quad \quad\quad \quad \quad
         \quad  \quad \quad \quad \quad\quad \quad \quad\quad \quad \quad\left.
        -\frac{(\ell{+}1)(m{+}\ell{-}2)+m}{2m} u  \partial_z^{\ell} T^{(3)}\right),
    \end{split}
\end{equation}
where we have relabeled so that $\ell$ again labels the ${\rm SL}(2,\mathbb{C})$ spin $\ell = h- \bar{h}$. These transform under ${\rm SL}(2, \mathbb{C})$ as
\begin{equation} \label{sl2c-op}
    \begin{split}
        -\delta_Y\mathcal{O}_{m, \ell}
            &=  \left( Y^{z}  \partial_{z} +\frac{m+2\ell}{2} \partial_z Y^z\right)\mathcal{O}_{m, \ell} 
     \\& \quad \quad 
       +\frac{\ell(\ell+1)(\ell+2)}{4}(m+\ell-3)(m+\ell-2)(m+\ell-1) \partial_z^2 Y^z
           \int du~ (u \partial_z \partial_{\bar{z}})^{m-1} \partial_z^{\ell-3} T_{zz}^{(2)},
    \\
      -\delta_{\bar{Y}}  \mathcal{O}_{m, \ell}
            &= \left(\bar{Y}^{\bar{z}}\partial_{\bar{z}}+ \frac{m}{2} \partial_{\bar{z}}\bar{Y}^{\bar{z}}\right)    \mathcal{O}_{m, \ell}.
    \end{split}
\end{equation}
Notice that for the same reason as in the above analysis, we find primaries at $\ell = 0, -1, -2$.  These are the primary descendants of the primaries in \eqref{def-l} and \eqref{def-x-and-y}. 

Since we are interested in identifying additional primaries, we next consider all remaining elements constructed from the basis \eqref{basis} with the same left and right ${\rm SL}(2, \mathbb{C})$ weight as $\mathcal{O}_{m, \ell}$.  In particular, it is useful to organize the elements into the combinations
\begin{equation} \label{general-combo-fixed-sl2c}
    \begin{split}
       & R_{m,\ell, n}[a,b,c,d,e,f]\\
            & \equiv \int du~ (u\partial_z \partial_{\bar{z}})^{m-n+1}
        \left (a_n\partial_z^{\ell} T_{rr}^{(n+2)} 
            + b_n \partial_z^{\ell-1}T_{rz}^{(n+1)}
            + c_n  \partial_z^{\ell-2} T_{zz}^{(n)}
            + d_n u\partial_z^{\ell} T^{(n+1)}
            + e_n u \partial_z^{\ell+1} T_{r \bar{z}}^{(n)}\right) 
        \\& \quad \quad +f_n \int du~
            u^2 (u\partial_z \partial_{\bar{z}})^{m-n} \partial_z^{\ell +2} T_{\bar{z}\bar{z}}^{(n-1)},
    \end{split}
\end{equation}
for varying values of $n$. To find an ${\rm SL}(2,\mathbb{C})$ primary of scaling dimension $\Delta = -m$ and anti-chiral ${\rm SL}(2, \mathbb{C})$ weight $\bar{h} = \frac{m}{2}$, we begin with the most general operator that could appear, which is 
\begin{equation}
\mathcal{O}_{m,\ell} + \sum_n R_{m,\ell,n}[a_n,b_n,c_n,d_n,e_n,f_n] + g \int du~(u\partial_z \partial_{\bar{z}})^{m-1}  u  \partial_z^{\ell} T^{(3)},
\end{equation}
and has chiral ${\rm SL}(2, \mathbb{C})$ weight $h = \ell+\frac{m}{2}$.    The coefficients $a_n$, $b_n$, $c_n$, $d_n$, $e_n$, $f_n$, and $g$ are fixed by enforcing left and right primary conditions.  In Appendix \ref{app:primary-derivation}, we recursively solve for the coefficients $a_n$ through $f_n$ and find that at general $m$, the left and right primary conditions can only be simultaneously satisfied for $\ell =1$ and $\ell =2$.  The coefficient $g$ is fixed last to ensure that the entire operator transforms as an ${\rm SL}(2, \mathbb{C})$ primary. The corresponding operators are interpreted as the primary descendants of higher-spin light-ray operators $\mathcal{V}_{m}$ and $\mathcal{W}_m$, respectively, where their primary descendants take the form 
\begin{equation} \label{def-v}
    \begin{split}
        &\partial_{\bar{z}}^{m-1} \mathcal{V}_{m}(z, \bar{z})\\
            &\equiv\frac{1}{2^{m+1}(m+2) m!} \Bigg[\partial_{\bar{z}}\int du~ u  (u\partial_z \partial_{\bar{z}})^{m-2}   
            \left(u^{ 2}\partial_z^{ 2} T_{uu}^{(2)} -  m u  \partial_z  T_{uz}^{(2)}
        + m (m-1) T_{zz}^{(2)}+\frac{m-2}{2m} u \partial_z^2 T^{(3)}\right)
        \\& \quad - \sum_{n=3}^{m+1}\frac{m! (n-3)!}{(m-n+1)!}
            \partial_{\bar{z}}\int du~ u(u\partial_z \partial_{\bar{z}})^{m-n}  \\& \quad \quad \quad \quad \times 
          \left ( \partial_z^{2} T_{rr}^{(n+2)} 
            + (2n-m-4) \partial_z T_{rz}^{(n+1)}
            - (n-2)(m-n+1)  T_{zz}^{(n)}
            -\frac{m+2}{2m} u\partial_z^{2} T^{(n+1)} \right)\Bigg],
    \end{split}
\end{equation}
and 
\begin{equation} \label{def-w} 
    \begin{split}
       & \partial_{\bar{z}}^{m-1} \mathcal{W}_{m}(z, \bar{z})\\
            & \equiv \frac{1}{2^{m+1}(m+1)!} \Bigg[\int du~ (u\partial_z \partial_{\bar{z}})^{m-1} 
            \left(u^{ 2}\partial_z^{2} T_{uu}^{(2)} 
                - 2(m+1)u  \partial_z  T_{uz}^{(2)}
                +3m(m+1) T_{zz}^{(2)}-(m+1) u \partial_z^2 T^{(3)}\right)
            \\& \quad + \sum_{n=3}^{m+1} \frac{(m+1)!(n-3)!}{(m-n+1)!}
                \int du~ (u\partial_z \partial_{\bar{z}})^{m-n+1}
        \left ( \partial_z^{2} T_{rr}^{(n+2)} 
            +2(n-1)\partial_z T_{rz}^{(n+1)}
            +(n+1)(n-2)   T_{zz}^{(n)} \right)\Bigg].  
    \end{split} 
\end{equation}
The explicit expression for these primaries is one of the major results of this paper.  Note that to present a more symmetric formula, we have not strictly reduced to the minimal basis \eqref{basis}.  The normalizations are chosen to simplify expressions for the transformation of these operators under translations. 

In a conformal field theory, note that the trace terms in $\mathcal{V}_m$ and $\mathcal{W}_m$ are absent, as is $T_{rr}^{(5)}$.  We observe that this only slightly modifies the expression for $\mathcal{W}_m$, while the modification to $\mathcal{V}_m$ is more significant since it contains contributions from the trace $T^{(n)}$ for $3 \leq n \leq m+2$. Perhaps the more drastic modification to the expression for $\mathcal{V}_m$ is related to the massless spin-1 degree of freedom in conformal gravity \cite{Riegert:1984hf}. 

Our expressions \eqref{def-v} and \eqref{def-w} reveal that the primaries $\mathcal{W}_m$ and $\mathcal{V}_m$ involve non-local (on the celestial sphere) contributions from the stress tensor.  In Section \ref{sec:freefields}, we show that in free field theory, this non-locality in $z, \bar{z}$ can be traded for additional non-locality in $u$, when $T_{\mu\nu}$ is expressed in terms of a free scalar.  Despite this intricate locality structure, the $\mathcal{W}_m$ and $\mathcal{V}_m$ light-ray operators admit a number of simple properties. 

First, by construction, $\mathcal{W}_m$ and $\mathcal{V}_m$ transform as ${\rm SL}(2, \mathbb{C})$ primaries of weight
\begin{equation}
    \begin{split}
        \mathcal{V}_m: \quad \left(h, \bar{h}\right) = \left(\frac{m+2}{2}, \frac{2-m}{2}\right), \quad \quad \quad 
        \mathcal{W}_m: \quad \left(h, \bar{h}\right) = \left(\frac{m+4}{2}, \frac{2-m}{2}\right).
    \end{split}
\end{equation}
Hence, they carry increasing ${\rm SL}(2, \mathbb{C})$ spin $m$ and $m+1$ for $\mathcal{V}_m$ and $\mathcal{W}_m$, respectively.  In the next section, we show that the $\mathcal{W}_m$ primaries generate the wedge algebra of ${\rm w}_{1+\infty}$.  Recall from the introduction \eqref{eq:wloop-intro}, the generators of ${\rm w}_{1+\infty}$ are labeled by their left-${\rm SL}(2, \mathbb{C})$ weight $h$, so to reproduce that notation, we introduce a superscript notation
\begin{equation} \label{eq:superscript}
    \begin{split}
        \mathcal{W}^{\frac{m+4}{2}} \equiv \mathcal{W}_m,
    \end{split}
\end{equation}
where the superscript labels the left ${\rm SL}(2, \mathbb{C})$ weight $h = \frac{m+4}{2}$ and the subscript labels the scaling dimension $m = -\Delta$. In Section \ref{sec:1pt}, we identify a precise relationship between observables in a conformal field theory involving the $\mathcal{W}^p$ generators and the tower of soft graviton theorems in asymptotically flat spacetimes.  

The primaries $\mathcal{W}_m$ and $\mathcal{V}_m$ also transform simply under translations, 
\begin{equation} \label{eq:Wtranslation}
    \begin{split}
         \delta_f  \mathcal{W}_m 
        & =  \frac{1}{2}\partial_z \left( f \mathcal{W}_{m-1} \right) + \frac{m+2}{2} \partial_z f  \mathcal{W}_{m-1} , 
    \end{split}
\end{equation}
and 
\begin{equation}
    \begin{split}
        \delta_f  \mathcal{V}_m&=
            \frac{1}{2}\partial_z  \left(f  \mathcal{V}_{m-1}\right) 
            +\frac{m}{2}  \partial_zf  \mathcal{V}_{m-1}  
            +\frac{1}{2(m+1)(m+2)}  f  \mathcal{W}_{m-1}  \\& \quad \quad+\frac{1}{2}   \partial_z\partial_{\bar{z}}  \left(f \mathcal{T}_{m-1}\right)-\frac{m}{2}\partial_z \left(\partial_{\bar{z}} f \mathcal{T}_{m-1} \right)
            +\frac{m}{2}   \partial_{\bar{z}}  \left( \partial_z f \mathcal{T}_{m-1}\right)-\frac{m^2}{2}\partial_z \partial_{\bar{z}} f \mathcal{T}_{m-1} . 
    \end{split}
\end{equation}
Here we observe that the $\mathcal{W}_m$ family transforms into itself under translations.  This transformation is the non-gravitational analog of the recursion relation relating different ${\rm w}_{1+\infty}$ charge aspects in the covariant phase space formalism approach to asymptotic symmetries \cite{Freidel:2021ytz,Kmec:2024nmu,Cresto:2024fhd,Cresto:2024mne}.  Notice in particular that in this non-gravitational context, the non-linearities in gravitational degrees of freedom that proliferate for high-spin ${\rm w}_{1+\infty}$ generators are absent. On the other hand, the $\mathcal{V}_m$ family mixes with the $\mathcal{W}_m$ family, together with another family $\mathcal{T}_m$, which are primaries constructed entirely from the trace:
\begin{equation}
    \begin{split}
         \partial_{\bar{z}}^{m+1} \mathcal{T}_m(z, \bar{z}) \equiv \frac{1}{2^{m+2}(m+1)!}\sum_{n=0}^m \frac{n!(m-1)!}{(m-n)!} \int du~ (u \partial_z \partial_{\bar{z}})^{m-n} T^{(n+3)}. 
    \end{split}
\end{equation}
The primary $\mathcal{T}_m$ carries ${\rm SL}(2, \mathbb{C})$ weight
\begin{equation}
    \begin{split}
        \mathcal{T}_m: \quad \left(h, \bar{h}\right) = \left(\frac{m+2}{2}, -\frac{m}{2}\right).
    \end{split}
\end{equation}
Note that the right conformal weight $\bar{h}$ of $\mathcal{T}_m$ differs from that of $\mathcal{W}_m$ and $\mathcal{V}_m$, which is why it was not identified in our analysis above.  Finally, this new class of primaries also transforms only among itself under translations 
\begin{equation}
    \begin{split}
        \delta_f  \mathcal{T}_m = \frac{1}{2}\frac{m-1}{m+1} \left( \partial_z   \left(f  \mathcal{T}_{m-1}\right)+ m   \partial_zf  \mathcal{T}_{m-1} \right) .
    \end{split}
\end{equation}

Like the general $\mathcal{X}_m$ and $\mathcal{Y}_m$, these high-spin primaries do not transform only among the classified light $\bar{h}$ primaries under special conformal transformations, so we do not include expressions for these transformations.  Indeed, even at leading scaling dimension where some of these high-spin operators coincide with the low spin operators (e.g.~$\mathcal{W}_1 \propto \mathcal{Y}_1$ and $\mathcal{V}_2 \propto \mathcal{Y}_2$), these operators do not admit simple transformations in terms of classified primaries under special conformal transformations.  

To close, we note that the local special conformal generator $\mathcal{K}$ also admits an analog at every order in scaling dimension.  These take the form 
\begin{equation} \label{def-K}
    \begin{split}
        \mathcal{K}_{m}(z, \bar{z}) &\equiv
             \int du~u^{m-1} \left( u^2 \partial_z \partial_{\bar{z}} T_{uu}^{(2)}+ (m-2)^2T_{z\bar{z}}^{(2)}+(m-2)u\partial_{\bar{z}} T_{uz}^{(2)}+(m-2)u \partial_z T_{u \bar{z}}^{(2)} \right)\\& \quad \quad 
             - \int du~u^{m-1} \left( \frac{(m-2)^2}{2} T^{(4)}+  \frac{1}{m}u \partial_z \partial_{\bar{z}}T^{(3)}\right),
    \end{split}
\end{equation}
and carry ${\rm SL}(2, \mathbb{C})$ weight
\begin{equation}
    \begin{split}
        \mathcal{K}_{m}: \quad \left(h, \bar{h}\right) = \left(\frac{4-m}{2}, \frac{4-m}{2}\right). 
    \end{split}
\end{equation}
There is in fact another degenerate primary built entirely from the trace:
\begin{equation}
    \widehat{\mathcal{K}}_m(z, \bar{z})\equiv \int du~u^{m-1} \left((m-2) T^{(4)}-u \partial_z \partial_{\bar{z}}T^{(3)}\right),
\end{equation}
so any linear combination of $\mathcal{K}_m$ and $\widehat{\mathcal{K}}_m$ is an ${\rm SL}(2, \mathbb{C})$ primary of weight $\left(h, \bar{h}\right) = \left(\frac{4-m}{2}, \frac{4-m}{2}\right)$.  However, the combination in \eqref{def-K} is a natural definition since it is the combination that appears in transformation of the $\mathcal{L}_m$ primaries under special conformal transformations \eqref{l-SCT}.  Furthermore, defined in this way, $\mathcal{K}_m$ does not mix with $\widehat{\mathcal{K}}_m$ under translations: 
\begin{equation}
    \begin{split}
        &\delta_f \mathcal{K}_m \\&=\frac{(m-1)(m-2)^2}{(m-3)^2}f \mathcal{K}_{m-1} 
        -\frac{2(m-2)}{(m-3)^2} \left(\partial_{\bar{z}} \left(f \mathcal{X}_{m-1}\right)+\partial_{z} \left(f \bar{\mathcal{X}}_{m-1}\right) \right)+ \frac{2(m-2)^2}{(m-3)^2} \left(\partial_{\bar{z}} f \mathcal{X}_{m-1} +\partial_{z} f \bar{\mathcal{X}}_{m-1} \right)\\& \quad \quad
        -\frac{m-5}{(m-3)^2}   \left( \partial_z \partial_{\bar{z}} \left(f\mathcal{L}_{m-1}\right)- (m-2) \left( \partial_z \left(\partial_{\bar{z}}f\mathcal{L}_{m-1}\right)+\partial_{\bar{z}} \left(\partial_zf\mathcal{L}_{m-1}\right) \right) +  (m-2)^2 \partial_z \partial_{\bar{z}} f\mathcal{L}_{m-1} \right) \\& \quad \quad 
        - \frac{1}{(m-3)^2}  \left(\partial_z \partial_{\bar{z}} \left(f\widehat{\mathcal{T}}_{m-1}\right)
        -(m-2)\left( \partial_z \left(\partial_{\bar{z}}f\widehat{\mathcal{T}}_{m-1}\right)+\partial_{\bar{z}} \left(\partial_zf\widehat{\mathcal{T}}_{m-1}\right) \right) 
        +(m-2)^2\partial_z \partial_{\bar{z}} f\widehat{\mathcal{T}}_{m-1}\right) 
        ,
    \end{split}
\end{equation} 
and 
\begin{equation}
    \begin{split}
        \delta_f \widehat{\mathcal{K}}_m &= \frac{(m-2)(m-1)}{m-3} f \widehat{\mathcal{K}}_{m-1}+ \frac{2}{m-3}\partial_z \partial_{\bar{z}} \left( f\widehat{\mathcal{T}}_{m-1}\right)-2 \frac{m-2}{m-3}\partial_z \left( \partial_{\bar{z}}f \widehat{\mathcal{T}}_{m-1}\right)\\& \quad \quad
        -2\frac{m-2}{m-3}\partial_{\bar{z}}  \left(\partial_zf \widehat{\mathcal{T}}_{m-1}\right) + 2\frac{(m-2)^2}{m-3}\partial_z \partial_{\bar{z}}f\widehat{\mathcal{T}}_{m-1}.
    \end{split}
\end{equation}

The $\mathcal{K}_m$ primaries do not transform in a simple way under special conformal transformations for generic values of $m$.  However,  the local version of the special conformal transformation generator \eqref{def-primaries-level-1} $\mathcal{K}_1  = \mathcal{K}$, corresponding to $m = 1$, transforms as
\begin{equation}
    \begin{split}
         -\delta_g  \mathcal{K} &= -\frac{1}{2} \partial_z^2 \left(g \bar{\mathcal{Y}}_2\right)-\frac{1}{2} \partial_{\bar{z}}^2 \left(g \mathcal{Y}_2\right)
        +\frac{1}{2} \partial_z^2 g \bar{\mathcal{Y}}_2 
        +\frac{1}{2} \partial_{\bar{z}}^2 g \mathcal{Y}_2 
        + \partial_z^2  \partial_{\bar{z}}^2 \left(g \mathcal{L}_2\right)
        -\partial_z^2  \partial_{\bar{z}}^2  g \mathcal{L}_2 ,
    \end{split}
\end{equation}
where we have assumed the conformal field theory fall-off conditions \eqref{cft-fall-off} to eliminate contributions from the trace. 

\section{Stress-tensor Light-ray Operator Algebra} \label{sec:LRAtensor}

In this section, we specialize exclusively to conformal field theories and derive an algebra generated by light-ray operators built from the stress tensor. In the course of this section, we comment on corrections to this algebra that can arise in relativistic (but not necessarily conformal) quantum field theories.  

\subsection{General Constraints on Light-ray Operator Commutators}\label{subsec:general-constraints}

We begin with a general discussion of constraints on commutators of light-ray operators constructed from the stress tensor in a conformal field theory.  Determining the algebra amounts to constraining all possible terms that appear in the commutator of two stress-tensor light-ray operators.  As discussed in \cite{Besken:2020snx}, the commutation relation between two stress tensors can be derived from the operator product expansion. In conformal field theories in spacetime dimensions $d>2$ the operator product expansion between two stress tensors generically involves the neutral operators in the theory and as a result, the commutator between a pair of stress-tensor operators is a priori theory-dependent.  However, it was also observed in \cite{Besken:2020snx} that in conformal field theories in $d=4$ spacetime dimensions, if there are no light neutral scalars with $1 \leq \Delta \leq 2$, then the identity and the stress tensor are the only operators that can contribute to commutators between two stress tensors \emph{on the same light sheet}, and these contributions are local (i.e.~contact terms) in the transverse directions and along the null direction of the lightsheet.\footnote{Note that although commutation relations of local operators are operator statements, commutation relations of corresponding integrated operators, including light-ray operators, can depend in subtle ways on the convergence properties of the integrals in correlation functions. Sufficient conditions for convergence in four-point functions were formulated for example in \cite{Kologlu:2019bco} and discrepancies when evaluating integrated operator algebras in correlation functions were observed in \cite{Belin:2020lsr}.  }  We thus work under the assumption that there are no light neutral scalars with scaling dimension in the range $1 \leq \Delta \leq 2$,\footnote{Free scalar field theory in fact violates this assumption.  However, in the free scalar case, the deviations from the structure of this section are fairly mild and are presented in Section \ref{sec:freefields}. It would be interesting to study the general possible corrections from light neutral scalars to the algebra derived in this section. We also comment below on which parts of the algebra hold without this assumption.} and further focus solely on the contributions from the stress tensor and not the identity.\footnote{In a future investigation, it would be interesting to determine the vacuum contributions to our algebra, which can depend on the central charges of the theory. Note that these contributions appear divergent by  dimensional analysis.} As a direct result, the operators appearing in the commutator of two light-ray stress-tensor operators are restricted to those constructed from null integrals of the stress tensor, with delta-function support in the transverse directions. These are precisely the operators that were classified in the previous section.  

We introduce the notation  $O_i(z, \bar{z})$ to represent light-ray operators of the form \begin{equation} \label{eq:schematic}
O_i(z,\bar{z}) \equiv \int du~ u^{k_i} \partial_z^{m_i}\partial_{\bar{z}}^{\bar{m}_i} T_{\mu_i\nu_i}^{(n_i)}(u,z,\bar{z}). 
\end{equation}
The operators $O_i$ are of definite scaling dimension and, while they are \textit{not} always ${\rm SL}(2, \mathbb{C})$ primaries, they carry definite ${\rm SL}(2, \mathbb{C})$ weights $(h_i,\bar{h}_i)$.    Here $i$ labels a full set of indices that take values in the ranges $k_i \in \mathbb{Z}_{\geq 0}, m_i, \bar{m}_i \in \mathbb{Z}, n_i \in \mathbb{Z}_{\geq 2},$ and $\mu_i,\nu_i \in \{u,r,z,\bar{z}\}$. The minimal basis for such operators is presented in Section \ref{sec:classification}, specifically equations \eqref{anec}, \eqref{subleading-basis}, \eqref{delta-1-basis}, \eqref{basis}, and \eqref{basis-caveate}.

In the following, we determine commutators between light-ray operators that satisfy the ${\rm SL}(2, \mathbb{C})$ primary condition and thus involve linear combinations of the operators $O_i$, as demonstrated in Section \ref{sec:classification}. We distinguish ${\rm SL}(2, \mathbb{C})$ primary operators with the script notation $\mathcal{O}_i$. Using the operator product expansion as discussed above, our commutators of interest thus take the following form:  
\begin{equation} 
\label{eq:generalCommutator}
\begin{aligned}
  \left[ \mathcal{O}_1(z_1,\bar{z}_1),\mathcal{O}_2(z_2,\bar{z}_2)\right] &= \sum_{i}\sum_{\ell,\bar{\ell}}a^i_{\ell,\bar{\ell}}\partial_{z_1}^{\ell}\partial_{\bar{z}_1}^{\bar{\ell}} \left(\delta^{(2)}(z_{12}) O_i(z_2,\bar{z}_2) \right).
  \end{aligned}
\end{equation}

The coefficients $a^i_{\ell, \bar{\ell} }$ are imaginary and constrained by the action of the conserved global conformal charges \eqref{charges-null-infinity}, which must act equivalently on the left and right-hand sides. The action of a conserved charge on the left-hand side can be  evaluated using the Jacobi identity
\begin{equation}
\left[Q, \left[\mathcal{O}_1(z_1,\bar{z}_1), \mathcal{O}_2(z_2,\bar{z}_2)\right]\right] = \left[\mathcal{O}_1(z_1,\bar{z}_1), \left[Q, \mathcal{O}_2(z_2,\bar{z}_2)\right]\right]  + \left[ \left[Q, \mathcal{O}_1(z_1,\bar{z}_1) \right] , \mathcal{O}_2(z_2,\bar{z}_2)\right]. 
\end{equation}
For instance, using the Jacobi identity to enforce covariance under the action of the dilation charge $Q_D$ \eqref{eq:globalD} on a commutator of operators of definite scaling dimensions $\Delta_{\mathcal{O}_1}$ and $\Delta_{\mathcal{O}_2}$ implies that only operators with scaling dimension $\Delta_{\mathcal{O}_1} + \Delta_{\mathcal{O}_2}$ will appear on the right-hand side. Note that the light-ray operators in the previous section are labeled by their length scaling dimension (equivalently negative mass dimension), such that the labels simply add in the commutator. 

In the following sections, we derive commutators using global Poincar\'e constraints, particularly by enforcing covariance under the action of ${\rm SL}(2,\mathbb{C})$ and time translations $P_u$ (i.e.~\eqref{eq:globalP} with $f=1$). Constraints from the other translations are then  automatically satisfied.  The ${\rm SL}(2,\mathbb{C})$ action is relatively intricate and its constraints are discussed in detail in the following subsection. 

We now briefly comment on the extension of this analysis to non-conformal quantum field theories. Our primary use of conformal symmetry is in the argument forbidding the appearance of operators other than null integrals of the stress tensor in the commutator.  In a generic quantum field theory, there is no general analogous argument. However, it may be possible to constrain commutators of the subset of operators that take the form of local versions of the Poincar\'e charges, similar to the way that commutation relations of spacelike-separated (unintegrated) stress tensors are constrained by their relation to the generators of Poincar\'e \cite{Deser:1967zzf}.\footnote{If light-ray operators integrate to global Poincar\'e charges, then consistency with the charge action and charge algebra would forbid certain terms from appearing.  Such an argument would for instance rule out a bare $\delta^{(2)}(z-z')$ term in the commutator of two ANEC operators \eqref{eq:ANECcomm}, since it would spoil the global translation algebra. Similar arguments would also apply to certain commutators of light-ray operators constructed from a conserved current in a generic QFT.  However, the integration of light-ray operators to form global Poincar\'e charges can be subtle, as is discussed for example in \cite{Hartman:2023ccw}. We thank Greg Mathys, Prahar Mitra, and Shu-Heng Shao for discussion and sharing unpublished notes on this subtlety. In a CFT setting, which is assumed in this section, none of the results in this work rely on integrating light-ray operators to form global charges.}  Conformal symmetry is also implicit in the arguments based on scaling dimension, but they may potentially generalize to generic quantum field theories since the dimension of the stress tensor is protected (although in principle the null integration kernel could run).  Finally, the remaining part of the analysis relies solely on Poincar\'e symmetry and thus extends to generic relativistic quantum field theories.  Even though our commutators in a conformal field theory are determined from Poincar\'e constraints and dilation symmetry alone, they respect the action of special conformal transformations, which relates linear combinations of commutators of ${\rm SL}(2,\mathbb{C})$ primaries.   

\subsection{${\rm SL}(2,\mathbb{C})$ Constraints} \label{subsec:sl2c-constraints}

Here, we derive constraints on commutators of the form \eqref{eq:generalCommutator} from the action of global ${\rm SL}(2,\mathbb{C})$ charges $Q_Y$ \eqref{eq:globalY} and $Q_{\bar{Y}}$  \eqref{eq:globalYbar}. 

\subsubsection{Basis of Primaries} \label{subsub:SL2cprimaries}

For simplicity, we begin with the case in which the right-hand side of \eqref{eq:generalCommutator} involves only ${\rm SL}(2,\mathbb{C})$ primaries and their descendants:
\begin{equation} 
\label{eq:generalCommutatorPrimaries}
\begin{aligned}
  \left[ \mathcal{O}_1(z_1,\bar{z}_1),\mathcal{O}_2(z_2,\bar{z}_2)\right] &= \sum_{P}\sum_{\ell,\bar{\ell}}a^P_{\ell,\bar{\ell}}\partial_{z_1}^{\ell}\partial_{\bar{z}_1}^{\bar{\ell}} \left(\delta^{(2)}(z_{12}) \partial_{z_2}^{\ell_P - \ell}\partial_{\bar{z}_2}^{\bar{\ell}_P - \bar{\ell}}\mathcal{O}_P(z_2,\bar{z}_2) \right) , 
  \end{aligned}
\end{equation} 
where the index $P$ now sums over primaries $\mathcal{O}_P$ of weight $(h_P, \bar{h}_P)$. As demonstrated in Sections \ref{sec:CSlro} and \ref{sec:m1lro},  there is a one-to-one map between the minimal basis and ${\rm SL}(2,\mathbb{C})$-primary light-ray operators at bulk dimension $\Delta \geq -1$.  For commutators with operators of these heavier scaling dimensions appearing on the right-hand side, which include many of our examples, the representations \eqref{eq:generalCommutatorPrimaries} and \eqref{eq:generalCommutator} are equivalent. The general case will be discussed further below. 

We consider the action of $Q_Y$ on  \eqref{eq:generalCommutatorPrimaries}. The constraints from $Q_{\bar{Y}}$ factorize and can be derived analogously. Recall that a primary light-ray operator at null infinity $\mathcal{O}_P$ of weight $(h_P,\bar{h}_P)$ transforms under the chiral half of ${\rm SL}(2, \mathbb{C})$ as 
\begin{equation}
    \begin{split}
        \left[Q_Y, \mathcal{O}_{P}(z, \bar{z})\right] &= -i \left(Y^z \partial_z + h_P \partial_z Y^z \right) \mathcal{O}_P(z, \bar{z}). 
    \end{split}
\end{equation}
Consider the ${\rm SL}(2,\mathbb{C})$ transformation on the left-hand side of \eqref{eq:generalCommutatorPrimaries}: 
\begin{equation}
  \begin{aligned}
  &\left[ Q_Y, \left[ \mathcal{O}_1(z_1,\bar{z}_1),\mathcal{O}_2(z_2,\bar{z}_2)\right] \right] =  \left[ \left[ Q_Y,\mathcal{O}_1(z_1,\bar{z}_1) \right],\mathcal{O}_2(z_2,\bar{z}_2)\right] +  \left[ \mathcal{O}_1(z_1,\bar{z}_1), \left[ Q_Y,\mathcal{O}_2(z_2,\bar{z}_2)\right] \right] \\
  &\qquad = -i \left(Y^{z_1}\partial_{z_1}  + h_1 \partial_{z_1}Y^{z_1}+ Y^{z_2}\partial_{z_2}  + h_2 \partial_{z_2}Y^{z_2}\right)  \left[ \mathcal{O}_1(z_1,\bar{z}_1),\mathcal{O}_2(z_2,\bar{z}_2)\right].
  \end{aligned}
\end{equation} 
Then using \eqref{eq:generalCommutatorPrimaries} to replace the commutator $\left[ \mathcal{O}_1(z_1,\bar{z}_1),\mathcal{O}_2(z_2,\bar{z}_2)\right]$ in the last line and simplifying gives 
\begin{equation} \label{eq:LHSexpanded}
  \begin{aligned}
    &\left[ Q_Y, \left[ \mathcal{O}_1(z_1,\bar{z}_1),\mathcal{O}_2(z_2,\bar{z}_2)\right] \right] \\
    &= -i\sum_{P,\ell, \bar{\ell}}a^P_{\ell,\bar{\ell}}\Bigg(\partial_{z_1}^{\ell}\partial_{\bar{z}_1}^{\bar{\ell}} \left(\delta^{(2)}(z_{12}) \left[ Y^{z_2}\partial_{z_2} + (h_1 {+} h_2 {-} (\ell{+}1))\partial_{z_2}Y^{z_2} \right]\partial_{z_2}^{\ell_P - \ell}\partial_{\bar{z}_2}^{\bar{\ell}_P - \bar{\ell}}\mathcal{O}_P(z_2,\bar{z}_2) \right) \\
    & \qquad \qquad \qquad \qquad + \partial_{z_1}^{\ell-1}\partial_{\bar{z}_1}^{\bar{\ell}} \left(\delta^{(2)}(z_{12}) \partial_{z_2}^2 Y^{z_2} \left(\frac{\ell(\ell+1)}{2} - h_1 \ell\right) \partial_{z_2}^{\ell_P - \ell}\partial_{\bar{z}_2}^{\bar{\ell}_P - \bar{\ell}}\mathcal{O}_P(z_2,\bar{z}_2)\right) \Bigg). \\
  \end{aligned}
\end{equation}
Here we have used that $Y$ is quadratic in $z$, which follows from \eqref{conf-param} and implies the following identity for any operator $O_i$:
\begin{equation}
  \begin{aligned}
    \partial_{z_1}Y^{z_1} \partial_{z_1}^{\ell} \left(\delta^{(2)}(z_{12}) O_i(z_2,\bar{z}_2) \right) &= \partial_{z_1}^{\ell} \left(\delta^{(2)}(z_{12}) \partial_{z_2}Y^{z_2} O_i(z_2,\bar{z}_2) \right) - \ell \partial_{z_1}^{\ell-1} \left(\delta^{(2)}(z_{12}) \partial_{z_2}^2Y^{z_2} O_i(z_2,\bar{z}_2) \right) \\
    Y^{z_1} \partial_{z_1}^{\ell+1} \left(\delta^{(2)}(z_{12}) O_i(z_2,\bar{z}_2) \right) &= \partial_{z_1}^{\ell+1} \left(\delta^{(2)}(z_{12}) Y^{z_2} O_i(z_2,\bar{z}_2) \right) - (\ell+1) \partial_{z_1}^{\ell} \left(\delta^{(2)}(z_{12}) \partial_{z_2}Y^{z_2} O_i(z_2,\bar{z}_2) \right) \\
    &\qquad + \frac{\ell(\ell+1)}{2}\partial_{z_1}^{\ell-1} \left(\delta^{(2)}(z_{12}) \partial_{z_2}^2Y^{z_2} O_i(z_2,\bar{z}_2) \right).
  \end{aligned}
\end{equation}
Now \eqref{eq:LHSexpanded} must match the action of ${\rm SL}(2,\mathbb{C})$ on the right-hand side of \eqref{eq:generalCommutatorPrimaries}, which is 
\begin{equation}
  \begin{aligned}
    &\sum_{P,\ell, \bar{\ell}}a^P_{\ell,\bar{\ell}}\partial_{z_1}^{\ell}\partial_{\bar{z}_1}^{\bar{\ell}} \left(\delta^{(2)}(z_{12}) \left[Q_Y,\partial_{z_2}^{\ell_P - \ell}\partial_{\bar{z}_2}^{\bar{\ell}_P - \bar{\ell}}\mathcal{O}_P(z_2,\bar{z}_2)\right] \right) \\
    &= -i \sum_{P,\ell, \bar{\ell}}a^P_{\ell,\bar{\ell}}\partial_{z_1}^{\ell}\partial_{\bar{z}_1}^{\bar{\ell}} \Bigg(\delta^{(2)}(z_{12}) \left[ Y^{z_2}\partial_{z_2} {+} \left(h_P + \ell_P - \ell\right)\partial_{z_2} Y^{z_2} \right] \partial_{z_2}^{\ell_P - \ell}\partial_{\bar{z}_2}^{\bar{\ell}_P - \bar{\ell}}\mathcal{O}_P(z_2,\bar{z}_2)  \\
    &\qquad \qquad \qquad \qquad \quad  + \delta^{(2)}(z_{12})  \partial_{z_2}^2Y^{z_2} (\ell_P-\ell) \left(h_P + \frac{\ell_P-\ell-1}{2}\right) \partial_{z_2}^{\ell_P - \ell-1}\partial_{\bar{z}_2}^{\bar{\ell}_P - \bar{\ell}}\mathcal{O}_P(z_2,\bar{z}_2) \Bigg) ,
  \end{aligned}
\end{equation}
and follows from the general transformation of ${\rm SL}(2,\mathbb{C})$ descendants \eqref{eq:transSLdesc}. 

Comparing the coefficients of $\partial_{z_2}Y^{z_2}$ at each order on the left and right-hand sides gives a constraint on $\ell_P$. Together with the analogous anti-holomorphic constraint, it implies
\begin{equation} \label{eq:SLweightsPrimary}
\ell_P = h_1 + h_2 - h_P-1, \qquad \bar{\ell}_P = \bar{h}_1 + \bar{h}_2 - \bar{h}_P -1. 
\end{equation}
Notice that this is consistent with assigning the delta function in transverse coordinates in \eqref{eq:generalCommutatorPrimaries} ${\rm SL}(2,\mathbb{C})$ weights $(1,1)$. 

Next, re-indexing and comparing the coefficients of $\partial_{z_2}^2Y^{z_2}$  at each order gives a recursion relation for the coefficients.  Similarly, one can find a recursion relation involving the anti-holomorphic weights and derivatives. Explicitly, the two recursion relations are 
\begin{equation}
\begin{aligned}
\label{eq:SLrecursionPrimaries}
  a^P_{\ell, \bar{\ell}}(\ell_P - \ell)\left(h_P + \frac{\ell_P-\ell-1}{2}\right) &= a^{P}_{\ell+1,\bar{\ell}} (\ell +1)\left(\frac{\ell+2}{2} - h_1 \right),  \\
  a^P_{\ell, \bar{\ell}}(\bar{\ell}_P - \bar{\ell})\left(\bar{h}_P + \frac{\bar{\ell}_P-\bar{\ell}-1}{2}\right) &= a^{P}_{\ell,\bar{\ell}+1} (\bar{\ell} +1)\left(\frac{\bar{\ell}+2}{2} - \bar{h}_1 \right),
  \end{aligned}
\end{equation}
which are solved by 
\begin{equation} \label{eq:SLrecursionPrimarySola}
\begin{aligned}
a^P_{\ell, \bar{\ell}} &= a^P_{\ell_P, \bar{\ell}_P} \frac{\Gamma(2h_P) \Gamma(\ell_P + 1)\Gamma(\ell_P-2h_1 + 2)}{\Gamma(\ell_P-\ell + 2h_P)\Gamma(\ell+1)\Gamma(\ell-2h_1 + 2)\Gamma(\ell_P-\ell +1)} \\
&\qquad \qquad \times \frac{\Gamma(2\bar{h}_P) \Gamma(\bar{\ell}_P + 1)\Gamma(\bar{\ell}_P-2\bar{h}_1 + 2)}{\Gamma(\bar{\ell}_P-\bar{\ell} + 2\bar{h}_P)\Gamma(\bar{\ell}+1)\Gamma(\bar{\ell}-2\bar{h}_1 + 2)\Gamma(\bar{\ell}_P-\bar{\ell} +1)}. 
\end{aligned}
\end{equation}
Thus, when the right-hand side involves only primaries and descendants, the coefficients of descendants are fully fixed by the coefficient of the primary as expected. 

\subsubsection{General Basis} \label{sec:GenBasis}

Now we consider constraints on the more general commutator \eqref{eq:generalCommutator}, where the two operators on the left-hand side are ${\rm SL}(2, \mathbb{C})$ primaries, but the operators $O_i$ on the right-hand side take the general form \eqref{eq:schematic}.  Operators of this form are not generally ${\rm SL}(2, \mathbb{C})$ primaries or descendants, and are relevant for commutators of light-ray operators of overall weight $m=- \Delta \geq 2$. 

A general light-ray operator \eqref{eq:schematic} at null infinity $O_i$ of weight $(h_i,\bar{h}_i)$  transforms under (a chiral half of) ${\rm SL}(2, \mathbb{C})$ as 
\begin{equation}
    \begin{split}
        \left[Q_Y, O_{i}(z, \bar{z})\right] &= -i \left(Y^z \partial_z + h_i \partial_z Y^z \right) O_i(z, \bar{z}) -  i\alpha_i^{i'} \partial_{z}^2Y^{z} O_{i'}(z,\bar{z}),
    \end{split}
\end{equation}
where the $i'$ index is summed and $O_{i'}$ are operators of ${\rm SL}(2,\mathbb{C})$ weight $(h_i-1,\bar{h}_i)$. The coefficients $\alpha_i^{i'}$ describe the mixing of $O_{i}$ with $O_{i'}$ under chiral ${\rm SL}(2,\mathbb{C})$ transformations. The action of $Q_{\bar{Y}}$ takes an analogous form:
\begin{equation}
    \begin{split}
        \left[Q_{\bar{Y}}, O_{i}(z, \bar{z})\right] &= -i \left(\bar{Y}^{\bar{z}} \partial_{\bar{z}} + \bar{h}_i \partial_{\bar{z}} \bar{Y}^{\bar{z}} \right) O_i(z, \bar{z}) -  i\bar{\alpha}_i^{i'} \partial_{\bar{z}}^2\bar{Y}^{\bar{z}} O_{i'}(z,\bar{z}),
    \end{split}
\end{equation}
where again the $i'$ index is summed. Here the operators $O_{i'}$ carry ${\rm SL}(2,\mathbb{C})$ weight $(h_i,\bar{h}_i-1)$ and mix according to the coefficients $\bar{\alpha}_i^{i'}$, which are independent from  $\alpha_i^{i'}$.

As before, the action of $Q_Y$ on the left-hand side of \eqref{eq:generalCommutator}, which consists of a commutator between two primaries, can be simplified with the Jacobi identity as follows:
\begin{equation}
  \begin{aligned}
    &\left[ Q_Y, \left[ \mathcal{O}_1(z_1,\bar{z}_1),\mathcal{O}_2(z_2,\bar{z}_2)\right] \right] \\
    &\qquad \qquad = -i\sum_{\ell, \bar{\ell}}a^i_{\ell,\bar{\ell}}\Bigg(\partial_{z_1}^{\ell}\partial_{\bar{z}_1}^{\bar{\ell}} \left(\delta^{(2)}(z_{12}) \left[ Y^{z_2}\partial_{z_2} + (h_1 {+} h_2 {-} (\ell{+}1))\partial_{z_2}Y^{z_2}\right]O_i(z_2,\bar{z}_2) \right) \\
    & \qquad \qquad \qquad \qquad \qquad + \partial_{z_1}^{\ell-1}\partial_{\bar{z}_1}^{\bar{\ell}} \left(\delta^{(2)}(z_{12}) \partial_{z_2}^2 Y^{z_2} \left(\frac{\ell(\ell+1)}{2} - h_1 \ell\right) O_i(z_2,\bar{z}_2) \right), \\
  \end{aligned}
\end{equation}
while the action of ${\rm SL}(2,\mathbb{C})$ on the right-hand side now takes the modified form 
\begin{equation}
  \begin{aligned}
    &\sum_{\ell, \bar{\ell}}a^i_{\ell,\bar{\ell}}\partial_{z_1}^{\ell}\partial_{\bar{z}_1}^{\bar{\ell}} \left(\delta^{(2)}(z_{12}) \left[Q_Y,O_i(z_2,\bar{z}_2)\right] \right) \\
    &\qquad = -i \sum_{\ell, \bar{\ell}}a^i_{\ell,\bar{\ell}}\partial_{z_1}^{\ell}\partial_{\bar{z}_1}^{\bar{\ell}} \Bigg(\delta^{(2)}(z_{12})  \left(  \left[ Y^{z_2}\partial_{z_2} {+} h_{i}\partial_{z_2} Y^{z_2} \right] O_i(z_2,\bar{z}_2) + \alpha_{i}^{i'}\partial_{z_2}^2Y^{z_2} O_{i'}(z_2,\bar{z}_2)  \right) \Bigg). \\
  \end{aligned}
\end{equation}
Comparing the left and right-hand sides, and combining with the results from analogous transformations under $Q_{\bar{Y}}$, we find 
\begin{equation} \label{eq:RHweightsNew}
  \begin{aligned}
    h_{i} &= h_1 + h_2 - (\ell+1), \qquad \bar{h}_{i} = \bar{h}_1 + \bar{h}_2 - (\bar{\ell} + 1),
  \end{aligned}
\end{equation} 
and the recursion relations
\begin{equation} 
\begin{aligned} \label{eq:SLrecursion}
  a^i_{\ell, \bar{\ell}}\alpha_i^{ i'} &= a^{i'}_{\ell+1,\bar{\ell}} (\ell +1)\left(\frac{\ell+2}{2} - h_1 \right), \\
  a^i_{\ell, \bar{\ell}}\bar{\alpha}_i^{i'} &= a^{i'}_{\ell,\bar{\ell}+1} (\bar{\ell} +1)\left(\frac{\bar{\ell}+2}{2} - \bar{h}_1 \right), \\
  \end{aligned}
\end{equation}
where $i$ is summed. From the holomorphic recursion relation, we observe that if $a^i_{\ell, \bar{\ell}}\alpha_i^{ i'} = 0$, $h_1 \neq \frac{\ell+2}{2}$, and $\ell \neq -1$,  then $a^{i'}_{\ell+1,\bar{\ell}} = 0$. If $a^i_{\ell, \bar{\ell}}\alpha_i^{ i'} = 0$, and  $h_1 = \frac{\ell+2}{2}$ or $\ell =-1$, the higher coefficient $a^{i'}_{\ell+1,\bar{\ell}}$ is independent from the lower set of $a^i_{\ell, \bar{\ell}}$ coefficients. 

\subsection{Extended BMS Algebra} \label{subsec:bms-commutators}

In this subsection, we re-derive the commutators of stress-tensor light-ray operators that were previously determined in \cite{Cordova:2018ygx,Kologlu:2019bco,Besken:2020snx}, but employ a more general analysis that extends to a larger class of stress-tensor light-ray operators. Beginning with the general form \eqref{eq:generalCommutator}, our strategy is to further constrain the right-hand side by enforcing covariance under dilation symmetry and Lorentz transformations, realized as ${\rm SL}(2,\mathbb{C})$, thus reducing the commutator to a few undetermined coefficients associated to the definite-scaling-dimension ${\rm SL}(2,\mathbb{C})$ primaries classified in the previous section.  These remaining coefficients can either be determined by enforcing translation covariance, which relates undetermined commutators to previously-determined commutators, or by direct calculation via the method in \cite{Besken:2020snx}. In a later subsection this logic will be used to determine the commutation of a generic pair of $\mathcal{W}_m$ generators via an induction proof.

First, recall at leading (i.e.~heaviest) order in scaling dimension $\Delta = 1$, the unique stress-tensor light-ray operator is the ANEC operator $\mathcal{W}_{-1}$ \eqref{anec}.  Since the commutator between two ANEC operators can only involve a stress-tensor light-ray operator of scaling dimension $\Delta = 2$, the commutator must vanish
\begin{equation} \label{eq:ANECcomm}
    \begin{split}
    \left[\mathcal{W}_{-1}(z, \bar{z}), \mathcal{W}_{-1}(z', \bar{z}') \right] =0,
    \end{split}
\end{equation}
because there is no such $\Delta = 2$ operator. This commutator has been derived from several different approaches \cite{Casini:2017roe,Cordova:2018ygx,Kologlu:2019bco,Besken:2020snx}. In \cite{Kologlu:2019bco} it was shown to hold generally in CFT correlation functions (without the assumption of no light neutral scalars).  

Next, we consider commutators also involving operators at bulk dimension $\Delta = 0$, namely $\mathcal{D}$ defined in \eqref{eq:Dlocal} and $\mathcal{W}_0$, $\mathcal{\overline{W}}_0$, defined in \eqref{eq:W0local}. The commutators of $\mathcal{W}_{-1}$ with these operators can be derived by direct integration of commutators derived from the stress-tensor OPE, as in \cite{Besken:2020snx}. The results are
\begin{equation}\label{eq:Wm1W0}
\left[\mathcal{W}_{-1}(z,\bar{z}), \mathcal{W}_{0}(z',\bar{z}')\right] =   i \left[  \frac{1}{2}\partial_{z'} - \partial_z\right] \left(\delta^{(2)}(z-z') \mathcal{W}_{-1}(z',\bar{z}')\right),
\end{equation}
and 
\begin{equation} \label{eq:Wm1D}
\left[\mathcal{W}_{-1}(z,\bar{z}), \mathcal{D}(z',\bar{z}')\right] = i  \delta^{(2)}(z-z') \mathcal{W}_{-1}(z',\bar{z}'). 
\end{equation}
As discussed in \cite{Cordova:2018ygx}, these results are consistent with the action of the global translation, Lorentz transformation, and dilation charges upon integration in the transverse directions. In fact, the above form is uniquely consistent with the full supertranslation action \eqref{level-0-translations} in which $f(z,\bar{z})$ is an arbitrary function.

Next, we consider the commutator of $\mathcal{W}_0$ with itself, which, given  $\left[\mathcal{W}_{-1}, \mathcal{W}_0\right]$, can be entirely constrained by ${\rm SL}(2,\mathbb{C})$ and the Jacobi identity with $P_u$, the translation generator \eqref{eq:globalP} with $f=1$. The $\left[\mathcal{W}_{0}, \mathcal{W}_0\right]$ commutator has overall dimension $\Delta = 0$ and the primaries $\mathcal{W}_0$, $\overline{\mathcal{W}}_0$, $\mathcal{D}$ provide a general basis for the right-hand side, thus allowing the use of the ${\rm SL}(2, \mathbb{C})$ constraints in Subsection \ref{subsub:SL2cprimaries}. 

First, we consider the action of $P_u$ on the commutator, which will allow us to rule out the appearance of inverse transverse derivatives on the right-hand side. Using the Jacobi identity, 
\begin{equation} \label{eq:W0W0Jacobi}
  \begin{aligned}
    \left[ P_u, \left[\mathcal{W}_{0}(z,\bar{z}), \mathcal{W}_0(z',\bar{z}')\right] \right] &= \left[\left[  P_u, \mathcal{W}_{0}(z,\bar{z}) \right], \mathcal{W}_0(z',\bar{z}') \right] +  \left[\mathcal{W}_0(z,\bar{z}), \left[ P_u, \mathcal{W}_0(z',\bar{z}')\right] \right] \\
    &= \frac{i}{2} \partial_z \left[\mathcal{W}_{-1}(z,\bar{z}), \mathcal{W}_0(z',\bar{z}')\right] + \frac{i}{2} \partial_{z'} \left[\mathcal{W}_{0}(z,\bar{z}), \mathcal{W}_{-1}(z',\bar{z}')\right] \\
    &= \frac{i^2}{2} \left[  \delta^{(2)}(z-z')\partial_{z'}^2\mathcal{W}_{-1}(z',\bar{z}') - 2\partial_z \left( \delta^{(2)}(z-z') \partial_{z'}\mathcal{W}_{-1}(z',\bar{z}')\right) \right], 
  \end{aligned}
\end{equation}
we see that the only operators that can appear on the right-hand-side of $\left[\mathcal{W}_{0}, \mathcal{W}_0\right]$ are those that satisfy $\left[P_u,\mathcal{O}\right] \propto \partial_z\mathcal{W}_{-1}$ or $ \partial_z^2\mathcal{W}_{-1}$, or are in the kernel of $P_u$. Given the basis at $\Delta = 0$, the operators satisfying these conditions are $\mathcal{W}_0, \partial_z\mathcal{W}_0, \partial_z\mathcal{D},$ and $\partial_z^2 \mathcal{D}$, because $P_u$ acts according to  \eqref{level-0-translations}  evaluated at $f=1$:
\begin{equation} \label{eq:PuEx}
\left[P_u, \mathcal{D}\right] = i \mathcal{W}_{-1}, \qquad \left[P_u, \mathcal{W}_0\right] = \frac{i}{2} \partial_z\mathcal{W}_{-1}. 
\end{equation}
Note that the conjugate of the second equation rules out the presence of $\overline{\mathcal{W}}_0$, which has $\ell_P=2$, $\bar{\ell}_P = -1$, and thus would always involve inverse $\partial_{\bar{z}}$ derivatives if it were to appear on the right-hand side of \eqref{eq:generalCommutatorPrimaries}.  

$\mathcal{D}$ has $\ell_P = 2$,  $\bar{\ell}_P = 0$ and $\mathcal{W}_0$ has $\ell_P=1$, $\bar{\ell}_P=0$, so the commutator is thus fixed by the ${\rm SL}(2, \mathbb{C})$ recursion relations \eqref{eq:SLrecursionPrimarySola} up to two undetermined imaginary coefficients, $c$ and $d$:
\begin{equation}
\begin{aligned}
\big[ \mathcal{W}_0&(z,\bar{z}), \mathcal{W}_0(z',\bar{z}') \big] \\
&= c \left[\delta^{(2)}(z-z')\partial_{z'}\mathcal{W}_{0}(z',\bar{z}') - 2 \partial_z\left(\delta^{(2)}(z-z')\mathcal{W}_{0}(z',\bar{z}')\right)\right] \\
&\quad + d \left[ \frac{1}{3}\delta^{(2)}(z - z')\partial_{z'}^2\mathcal{D}(z',\bar{z}') -  \partial_z\left(\delta^{(2)}(z-z')\partial_{z'}\mathcal{D}(z',\bar{z}')\right) + \partial_z^2\left(\delta^{(2)}(z-z')\mathcal{D}(z',\bar{z}')\right)\right].
\end{aligned}
\end{equation}
These coefficients are fixed by covariance of the commutator under the action of $P_u$.  First, \eqref{eq:PuEx} sets $d=0$ because a nonzero $d$ coefficient implies the appearance of a  $\partial_z^2 \left(\delta^{(2)}(z-z') \mathcal{W}_{-1} \right)$  term in contradiction with the form of \eqref{eq:W0W0Jacobi}.  Then, we compare 
\begin{equation}
\begin{aligned}
\left[P_u, \left[ \mathcal{W}_0(z,\bar{z}), \mathcal{W}_0(z',\bar{z}') \right] \right] 
&=  \frac{i}{2} c \left[\delta^{(2)}(z-z')\partial_{z'}^2\mathcal{W}_{-1}(z',\bar{z}') - 2 \partial_z\left(\delta^{(2)}(z-z')\partial_{z'}\mathcal{W}_{-1}(z',\bar{z}')\right)\right] \\
\end{aligned}
\end{equation}
with \eqref{eq:W0W0Jacobi} to determine that $c=i$. For the purpose of the ${\rm w}_{1+\infty}$ algebra discussion below, it is convenient to present the result for $\left[\mathcal{W}_{0}, \mathcal{W}_0\right]$ in the equivalent form
\begin{equation} \label{eq:W0W0}
\left[ \mathcal{W}_0(z,\bar{z}), \mathcal{W}_0(z',\bar{z}') \right] = i \left[\partial_{z'} - \partial_z \right]\left( \delta^{(2)}(z-z') \mathcal{W}_0(z',\bar{z}')\right). 
\end{equation}
Similarly, we can compute the commutators of all of the light-ray operators of bulk dimensions $\Delta = 1,0$, entirely from \eqref{eq:Wm1W0}, its conjugate, and \eqref{eq:Wm1D} using ${\rm SL}(2,\mathbb{C})$ constraints and the Jacobi identity with $P_u$. We summarize the results here: 
\begin{equation} \label{eq:CordovaShaoComms}
\begin{aligned}
\left[\mathcal{W}_{-1}(z,\bar{z}), \mathcal{W}_{-1}(z',\bar{z}')\right]  &= 0, \\
\left[\mathcal{W}_{-1}(z,\bar{z}), \mathcal{W}_{0}(z',\bar{z}')\right] &=   i \left[  \frac{1}{2}\partial_{z'} - \partial_z\right] \left(\delta^{(2)}(z-z') \mathcal{W}_{-1} \right), \\
\left[\mathcal{W}_{-1}(z,\bar{z}), \mathcal{D}(z',\bar{z}')\right] &= i \delta^{(2)}(z-z') \mathcal{W}_{-1} , \\
\left[ \mathcal{D}(z,\bar{z}), \mathcal{W}_0(z',\bar{z}') \right] &= - i\partial_z \left( \delta^{(2)}(z-z') \mathcal{D} \right), \\
\left[ \mathcal{D}(z,\bar{z}), \mathcal{D}(z',\bar{z}') \right] &= 0 ,\\
\left[ \mathcal{W}_0(z,\bar{z}), \mathcal{W}_0(z',\bar{z}') \right] &= i \left[\partial_{z'} - \partial_z \right]\left( \delta^{(2)}(z-z') \mathcal{W}_0 \right) ,\\
\left[ \mathcal{W}_0(z,\bar{z}), \overline{\mathcal{W}}_0(z',\bar{z}') \right] &= i \partial_{z'}\left( \delta^{(2)}(z-z') \overline{\mathcal{W}}_0 \right) - i \partial_{\bar{z}} \left( \delta^{(2)}(z-z') \mathcal{W}_0 \right).
\end{aligned}
\end{equation} 
These reproduce the results of \cite{Casini:2017roe,Cordova:2018ygx,Kologlu:2019bco,Besken:2020snx,Belin:2020lsr},  and also are consistent with the corresponding global charge action and algebra upon integrating in the transverse directions.

\subsection{Local Conformal Algebra} \label{sec:LocalConfAlg}

The commutators of all light-ray operators of bulk dimension $\Delta =1,0$ can be supplemented by their commutation relations with the $\Delta = -1$ operator $\mathcal{K}$ to give a local version of the global conformal algebra. As before, all of these commutators are fully determined by ${\rm SL}(2,\mathbb{C})$ constraints and the Jacobi identity with $P_u$. Together with \eqref{eq:CordovaShaoComms}, this algebra includes
\begin{equation}
\begin{aligned}
    \left[\mathcal{W}_{-1}(z,\bar{z}), \mathcal{K}(z', \bar{z}')\right]
    & = i \left[\partial_{z'} -\partial_z\right] \left[\partial_{\bar{z}'} -\partial_{\bar{z}}\right] \left(\delta^{(2)}(z-z')  \mathcal{D}     \right) \\
    & \quad   + i \left[\partial_{z'} -\partial_z\right]\left(\delta^{(2)}(z-z') \overline{\mathcal{W}}_0  \right)  + i  \left[\partial_{\bar{z}'} -\partial_{\bar{z}}\right]\left(\delta^{(2)}(z-z')  \mathcal{W}_0  \right), \\
    \left[\mathcal{D}(z,\bar{z}), \mathcal{K}(z',\bar{z}')\right]
    &=i   \delta^{(2)}(z-z')\mathcal{K} -2 i   \partial_z \left(\delta^{(2)}(z-z') \bar{\mathcal{X}}_1\right) -2 i  \partial_{\bar{z}} \left(\delta^{(2)}(z-z')  \mathcal{X}_1\right) \\
    & \quad \quad  + 3 i \partial_z \partial_{\bar{z}} \left(\delta^{(2)}(z-z')  \mathcal{L}_1 \right), \\
    \left[\mathcal{W}_0(z,\bar{z}),\mathcal{K}(z',\bar{z}')\right] &= i \left[ \partial_{z'} - \frac{1}{2}\partial_z\right]\left(\delta^{(2)}(z-z') \mathcal{K}  \right) -i  \frac{4}{3} \partial_{\bar{z}} \left(\delta^{(2)}(z-z') \mathcal{W}_1  \right)\\
    & \quad   -i \frac{4}{3}\left[ \partial_{z'} - \frac{1}{2}\partial_z\right]\partial_{\bar{z}}\left(\delta^{(2)}(z-z') \mathcal{X}_1  \right) + \frac{i}{2} \partial_{z'}^2 \partial_{\bar{z}} \left(\delta^{(2)}(z-z')  \mathcal{L}_1 \right), \\
    \left[\mathcal{K}(z, \bar{z}), \mathcal{K}(z', \bar{z}')\right] &=   \frac{i}{2} \left[\partial_{z'}^2-\partial_{z}^2\right] \left(\delta^{(2)}(z-z') \bar{\mathcal{Y}}_2 \right) +  \frac{i}{2} \left[\partial_{\bar{z}'}^2-\partial_{\bar{z}}^2\right] \left(\delta^{(2)}(z-z') \mathcal{Y}_2 \right) \\
    & \quad \quad -i \left[\partial_{z'}^2\partial_{\bar{z}'}^2-\partial_{z}^2\partial_{\bar{z}}^2\right]  \left(\delta^{(2)}(z-z') \mathcal{L}_2 \right). 
\end{aligned}
\end{equation}
These commutators can be integrated and reproduce the global conformal transformations of local operators presented in the previous subsections, including the transformation under dilations and Lorentz transformations, as well as \eqref{level--1-translations}, \eqref{level-0-translations}, \eqref{level--1-sct}, \eqref{level-1-translations}, \eqref{SCT-level-0}, and the global charge algebra \eqref{eq:globalg}.  Notice in particular that light-ray operators such as $\mathcal{W}_1$, $ \mathcal{X}_1$ and $\mathcal{L}_2$ are present in the local conformal algebra and are projected out by the appropriate integration to form the global conformal generators. A similar fact was observed and exploited in the analyses of \cite{Sheta:2025oep,Strominger:2026yrh}. 

\subsection{${\rm w}_{1+\infty}$ Algebra} \label{sec:wAlg}

Here we prove that modes of the  $\mathcal{W}_m = \mathcal{W}^{\frac{m+4}{2}}$ operators defined in Section \ref{subsec:classification-all-orders} obey the wedge subalgebra of the ${\rm w}_{1+\infty}$ algebra. More precisely, we prove that 
\begin{equation} \label{eq:InductiveAssumption}
  \begin{aligned}
\left[\mathcal{W}^p(z,\bar{z}), \mathcal{W}^q(z',\bar{z}')\right] &= i \left[(p-1) \partial_{z'} - (q-1)\partial_z\right] \delta^{(2)}(z-z') \mathcal{W}^{p+q-2}(z',\bar{z}') \\
  &\ + \sum_{\ell = 2p-1} a_{\ell}^{i}\partial_z^{\ell}\left(\delta^{(2)}(z-z') O_i(z',\bar{z}') \right) + \sum_{k = 2q-1} b_k^j\partial_{z'}^k\left(\delta^{(2)}(z-z') O_j(z',\bar{z}') \right) ,
  \end{aligned}
\end{equation}
where contributions from the identity are neglected and the terms on the second line do not contribute to the wedge subalgebra. In particular, the wedge modes are constructed in the following way: 
\begin{equation} \label{eq:Wmodes}
{\rm w}^{p}_{m,\bar{m}} \equiv \int  d^2z~ z^{p+ m - 1} \bar{z}^{2-p+\bar{m}} \mathcal{W}^{p}(z,\bar{z}), \qquad  1-p \leq m \leq p-1.
\end{equation}
Integrating \eqref{eq:InductiveAssumption} thus projects out the terms in the second line and reproduces the wedge subalgebra of the loop algebra of ${\rm w}_{1+\infty}$: 
\begin{equation} \label{eq:sec4loop}
\left[{\rm w}^{p}_{m,\bar{m}}, {\rm w}^{q}_{n,\bar{n}} \right] = i \left[m(q-1) - n(p-1)\right] {\rm w}^{p+q-2}_{m+n,\bar{m}+\bar{n}}. 
\end{equation}

\subsubsection{Proof}

Our proof proceeds by induction in $p+q$ with a base case of $p+q=\frac{11}{2}$, which corresponds to $\Delta = -3$ on the right-hand side of \eqref{eq:InductiveAssumption}. The commutators with $p+q \leq \frac{11}{2}$ can be directly calculated  using Poincar\'e constraints order-by-order from the lowest nontrivial case $p+q=\frac{7}{2}$.  They respect the form of the inductive assumption \eqref{eq:InductiveAssumption} and are presented in Appendix \ref{app:Wderivation}.  Note that an inductive proof could have been formulated with a base case of  $p+q=\frac{7}{2}$, but as explained below, the inductive argument takes a generic form once $p+q = \frac{11}{2}$. The base case $p+q = \frac{11}{2}$ consists of the commutators $\left[\mathcal{W}^{\frac{5}{2}} , \mathcal{W}^{3} \right]$ in \eqref{eq:W1W2}, $\left[\mathcal{W}^2 ,\mathcal W^{\frac{7}{2}} \right]$ in \eqref{eq:W0W3}, and $\left[\mathcal{W}^{\frac{3}{2}} , \mathcal{W}^{4} \right]$ in \eqref{eq:Wm1W4}.

Now we perform the inductive step.  Assuming that \eqref{eq:InductiveAssumption} holds for $p+q = x-\frac{1}{2}$ for some $x \geq 6$, our goal is to show that it holds for $p+q=x$. From our discussion in Subsection \ref{subsec:general-constraints}, the commutator at $p+q=x$ takes the general form \eqref{eq:generalCommutator}
\begin{equation}
    \begin{split}
        \left[ \mathcal{W}^p(z,\bar{z}),\mathcal{W}^q(z',\bar{z}')\right]
            & = \sum_{\ell, \bar{\ell} }a^i_{\ell,\bar{\ell}}\partial_{z}^{\ell}\partial_{\bar{z}}^{\bar{\ell}} \left(\delta^{(2)}(z-z') O_i(z',\bar{z}') \right).
    \end{split}
\end{equation}
Here and through the remainder of this section, there is an implicit sum over the index $i$ that is suppressed for notational simplicity. As before, covariance under dilations restricts the operators on the right-hand side to have scaling dimension $\Delta = 8-2x$.  For the purpose of the proof, it is helpful to iteratively use the identity
\begin{equation} \label{delta-function-identity}
    \begin{split}
        \partial_z \left(\delta^{(2)}(z-z') ~\cdot\right)
        =\delta^{(2)}(z-z') \partial_{z'} (\cdot) - \partial_{z'} \left(\delta^{(2)}(z-z') ~\cdot\right)
     \end{split}
\end{equation}
to re-express the right-hand side in the more symmetric form
\begin{equation} \label{wp-wq-modified-ansatz}
  \begin{aligned}
    \left[\mathcal{W}^p(z,\bar{z}), \mathcal{W}^q(z',\bar{z}')\right]&=  i \left[(p-1) \partial_{z'} - (q-1)\partial_z\right] \delta^{(2)}(z-z') \mathcal{W}^{p+q-2}(z',\bar{z}') \\
    & \  + \sum_{ \ell, \bar{\ell}}a^i_{\ell,\bar{\ell}}\partial_{z}^{\ell}\partial_{\bar{z}}^{\bar{\ell}} \left(\delta^{(2)}(z-z') O_i(z',\bar{z}') \right) + \sum_{ k, \bar{k}}b^j_{k,\bar{k}}\partial_{z'}^{k}\partial_{\bar{z}'}^{\bar{k}} \left(\delta^{(2)}(z-z') O_j(z',\bar{z}') \right). 
  \end{aligned}
\end{equation}
Without loss of generality, we have additionally added and subtracted a contribution from $\mathcal{W}^{p+q-2}$ and redefined the coefficients $a^i_{\ell,\bar{\ell}}$ and $b^j_{k,\bar{k}}$ to absorb the subtraction. 

Next, we use Poincar\'e symmetry to rule out terms not of the form \eqref{eq:InductiveAssumption}.  First, we consider constraints from  translations $P_u$, where using \eqref{eq:Wtranslation} with $f=1$, we have
\begin{equation}
\left[P_u, \mathcal{W}^{p}(z,\bar{z})\right] = \frac{i}{2} \partial_z\mathcal{W}^{p-\frac{1}{2}}(z,\bar{z}). 
\end{equation}
Then, acting on the left-hand side of \eqref{wp-wq-modified-ansatz} and simplifying with the Jacobi identity, we find
\begin{equation} \label{eq:PuInductiveAssumption}
  \begin{aligned}
    \left[ P_u,  \left[\mathcal{W}^p(z,\bar{z}), \mathcal{W}^q(z',\bar{z}')\right] \right] 
    &= \frac{i}{2} \left( \partial_{z'}\left[\mathcal{W}^p(z,\bar{z}), \mathcal{W}^{q-\frac{1}{2}}(z',\bar{z}')\right] + \partial_z \left[\mathcal{W}^{p-\frac{1}{2}}(z,\bar{z}), \mathcal{W}^q(z',\bar{z}')\right]\right) \\
    &= \frac{i^2}{2} \left[(p-1)\partial_{z'} - (q-1)\partial_z\right]  \delta^{(2)}(z-z')\partial_{z'}\mathcal{W}^{p+q-\frac{5}{2}}(z',\bar{z}') + X, \\
  \end{aligned}
\end{equation}
where, using ${a^i_{\ell},b^j_{k}}$ to denote coefficients in $\left[\mathcal{W}^p,\mathcal{W}^{q-\frac{1}{2}}\right]$ and ${c^{i'}_{\ell},d^{j'}_{k}}$ in $\left[\mathcal{W}^{p-\frac{1}{2}},\mathcal{W}^{q}\right]$, the remaining term $X$ is given by 
\begin{equation} \label{eq:Xterms}
  \begin{aligned}
    X &= \frac{i}{2} \partial_{z'} \left[\sum_{\ell = 2p-1} a^i_{\ell}\partial_z^{\ell}\left(\delta^{(2)}(z-z')O_i(z',\bar{z}')\right) + \sum_{k = 2q-2} b_k^j\partial_{z'}^k\left(\delta^{(2)}(z-z') O_j(z',\bar{z}') \right)  \right] \\
    &\ \ + \frac{i}{2} \partial_z \left[\sum_{\ell = 2p-2} c^{i'}_{\ell}\partial_z^{\ell}\left(\delta^{(2)}(z-z')O_{i'}(z',\bar{z}')\right) + \sum_{k = 2q-1} d_{k}^{j'}\partial_{z'}^{k}\left(\delta^{(2)}(z-z') O_{j'}(z',\bar{z}') \right)  \right] \\
    &= \frac{i}{2} \sum_{\ell = 2p-1} \left(\partial_z^{\ell}\left(\delta^{(2)}(z{-}z')\left[a^i_{\ell}\partial_{z'}O_i(z',\bar{z}') + c_{\ell-1}^{i'} O_{i'}(z',\bar{z}')\right]\right) - \partial_z^{\ell+1}\left(\delta^{(2)}(z{-}z')a^i_{\ell}O_i(z',\bar{z}')\right)\right) \\
    & + \frac{i}{2} \sum_{k = 2q-1} \left(\partial_{z'}^{k}\left(\delta^{(2)}(z{-}z')\left[d^{j'}_{k}\partial_{z'}O_{j'}(z',\bar{z}') + b_{k-1}^{j} O_{j}(z',\bar{z}')\right]\right) - \partial_{z'}^{k+1}\left(\delta^{(2)}(z{-}z')d^{j'}_{k}O_{j'}(z',\bar{z}')\right)\right).
  \end{aligned}
\end{equation}
Here we have assumed that $p,q \neq \frac{3}{2}$, so that $\mathcal{W}^{p-\frac{1}{2}}$ and $\mathcal{W}^{q-\frac{1}{2}}$ are both non-trivial operators.\footnote{When $p=\frac{3}{2}$ and $q > \frac{3}{2}$, $\mathcal{W}^{\frac{3}{2}}$ is annihilated by $P_u$ and we instead find 
\begin{equation*}
  \begin{aligned}
  &\left[ P_u,  \left[\mathcal{W}^{\frac{3}{2}}(z,\bar{z}), \mathcal{W}^q(z',\bar{z}')\right] \right]  \\    &= \frac{i}{2} \Bigg( i\left[ \frac{1}{2} \partial_{z'} - \left(q-1\right) \partial_z \right] \delta^{(2)}(z-z')\partial_{z'}\mathcal{W}^{q-1}(z',\bar{z}') + i(q-1)\partial_z^2 \left(\delta^{(2)}(z-z')\mathcal{W}^{q-1}(z',\bar{z}')\right) \\
    &\qquad + \sum_{\ell=2} a_{\ell}^i\left[\partial_z^{\ell}\left(\delta^{(2)}(z{-}z')\partial_{z'}O_i(z',\bar{z}')\right) - \partial_z^{\ell+1}\left(\delta^{(2)}(z{-}z')O_i(z',\bar{z}')\right) \right] + \sum_{k = 2q-1} b_{k-1}^j\partial_{z'}^k\left(\delta^{(2)}(z{-}z') O_j(z',\bar{z}') \right)\Bigg).
  \end{aligned}
\end{equation*}
Note that in this case an additional $\mathcal{W}$ term appears outside the $p=\frac{3}{2}$ wedge.} 

Next, we act on the right-hand side of \eqref{wp-wq-modified-ansatz} with $P_u$ and recover the first term in \eqref{eq:PuInductiveAssumption} as well as additional terms:
\begin{equation}
  \begin{aligned}
    \left[ P_u,  \left[\mathcal{W}^p(z,\bar{z}), \mathcal{W}^q(z',\bar{z}')\right] \right] &= \frac{i^2}{2}\left( (p{-}1)\partial_{z'} - (q{-}1)\partial_z\right) \delta^{(2)}(z-z')\partial_{z'}\mathcal{W}^{p+q-\frac{5}{2}}(z',\bar{z}')  \\
    &\qquad  + \sum_{\ell, \bar{\ell}}a^i_{\ell,\bar{\ell}}\partial_{z}^{\ell}\partial_{\bar{z}}^{\bar{\ell}} \left(\delta^{(2)}(z-z') \left[P_u,O_i(z',\bar{z}')\right] \right) \\
     &\qquad + \sum_{k, \bar{k}}b^j_{k,\bar{k}}\partial_{z'}^{k}\partial_{\bar{z}'}^{\bar{k}} \left(\delta^{(2)}(z-z') \left[P_u,O_j(z',\bar{z}')\right] \right).
  \end{aligned}
\end{equation}
Since the action of translations is local in the transverse coordinates, the operators appearing in $\left[P_u, \mathcal{O}_i\right]$ and $\left[P_u, \mathcal{O}_j\right]$ will not decrease the powers of $\ell, \bar{\ell}, k, \bar{k}$  appearing in the sums.  Thus comparing with \eqref{eq:Xterms}, we find that for operators $O_i$, $O_j$ in \eqref{wp-wq-modified-ansatz} that are \emph{not} annihilated by $P_u$, $a^i_{\ell, \bar{\ell}}$ and $b^j_{k, \bar{k}}$ are only non-zero when $\bar{\ell} = \bar{k} = 0$ (i.e.~no anti-holomorphic derivatives) and $\ell \geq 2p-1$ or $k \geq 2q-1$.  Thus, apart from additional terms in the kernel of $P_u$, which will be considered below, the general commutator at $p+q =x$ \eqref{wp-wq-modified-ansatz} is constrained to take the desired form \eqref{eq:InductiveAssumption}.

\subsubsection*{Proof that the Kernel of $P_u$ Respects \eqref{eq:InductiveAssumption}} 

We now prove that any operator in the kernel of $P_u$ respects the form of the commutator \eqref{eq:InductiveAssumption} at $p+q=x$. Specifically, there are no further contributions to \eqref{wp-wq-modified-ansatz} from the kernel of $P_u$ with any non-zero number of anti-holomorphic derivatives or few enough holomorphic transverse derivatives to violate the form of \eqref{eq:InductiveAssumption}. Any such terms would spoil the algebra  \eqref{eq:sec4loop}. 

The most general operator that can appear in the kernel of $P_u$ is
\begin{equation} \label{eq:KernelOp}
\mathcal{N}_{m,j,\bar{j}} = \int du~ \partial_z^{j} \partial_{\bar{z}}^{\bar{j}}\left( \alpha \partial_{\bar{z}}^2 T_{zz}^{(m+1)} + \beta \partial_{\bar{z}}^2\partial_z T_{rz}^{(m+2)} + \gamma \partial_{\bar{z}}^2 \partial_z^2 T_{rr}^{(m+3)}+ \mu \partial_{\bar{z}}\partial_z^2 T_{r\bar{z}}^{(m+2)} + \nu \partial_z^2 T_{\bar{z}\bar{z}}^{(m+1)}\right), 
\end{equation}
which has ${\rm SL}(2, \mathbb{C})$ weights
$\left(\frac{m+4}{2}+j, \frac{m+4}{2} + \bar{j}\right)$.  These operators can thus contribute to the $[\mathcal{W}^p, \mathcal{W}^q]$ commutator in the form  
\begin{equation}
    \begin{split}
         \left[\mathcal{W}^p (z, \bar{z}), \mathcal{W}^q(z', \bar{z}') \right]
           &=i \left[(p-1) \partial_{z'} - (q-1)\partial_z\right] \delta^{(2)}(z-z') \mathcal{W}^{p+q-2}(z',\bar{z}') \\
  &\ + \sum_{\ell = 2p-1} a_{\ell}^{i}\partial_z^{\ell}\left(\delta^{(2)}(z-z') O_i(z',\bar{z}') \right) + \sum_{k = 2q-1} b_k^j\partial_{z'}^k\left(\delta^{(2)}(z-z') O_j(z',\bar{z}') \right)\\
           &\ + \sum_{\ell, \bar{\ell}} a_{\ell, \bar{\ell}}^{\mathcal{N}_{m,j,\bar{j}}} \partial_z^\ell \partial_{\bar{z}}^{\bar{\ell}} \left( \delta^{(2)}(z-z')\mathcal{N}_{m,j,\bar{j}}\right)
                +\sum_{k, \bar{k}} b_{k, \bar{k}}^{\mathcal{N}_{m,j,\bar{j}}} \partial_{z'}^k \partial_{\bar{z}'}^{\bar{k}} \left( \delta^{(2)}(z-z')\mathcal{N}_{m,j,\bar{j}}\right),
    \end{split} 
\end{equation}
where $m = 2x-8$. We want to rule out coefficients $a_{\ell, \bar{\ell}}^{\mathcal{N}_{m,j,\bar{j}}}$, $b_{k, \bar{k}}^{\mathcal{N}_{m,j,\bar{j}}}$ with any of the following values: 
\begin{equation} \label{eq:wedge-condition}
    \begin{split} 
        \ell < 2p-1, \quad k < 2q-1, \quad   \bar{\ell}\neq 0,\quad \text{or} \quad \bar{k}\neq 0,
    \end{split}
\end{equation}
and an effective way to do this is to consider their transformation under ${\rm SL}(2, \mathbb{C})$.

When $p+q = x > \frac{11}{2}$, equivalently $m>3$, all of the operators in \eqref{eq:KernelOp} are non-zero and their transformation under ${\rm SL}(2,\mathbb{C})$ takes the most generic form. Importantly, when $m = -\Delta > 3$, every non-trivial operator $\mathcal{N}_{m,j,\bar{j}}$ mixes under the action of $Q_{\bar{Y}}$ with operators that are not in the kernel of $P_u$:
\begin{equation}
    \begin{split}
        \delta_{\bar{Y}} \mathcal{N}_{m,j,\bar{j}} \supset \partial_{\bar{z}}^2 \bar{Y}^{\bar{z}} O_{i'}, \quad \text{such that} \quad \left[P_u, O_{i'}\right] \neq 0.
    \end{split}
\end{equation}
An explicit formula for this transformation is provided in Appendix \ref{app:KerP-w}.  

Now consider the ${\rm SL}(2, \mathbb{C})$ constraint \eqref{eq:SLrecursion} and take $i'$ to be one of the operators that appears in $\delta_{\bar{Y}} \mathcal{N}_{m,j,\bar{j}}$, satisfying $\left[P_u, O_{i'}\right] \neq 0$.  Previously, we established that 
\begin{equation}
    \begin{split}
         \left[P_u, O_{i'}\right] \neq 0 \quad \quad & \Rightarrow \quad \quad  a^{i'}_{\ell, \bar{\ell}} = 0, \quad \forall~ \bar{\ell} \neq 0,\\
    \end{split}
\end{equation}
which means that the second constraint in \eqref{eq:SLrecursion} reduces to 
\begin{equation} \label{eq:Slrecursion-revist1}
    a^i_{\ell, \bar{\ell}} \bar{\alpha}^{i'}_i = 0, \quad \forall ~ \ell, \bar{\ell}. 
\end{equation} 
This constraint allows us to conclude that $a_{\ell, \bar{\ell}}^{\mathcal{N}_{m,j,\bar{j}}}$ must vanish for all values of $\bar{\ell} \neq 0$ because all other $a^i_{\ell, \bar{\ell}}$ coefficients (corresponding to operators not in the kernel of $P_u$) vanish when $\bar{\ell} \neq 0$.  For the same reason, $a_{\ell, \bar{\ell}}^{\mathcal{N}_{m,j,\bar{j}}}$ must vanish whenever $\ell < 2p-1$.   When $\bar{\ell} =0$ and $\ell \geq 2p-1$, $a_{\ell, \bar{\ell}}^{\mathcal{N}_{m,j,\bar{j}}}$ can be non-vanishing because its contribution can cancel in \eqref{eq:Slrecursion-revist1} against coefficients $a^i_{\ell, \bar{\ell}}$ of other operators that are not in the kernel of $P_u$.   The same arguments apply to $b_{k, \bar{k}}^{\mathcal{N}_{m,j,\bar{j}}}$, implying it must vanish whenever $\bar{k} \neq 0$ or $k< 2q-1$.  Therefore, we have ruled out all additional $\mathcal{N}_{m,j,\bar{j}}$ contributions that satisfy the conditions \eqref{eq:wedge-condition}, completing the proof of \eqref{eq:InductiveAssumption}.  

As explained in Appendix \ref{app:KerP-w}, the mixing structure of \eqref{eq:KernelOp} changes at $m\leq 3$.  Hence, for an inductive proof that starts from a base case of $p+q < \frac{11}{2}$, operators in the kernel of $P_u$ must be treated  case by case. Note also that although we only prove the form \eqref{eq:InductiveAssumption} in this paper, all of the higher $\left[\mathcal{W}^p, \mathcal{W}^q\right]$ commutators are in fact uniquely determined order-by-order by the ${\rm SL}(2,\mathbb{C})$ constraints and the Jacobi identity with $P_u$ (up to equivalence under the identity \eqref{delta-function-identity}).

\section{Light-ray Operator Algebra of Spin-one Currents} \label{sec:currents}
\subsection{Classification of Spin-one Current Light-ray Operators}\label{subsec:currents-classification}

The classification of light-ray operators built from a conserved current proceeds in analogy with the classification of stress-tensor light-ray operators, with the additional simplification that there is no distinction between expressions in quantum and conformal field theory.  The reason is that in an asymptotic symmetry analysis, conformal covariance of gauge theories at a classical level ensures that the resulting fall-off conditions are consistent with conformal symmetry.  

We begin by determining the fall-off behavior of currents at null infinity.  In the study of asymptotic symmetries, these are typically required to be consistent with finite charge and energy configurations and can be derived in free field theory from a saddle point approximation at large $r$ \cite{Strominger:2017zoo}.  However, as in our analysis of the stress tensor in \eqref{subsec:stress-tensor-falloff}, they can equivalently be derived in conformal field theory from the transformation of correlation functions under conformal symmetry. Namely, under an inversion  
\begin{equation}
    \begin{split}
        x^\mu \mapsto x'^\mu  = \frac{x^\mu}{x^2}, 
    \end{split}
\end{equation}
a conserved current transforms as
\begin{equation}
    \begin{split}
        j_\mu(x) &\mapsto j'_\mu(x') = (x^2)^{\Delta} I_\mu{}^\nu (x) j_\nu(x),
    \end{split}
\end{equation}
where again $I^\mu{}_\nu(x) = \delta^\mu_\nu - \frac{2 x^\mu x_\nu}{x^2}$  and here $\Delta = d-1=3$.  Under this transformation, 
\begin{equation} \label{j-correlation-inversion}
\begin{split}
        \langle j_\mu(x') \mathcal{O}_1(x'_1) \cdots \mathcal{O}_n(x'_n)\rangle &= (x^2)^3 I_\mu{}^\nu (x) \langle j_\nu(x) \mathcal{O}'_1(x'_1)\cdots \mathcal{O}'_n(x'_n)\rangle.
    \end{split}
\end{equation}
Using the transformation of the retarded coordinates 
\begin{equation}
    \begin{split}
        u' = -\frac{1}{r}, \quad \quad r' = -\frac{1}{u}, \quad \quad z' = z, \quad \quad \bar{z}' = \bar{z},
    \end{split}
\end{equation}
and noting that  the left-hand side of \eqref{j-correlation-inversion} is manifestly finite in the limit $r \to \infty$ leads to the fall-off behavior
\begin{equation}
    \begin{split}
        j_u, j_z, j_{\bar{z}} \sim \frac{1}{r^2}, \quad \quad j_r \sim \frac{1}{r^4}. 
    \end{split}
\end{equation}

With these fall-off conditions, there is again a unique light-ray operator with the heaviest scaling dimension $\Delta = 0$:
\begin{equation} \label{eq:S0}
    \begin{split}
        \mathcal{S}_0 (z, \bar{z}) \equiv \int du~ j_u^{(2)}, 
    \end{split}
\end{equation}
which was previously studied in field theory in \cite{Hofman:2008ar,Belitsky:2013xxa,Belitsky:2013bja,Beane:2015ufo,Cordova:2018ygx,Kologlu:2019mfz} and appears in the context of asymptotic symmetries as the ``hard'' part of the generator of large gauge transformations \cite{He:2014cra}.  It transforms as a primary under ${\rm SL}(2, \mathbb{C})$ with weights
\begin{equation}
    \begin{split}
        \mathcal{S}_0: \quad \left(h, \bar{h} \right) = \left( 1,1 \right),
    \end{split}
\end{equation}
and it is invariant under translations
\begin{equation}
    \begin{split}
        \delta_f \mathcal{S}_0 = 0.
    \end{split}
\end{equation}
Its transformation under special conformal transformations
\begin{equation} \label{eq:S0globalK}
    \begin{split}
         \delta_g \mathcal{S}_0&=   2\partial_{\bar{z}} \left(g \mathcal{S}_1\right)+ 2\partial_z \left(g \bar{\mathcal{S}}_1\right)- 2 \partial_z \partial_{\bar{z}} \left(g \mathcal{J}_1\right),
    \end{split}
\end{equation}
reveals new primaries at scaling dimension $\Delta = -1$:
\begin{equation}
    \begin{split}
        \mathcal{S}_1 (z, \bar{z})& \equiv \frac{1}{2} \int du ~\left(u \partial_z j_u^{(2)} - j_z^{(2)} \right)
        ,\\
        \bar{\mathcal{S}}_1 (z, \bar{z}) &\equiv \frac{1}{2} \int du ~\left(u \partial_{\bar{z}} j_u^{(2)} - j_{\bar{z}}^{(2)} \right)
        ,\\
        \mathcal{J}_1(z, \bar{z})& \equiv \int du ~ uj_u^{(2)}. 
    \end{split}
\end{equation} 

These primaries complete the classification of light-ray operators at dimension $\Delta = -1$. $\mathcal{S}_1$ appears in the generator of the asymptotic symmetry associated to the subleading soft photon/gluon theorem  \cite{GellMann1954,Low1954,Low1958,Burnett1968}, which was first established in QED in \cite{Lysov:2014csa}. Note that \eqref{eq:S0globalK} implies that the \textit{modes} of $\mathcal{S}_1$ can be constructed from the the modes of $\mathcal{S}_0$ by acting with special conformal transformations, as noted by \cite{Larkoski:2014hta,Sheta:2025oep}, but the $\mathcal{J}_1$ contribution to \eqref{eq:S0globalK} prohibits the full operator $\mathcal{S}_1$ strictly from being a conformal descendant of $\mathcal{S}_0$. 

To extend the classification to all orders in scaling dimension, we use current conservation, 
\begin{equation} \label{current-conservation}
    \begin{split}
        \partial_\mu j^\mu = 0,
    \end{split}
\end{equation}
to find a minimal set of independent light-ray operators.  Expanding \eqref{current-conservation} in $\frac{1}{r}$ gives
\begin{equation}
    \begin{split}
       -(n-2) \int du ~u^pj_{u}^{(n)}=    p  \int du~ u^{p-1}  j_{r}^{(n+1)}  
           +  \partial_z \int du ~u^p j_{\bar{z}}^{(n-1)}   +  \partial_{\bar{z}} \int du ~u^p j_{z}^{(n-1)} ,
    \end{split}
\end{equation}
from which we find that all $\int du ~u^pj_{u}^{(n)}$ for $n> 2$ can be traded for other elements of the basis.  Thus, a minimal basis at scaling dimension $\Delta = -m$  is given by
\begin{equation} \label{eq:basisCurrents}
    \begin{split}
        \int du~u^m j_u^{(2)}, 
        \quad   \int du~ u^{m-n}j_r^{(n+2)}, 
         \quad  \int du ~u^{m-n+1}j_z^{(n)},
          \quad   \int du ~u^{m-n+1}j_{\bar{z}}^{(n)}, \quad n \geq 2. 
    \end{split}
\end{equation}

The classification of primaries with light $\bar{h}= \frac{2-m}{2}$ weight proceeds in analogy with the classification for stress-tensor light-ray primaries. We find that the combination
\begin{equation}
    \begin{split}
        \mathcal{M}_{m, \ell}(z, \bar{z}) = \int du~ u^{m-1} \left(u \partial_z^\ell j_u^{(2)}+\ell(m-\ell-1) \partial_z^{\ell-1} j_z^{(2)} \right)
    \end{split}
\end{equation}
transforms as
\begin{equation} \label{sl2c-m}
    \begin{split}
        -\delta_Y \mathcal{M}_{m, \ell}
        & = \left(Y^z \partial_z + \frac{2-m+2 \ell}{2} \partial_z Y^z\right)\mathcal{M}_{m, \ell} \\
         & \quad\quad  -\frac{1}{2}\ell (\ell-1)(m-\ell-1) (m-\ell-2) \partial_z^2 Y^z  \int du~u^{m-1} \partial_z^{\ell-2} j_z^{(2)}, \\
        -\delta_{\bar{Y}} \mathcal{M}_{m, \ell}
        & = \left(\bar{Y}^{\bar{z}} \partial_{\bar{z}} + \frac{2-m}{2} \partial_{\bar{z}} \bar{Y}^{\bar{z}}\right)  \mathcal{M}_{m, \ell}.
    \end{split}
\end{equation}
We observe that $ \mathcal{M}_{m, \ell}$ is a primary when $\ell = 0,1$ and we denote these primaries as follows:
\begin{equation}
    \begin{split}
        \mathcal{J}_m (z, \bar{z})& \equiv \int du ~u^mj_u^{(2)}, \\
        \mathcal{R}_m (z, \bar{z})& \equiv \int du~u^{m-1} \left(u \partial_z j_u^{(2)} +(m-2) j_z^{(2)}\right).
    \end{split}
\end{equation}
From \eqref{sl2c-m}, we observe that these primaries transform with ${\rm SL}(2, \mathbb{C})$ weight
\begin{equation}
    \begin{split}
        \mathcal{J}_m: \quad \left(h, \bar{h}\right) =\left(\frac{2-m}{2}, \frac{2-m}{2}\right), 
        \quad \quad \quad 
        \mathcal{R}_m: \quad \left(h, \bar{h}\right) =\left(\frac{4-m}{2}, \frac{2-m}{2}\right), 
    \end{split}
\end{equation}
and they transform under translations as
\begin{equation}
    \begin{split}
        \delta_f \mathcal{J}_m &=m f \mathcal{J}_{m-1}
        ,\\
        \delta_f \mathcal{R}_m&= \frac{1}{m-3} \left((m-1)(m-2) f \mathcal{R}_{m-1} - 2 \partial_z \left(f \mathcal{J}_{m-1}\right)+2(m-2) \partial_z  f \mathcal{J}_{m-1}\right).
    \end{split}
\end{equation}
The transformation of these operators under special conformal transformations (apart from $\mathcal{J}_0 = \mathcal{S}_0$) involves operators that fall outside of the classification of light-$\bar{h}$ primaries and thus we do not include explicit expressions for these transformations. 

To find the higher-spin primaries, we consider the classification of primary descendants that involve linear combinations of 
\begin{equation}
    \begin{split}
          \partial_{\bar{z}}^{m-1} \mathcal{M}_{m, \ell+m-1}
              = \int du~ (u \partial_z\partial_{\bar{z}})^{m-1}  \left(u\partial_z^\ell j_{u}^{(2)} - \ell (\ell+m-1) \partial_z^{\ell-1} j_{z}^{(2)}\right),
    \end{split}
\end{equation}
and
\begin{equation}
    \begin{split}
        \int du~(u\partial_z \partial_{\bar{z}})^{m-n+1}  \left (a_n\partial_z^{\ell}j_{r}^{(n+1)}+ b_n\partial_z^{\ell-1}j_{z}^{(n)}
             +c_n u\partial_z^{\ell+1} j_{\bar{z}}^{(n-1)}\right),
    \end{split}
\end{equation}
for varying values of $n$. The coefficients $a_n$, $b_n$ and $c_n$ are fixed by requiring the resulting linear combination of operators to form an ${\rm SL}(2, \mathbb{C})$ primary.  In Appendix \ref{app:primary-derivation-current}, we solve these constraints and find that the only solution for general $m$ requires that $\ell=1$ and takes the form 
\begin{equation} \label{def-s-operator}
    \begin{split}
        \partial_{\bar{z}}^{m-1} \mathcal{S}_m(z, \bar{z})
            \equiv\frac{1}{2^m m!}& \Bigg[ \int du~ (u \partial_z\partial_{\bar{z}})^{m-1}  \left(u\partial_z  j_{u}^{(2)} -  m  j_{z}^{(2)}\right)\\
           &   - \sum_{n=3}^{m+1} \frac{m!(n-3)!}{(m-n+1)!}
            \int du~(u\partial_z \partial_{\bar{z}})^{m-n+1}  \left ( \partial_zj_{r}^{(n+1)}+ (n-2) j_{z}^{(n)} \right)\Bigg].
    \end{split}
\end{equation}
Here the overall coefficient is chosen to simplify expressions for the transformation of these operators under translations. Strikingly, this expression has previously appeared in the literature. Specifically, through an asymptotic symmetry analysis of ${\rm U}(1)$ gauge theory, \cite{Campiglia:2018dyi} derived this combination conserved currents as the ``hard'' part of conserved charges that generate the infinite tower of soft photon theorems in \cite{Li:2018gnc,Hamada:2018vrw}. Our analysis reveals that this combination is entirely fixed by its overall scaling dimension and transformation under the Lorentz group ${\rm SL}(2, \mathbb{C})$.  It would be interesting if this observation could be exploited to simplify or even circumvent the rather involved expansions about null infinity that underlie many asymptotic symmetry investigations. 

The resulting primaries $\mathcal{S}_m$ transform  with ${\rm SL}(2, \mathbb{C})$ weights 
\begin{equation}
    \begin{split}
         \mathcal{S}_m: \quad \left(h, \bar{h}\right) = \left(\frac{m+2}{2},\frac{2-m}{2} \right),
    \end{split}
\end{equation}
and are the desired generators of the $S$ algebra. For comparison with the literature and review of the algebra \eqref{eq:Salg-intro}, we note that the $S$ algebra generators are typically labeled by their left conformal weight $h$.  To conform with that notation, we introduce the following superscript notation:
\begin{equation} \label{eq:Ssuper}
    \begin{split}
        \mathcal{S}^{\frac{m+2}{2}} \equiv \mathcal{S}_m. 
    \end{split}
\end{equation}
Notably, this tower of primaries transforms into itself under translations:
\begin{equation} \label{eq:Strans}
    \begin{split}
        \delta_f   \mathcal{S}_m &=  \frac{1}{2} \partial_z   \left(f  \mathcal{S}_{m-1} \right) 
         + \frac{m}{2}  \partial_zf  \mathcal{S}_{m-1}.
    \end{split}
\end{equation}
As in the analysis of light-ray operators built from the stress tensor, special conformal transformations do not act in a simple way on $\mathcal{S}_m$ generators for generic values of $m$, so explicit expressions are not included.   

\subsection{Current Light-ray Operator Algebra} \label{subsec:current-algebra}

In this subsection, we study commutators of light-ray operators constructed from a conserved current $j_\mu = T^a j^a_\mu$ that generates a continuous global symmetry $G$. We begin at the highest order in  scaling dimension $\Delta = 0$, where the unique current light-ray operator is $\mathcal{S}_{0}^a$ \eqref{eq:S0}. The commutator between two $\mathcal{S}_0^a$ operators can be computed using the light-ray OPE \cite{Kologlu:2019mfz}, with the result
\begin{equation} \label{local-leading}
\left[\mathcal{S}_0^{a}(z,\bar{z}), \mathcal{S}_0^{b}(z',\bar{z}')\right] = i f^{abc} \delta^{(2)}(z-z') \mathcal{S}_0^{c}(z',\bar{z}'), 
\end{equation}
where $f^{abc}$ are the structure constants of $G$. As described by \cite{Cordova:2018ygx}, for an arbitrary function $f(z,\bar{z})$, the smeared operators
\begin{equation}
Q^a_f = \int d^2z~ f(z,\bar{z})  \mathcal{S}_0^{a}(z,\bar{z})
\end{equation}
obey the algebra 
\begin{equation} \label{eq:LocalQalg}
\left[Q^a_{f_1}, Q^b_{f_2}\right] = i f^{abc} Q^c_{f_1f_2}, 
\end{equation}
and thus reproduce the large gauge asymptotic symmetry algebra that is related to the leading tree-level soft gluon theorem \cite{Weinberg:1965nx,Strominger:2013lka}.  The global subalgebra is recovered by taking $f_1=f_2 = 1$ so that \eqref{eq:LocalQalg} becomes
\begin{equation}
\left[Q^a, Q^b\right] = i f^{abc} Q^c.
\end{equation}

\subsubsection{$S$ Algebra} \label{sec:Salg}

In this section, we switch to the superscript notation \eqref{eq:Ssuper} to match the standard $S$ algebra notation, reviewed in \eqref{eq:Salg-intro}. In analogy with \eqref{eq:schematic}, we use the following notation for generic operators formed from null integrals of the conserved current, with group indices suppressed:
\begin{equation} \label{eq:schematicJ}
O_i(z,\bar{z}) = \int du~ u^{k_i} \partial_z^{m_i}\partial_{\bar{z}}^{\bar{m}_i} j_{\mu_i}^{(n_i)}(u,z,\bar{z}). 
\end{equation}
The operators $O_i$ are not always ${\rm SL}(2, \mathbb{C})$ primaries, but still carry definite ${\rm SL}(2, \mathbb{C})$ weights $(h_i,\bar{h}_i)$.  Here $i$ again labels a full set of indices $k_i \in \mathbb{Z}_{\geq 0}$, $m_i, \bar{m}_i \in \mathbb{Z}$, $n_i \in \mathbb{Z}_{\geq 2}$, and $\mu_i \in \{u,r,z,\bar{z}\}$. The minimal basis for such operators is presented in \eqref{eq:basisCurrents}.

In analogy with  Section \ref{sec:wAlg}, we prove by induction in $p+q$ that the light-ray current operator contributions to the local commutator of the $\mathcal{S}^{p,a}$ operators take the form 
\begin{equation} \label{eq:localS}
\begin{aligned}
\Big[\mathcal{S}^{p,a}(z,\bar{z}), \mathcal{S}^{q,b}(z',\bar{z}')\Big] &= i f^{abc} \delta^{(2)}(z-z') \mathcal{S}^{p+q-1,c}(z',\bar{z}')  \\ 
&\ +  \sum_{\ell = 2p-1} a_{\ell}^{i}\partial_z^{\ell}\left(\delta^{(2)}(z-z') O_i(z',\bar{z}') \right) + \sum_{k = 2q-1} b_k^j\partial_{z'}^k\left(\delta^{(2)}(z-z') O_{j}(z',\bar{z}') \right), \\
\end{aligned}
\end{equation}
where terms on the second line do not contribute to the wedge subalgebra defined by the modes
\begin{equation} \label{eq:Smodes}
{S}^{p,a}_{m,\bar{m}} \equiv \int  d^2z~ z^{p+ m - 1}\bar{z}^{1-p+\bar{m}} \mathcal{S}^{p,a}(z,\bar{z}), \qquad  1-p \leq m \leq p-1. 
\end{equation}
Thus integrating \eqref{eq:localS} gives the wedge subalgebra of the $S$ loop algebra:
\begin{equation}
\left[S^{p,a}_{m,\bar{m}}, S^{q,b}_{n,\bar{n}}\right] = i f^{abc} S^{p+q-1,c}_{m+n, \bar{m}+\bar{n}}. 
\end{equation} 

We prove \eqref{eq:localS} by induction using the base case $p+q = 3$, corresponding to $\Delta = -2$ on the right-hand side. The commutators with $p+q \leq 3$ can be directly computed from Poincar\'e constraints and the leading case \eqref{local-leading} with $p+q = 2$. They also respect the form \eqref{eq:localS} and are presented in \eqref{eq:S0S1}, \eqref{eq:S1S1}, and \eqref{eq:S0S2}.

Now we perform the inductive step. Assume that \eqref{eq:localS} holds for $p+q = x-\frac{1}{2}$ for some $x >3$. We want to show that the general commutator  \eqref{eq:generalCommutator} at $p+q=x$,
\begin{equation}
    \begin{split}
        \left[ \mathcal{S}^{p,a}(z,\bar{z}),\mathcal{S}^{q,b}(z',\bar{z}')\right]
            & = \sum_{\ell, \bar{\ell} }a^i_{\ell,\bar{\ell}}\partial_{z}^{\ell}\partial_{\bar{z}}^{\bar{\ell}} \left(\delta^{(2)}(z-z') O_i(z',\bar{z}') \right) , 
    \end{split}
\end{equation}
must also take the form \eqref{eq:localS}. Covariance under dilations restricts operators on the right-hand side to have dimension $\Delta = 4-2x$. Again we use \eqref{delta-function-identity} to express the commutator in the more symmetric form
\begin{equation} \label{sp-sq-modified-ansatz}
\begin{aligned}
    \left[\mathcal{S}^{p,a}(z,\bar{z}), \mathcal{S}^{q,b}(z',\bar{z}')\right] 
    &= i f^{abc}\delta^{(2)}(z-z')\mathcal{S}^{p+q-1, c}(z',\bar{z}')  \\
    &\ + \sum_{\ell,\bar{\ell}}a^i_{\ell,\bar{\ell}}\partial_z^{\ell}\partial_{\bar{z}}^{\bar{\ell}}\left( \delta^{(2)}(z{-}z')  O_i(z',\bar{z}')\right) + \sum_{k,\bar{k}}b_{k,\bar{k}}^j\partial_{z'}^{k}\partial_{\bar{z}'}^{\bar{k}}\left( \delta^{(2)}(z{-}z') O_j(z',\bar{z}')\right), 
  \end{aligned}
\end{equation}
where without loss of generality we have added and subtracted a contribution from $\mathcal{S}^{p+q-1}$ and absorbed the subtraction into the $a^i_{\ell,\bar{\ell}}$ and $b_{k,\bar{k}}^j$ coefficients.

Next, we use Poincar\'e constraints to rule out terms not of the form \eqref{eq:localS}. Using \eqref{eq:Strans} with $f = 1$, we have 
\begin{equation}
\left[P_u, \mathcal{S}^{p,a}(z,\bar{z}) \right] = \frac{i}{2} \partial_z \mathcal{S}^{p-\frac{1}{2},a}(z,\bar{z}).  
\end{equation}
Acting on the left-hand side of \eqref{sp-sq-modified-ansatz} with $P_u$ and using the Jacobi identity  gives 
\begin{equation} \label{eq:PuLocalS}
\begin{aligned}
    \left[ P_u,  \left[\mathcal{S}^{p,a}(z,\bar{z}), \mathcal{S}^{q,b}(z',\bar{z}')\right] \right] 
    &= \frac{i}{2} \left( \partial_{z'}\left[\mathcal{S}^{p,a}(z,\bar{z}), \mathcal{S}^{q-\frac{1}{2},b}(z',\bar{z}')\right] + \partial_z \left[\mathcal{S}^{p-\frac{1}{2},a}(z,\bar{z}), \mathcal{S}^{q,b}(z',\bar{z}')\right]\right) \\
    &= \frac{i^2}{2} f^{abc}\delta^{(2)}(z-z')\partial_{z'}\mathcal{S}^{p+q-\frac{3}{2}, c}(z',\bar{z}') + X ,\\
  \end{aligned}
\end{equation}
where the $X$ terms are analogous  to the ones in Section \ref{sec:wAlg}, namely  \eqref{eq:Xterms} with \eqref{eq:schematic} replaced by \eqref{eq:schematicJ}, and we have assumed that $p,q \neq 1$ so that $\mathcal{S}^{p-\frac{1}{2}}$ and $\mathcal{S}^{q-\frac{1}{2}}$ are both nontrivial operators.\footnote{In the case $p=1$, $q > 1$, $\mathcal{S}^{1,a}$ is annihilated by $P_u$ and we have 
\begin{equation*}
\begin{aligned}
&\left[ P_u,  \left[\mathcal{S}^{1,a}(z,\bar{z}), \mathcal{S}^{q,b}(z',\bar{z}')\right] \right]  \\
    &= \frac{i}{2} f^{abc}\Bigg(  i\delta^{(2)}(z-z')\partial_{z'}\mathcal{S}^{q-\frac{1}{2},c}(z',\bar{z}') -i\partial_z \left(\delta^{(2)}(z-z')\mathcal{S}^{q-\frac{1}{2},c}(z',\bar{z}')\right) \\
    &\qquad + \sum_{\ell=1} a_{\ell}^i\left[\partial_z^{\ell}\left(\delta^{(2)}(z{-}z')\partial_{z'}O_i(z',\bar{z}')\right) - \partial_z^{\ell+1}\left(\delta^{(2)}(z{-}z')O_i(z',\bar{z}')\right) \right] + \sum_{k = 2q-1} b_{k-1}^j\partial_{z'}^k\left(\delta^{(2)}(z{-}z') O_j(z',\bar{z}') \right)\Bigg). 
\end{aligned}
\end{equation*}
Note that here an $\mathcal{S}$ operator appears outside the $p=1$ wedge.} Next, we act with $P_u$ on the right-hand side of \eqref{sp-sq-modified-ansatz} and recover the first term of \eqref{eq:PuLocalS} as well as additional terms:
\begin{equation} 
\begin{aligned}
    \left[ P_u,  \left[\mathcal{S}^{p,a}(z,\bar{z}), \mathcal{S}^{q,b}(z',\bar{z}')\right] \right]  
    &= \frac{i^2}{2} f^{abc}\delta^{(2)}(z-z')\partial_{z'}\mathcal{S}^{p+q-\frac{3}{2}, c}(z',\bar{z}')  \\
    &\qquad + \sum_{\ell,\bar{\ell}}a^i_{\ell,\bar{\ell}}\partial_z^{\ell}\partial_{\bar{z}}^{\bar{\ell}}\left( \delta^{(2)}(z-z') \left[P_u, O_i(z',\bar{z}')\right]\right) \\
    &\qquad + \sum_{k,\bar{k}}b_{k,\bar{k}}^j\partial_{z'}^{k}\partial_{\bar{z}'}^{\bar{k}}\left( \delta^{(2)}(z-z') \left[P_u, O_j(z',\bar{z}')\right]\right). 
  \end{aligned}
\end{equation}
Just as in Section \ref{sec:wAlg},  we find that for operators $O_i$, $O_j$ in \eqref{sp-sq-modified-ansatz} that are \emph{not} annihilated by $P_u$, $a^i_{\ell, \bar{\ell}}$ and $b^j_{k, \bar{k}}$ are only non-zero when $\bar{\ell} = \bar{k} = 0$ (i.e.~no anti-holomorphic derivatives) and $\ell \geq 2p-1$ or $k \geq 2q-1$.  Thus, apart from additional terms in the kernel of $P_u$, which will be considered next, the general commutator at $p+q =x$ \eqref{sp-sq-modified-ansatz} is constrained to take the desired form \eqref{eq:localS}.

The most general light-ray current operator that can appear in the kernel of $P_u$ is 
\begin{equation} \label{eq:KerPS}
\widetilde{\mathcal{N}}_{m,j,\bar{j}} = \partial_z^{j}\partial_{\bar{z}}^{\bar{j}} \int du~ \left( \alpha \partial_{\bar{z}}j_z^{(m+1)} + \gamma \partial_z\partial_{\bar{z}} j_r^{(m+2)} + \nu \partial_z j_{\bar{z}}^{(m+1)}\right),
\end{equation} 
which has weights $\left( \frac{m+2}{2} + j, \frac{m+2}{2}+ \bar{j}\right)$. These operators can contribute to the commutator as
\begin{equation} 
\begin{aligned}
\left[\mathcal{S}^{p,a}(z,\bar{z}), \mathcal{S}^{q,b}(z',\bar{z}')\right] &= i f^{abc} \delta^{(2)}(z-z') \mathcal{S}^{p+q-1,c}(z',\bar{z}')  \\ 
&\ + \sum_{\ell = 2p-1} a_{\ell}^{i}\partial_z^{\ell}\left(\delta^{(2)}(z-z') O_i(z',\bar{z}') \right) + \sum_{k = 2q-1} b_k^j\partial_{z'}^k\left(\delta^{(2)}(z-z') O_{j}(z',\bar{z}') \right), \\
&\ + \sum_{\ell, \bar{\ell}} a_{\ell, \bar{\ell}}^{\widetilde{\mathcal{N}}_{m,j,\bar{j}}} \partial_z^\ell \partial_{\bar{z}}^{\bar{\ell}} \left( \delta^{(2)}(z-z')\widetilde{\mathcal{N}}_{m,j,\bar{j}}\right)
                +\sum_{k, \bar{k}} b_{k, \bar{k}}^{\widetilde{\mathcal{N}}_{m,j,\bar{j}}} \partial_{z'}^k \partial_{\bar{z}'}^{\bar{k}} \left( \delta^{(2)}(z-z')\widetilde{\mathcal{N}}_{m,j,\bar{j}}\right),
\end{aligned}
\end{equation}
where $m=2x-4$. As in Section \ref{sec:wAlg}, we want to rule out coefficients $a_{\ell, \bar{\ell}}^{\widetilde{\mathcal{N}}_{m,j,\bar{j}}}$, $b_{k, \bar{k}}^{\widetilde{\mathcal{N}}_{m,j,\bar{j}}}$ with values in \eqref{eq:wedge-condition}, and an effective way to do so is by considering the transformation of $\widetilde{\mathcal{N}}_{m,j,\bar{j}}$ under  $Q_{\bar{Y}}$. When $p+q > 3$, equivalently  $m>2$, every operator of the form \eqref{eq:KerPS} mixes under the action of $Q_{\bar{Y}}$ with operators that are \textit{not} annihilated by $P_u$, as demonstrated explicitly in Appendix \ref{app:KerP-s}. Invoking an argument involving \eqref{eq:SLrecursion} in direct analogy with the one in Subsection \ref{sec:wAlg} rules out all additional contributions of the form \eqref{eq:KerPS} that satisfy \eqref{eq:wedge-condition} and thereby violate the form of \eqref{eq:localS}, completing the proof of \eqref{eq:localS}.

The mixing structure of \eqref{eq:KerPS} at $m=1$  and $m=2$ are also discussed in Appendix \ref{app:KerP-s} and  can be treated separately for an induction proof that starts from a base case of $p+q < 3$. Again all of the contributions from light-ray operators formed from the current to higher $\left[\mathcal{S}^{p,a}, \mathcal{S}^{q,b}\right]$ commutators are uniquely determined order-by-order by the ${\rm SL}(2,\mathbb{C})$ constraints and the Jacobi identity with $P_u$ (up to equivalence under the identity \eqref{delta-function-identity}). 

\subsection{Mixed Algebra of Current and Stress-tensor Light-ray Operators} \label{subsec:w-s-commutation}

We now consider mixed commutators of current and stress-tensor light-ray operators. 

\subsubsection{$\left[\mathcal{W},\mathcal{S}^a\right]$ Algebra} \label{sec:wSalg}

Here we prove that the $\mathcal{S}$ operators are in an adjoint representation of the ${\rm w}_{1+\infty}$ algebra. In particular, we prove that the contribution of light-ray current operators to the commutator is 
\begin{equation} \label{eq:InductiveAssumptionWS}
  \begin{aligned}
\left[\mathcal{W}^p(z,\bar{z}), \mathcal{S}^{q,a}(z',\bar{z}')\right] &= i   \left[(p-1) \partial_{z'} - (q-1)\partial_z\right] \delta^{(2)}(z-z') \mathcal{S}^{p+q-2,a
}(z',\bar{z}') \\
  &\ + \sum_{\ell = 2p-1} a^i_{\ell}\partial_z^{\ell}\left(\delta^{(2)}(z-z') O_{i}(z',\bar{z}') \right) + \sum_{k = 2q-1} b_k^j\partial_{z'}^k\left(\delta^{(2)}(z-z') O_{j}(z',\bar{z}') \right), \\
  \end{aligned}
\end{equation}
where terms on the second line have the form \eqref{eq:schematicJ} and do not contribute to the wedge subalgebra defined by the modes \eqref{eq:Wmodes} and \eqref{eq:Smodes}. Integrating \eqref{eq:InductiveAssumptionWS} gives the adjoint action of ${\rm w}_{1+\infty}$ on $S$, namely
\begin{equation}
\left[{\rm w}^{p}_{m, \bar{m}}, {S}^{q,a}_{n, \bar{n}} \right] = i \left[m(q-1) - n(p-1)\right] {S}^{p+q-2,a}_{m+n, \bar{m}+\bar{n}}. 
\end{equation}

As in the previous ${\rm w}_{1+\infty}$-algebra Section \ref{sec:wAlg} and $S$-algebra Section \ref{sec:Salg}, the proof proceeds by induction, here with a base case of $p+q=\frac{7}{2}$ (corresponding to $\Delta = -1$ on the right-hand side of \eqref{eq:InductiveAssumptionWS}). The lowest-order commutator vanishes,
\begin{equation} \label{eq:Wm1S0}
\left[\mathcal{W}^{\frac{3}{2}}(z,\bar{z}), \mathcal{S}^{1,a}(z',\bar{z}')\right] = 0,
\end{equation}
due to the absence of operators with $\Delta = 1$ that can be constructed out of null integrals of the conserved current. The other cases with $p+q\leq\frac{7}{2}$ are provided in Appendix \ref{app:MixedSW} and all of the commutators at $p+q\leq \frac{7}{2}$ are of the form \eqref{eq:InductiveAssumptionWS}.

We use $p+q=\frac{7}{2}$ as the base case for induction. This corresponds to commutators with overall dimension $\Delta = -1$, namely $\left[ \mathcal{W}^{\frac{3}{2}}, \mathcal{S}^{2,a}\right]$ in \eqref{eq:Wm1S2},  $\left[\mathcal{W}^{2}, \mathcal{S}^{\frac{3}{2},a}\right]$ in \eqref{eq:W0S1}, and $\left[\mathcal{W}^{\frac{5}{2}}, \mathcal{S}^{1,a}\right]$ in \eqref{eq:W1S0}. For the inductive step, assume \eqref{eq:InductiveAssumptionWS} holds at $p+q = x - \frac{1}{2}$ for some $x > \frac{7}{2}$.  Arguments analogous to those in Section \ref{sec:wAlg} and \ref{sec:Salg}  lead to the conclusion that the Jacobi identity with $P_u$ constrains the commutator at $p+q= x$ to take the correct form, up to additional terms in the kernel of $P_u$. The absence of anti-holomorphic derivatives in \eqref{eq:InductiveAssumptionWS} and, as discussed in Appendix \ref{app:KerP-s}, the transformation of $\widetilde{\mathcal{N}}_{m>2,j,\bar{j}}$ (corresponding to $x > 4$) under $Q_{\bar{Y}}$ and $\widetilde{\mathcal{N}}_{m=2,j,\bar{j}}$ (corresponding to $x=4$) under $Q_{Y}$, $Q_{\bar{Y}}$  imply that all additional contributions from the kernel of $P_u$ that satisfy \eqref{eq:wedge-condition} are ruled out. This completes the proof. 

\subsubsection{Algebra with Local Conformal Generators}

In this subsection, we consider the algebra of the local conformal generators with the local charge generator $\mathcal{S}_0^a$. The commutators $\left[\mathcal{W}_{-1},\mathcal{S}_0\right]$ and $\left[\mathcal{W}_0,\mathcal{S}_0\right]$ are presented in \eqref{eq:Wm1S0} and \eqref{eq:W0S0}. We find the additional commutators
\begin{equation}
\left[\mathcal{D}(z,\bar{z}), \mathcal{S}^{a}_0(z',\bar{z}')\right] = 0,
\end{equation}
which reproduces the transformation of $\mathcal{S}_0$ under $Q_D$ with $\Delta = 0$ upon  integrating $\mathcal{D}$ in the transverse directions to form $Q_D$, and 
\begin{equation}
    \begin{split}
        \left[\mathcal{K}(z,\bar{z}), \mathcal{S}^{a}_0(z',\bar{z}')\right]
            &= 2 i \partial_{\bar{z}'} \left(\delta^{(2)}(z-z')  \mathcal{S}^a_1\right) +2 i \partial_{z'} \left(\delta^{(2)}(z-z') \bar{\mathcal{S}}^a_1\right)-2i \partial_{z'} \partial_{\bar{z}'}  \left(\delta^{(2)}(z-z')  \mathcal{J}^a_1\right),
    \end{split}
\end{equation}
which can be integrated against $g(z,\bar{z})$ to reproduce the global special conformal transformation of $\mathcal{S}_0$ \eqref{eq:S0globalK}. 

\section{One-point Functions and Soft Factors} \label{sec:1pt}

In this section, we consider a simple set of observables involving the $\mathcal{W}$ and $\mathcal{S}$ light-ray operators, namely one-point correlation functions in the background of a scalar operator.  Interestingly, our results reveal a new connection to the infrared physics of gauge and gravitational theories, which was an original source of inspiration for this work. Specifically, we find that the values of the one-point functions of $\mathcal{W}$ and $\mathcal{S}$ exactly reproduce (derivatives of) universal soft factors associated to massive scalar particles that arise at every order in the soft expansion of an amplitude for the emission of a graviton or gauge boson \cite{Himwich:2023njb}.\footnote{Physically, the role of these derivatives is to turn the soft factor, which is essentially a local version of the source multipole moment, into a local version of the field multipole moment \cite{Hoskins:2026new}. The soft factor is an ${\rm SL}(2,\mathbb{C})$ primary and the derivatives produce its ${\rm SL}(2,\mathbb{C})$ primary descendant.}  Although these one-point functions are the only observables explicitly studied in this paper, the connection to soft factors may provide some insight into how the algebra holds in higher-point correlation functions. In particular, the soft factors themselves were shown to respect the correct algebras \cite{Himwich:2023njb}. In the large-$N$ limit of holographic conformal field theories, the $\langle \mathcal{O} T T \mathcal{O} \rangle$ four-point function is dominated by the product of two three-point functions and it would be instructive to study whether, in parallel, the $\mathcal{W}_m$ algebra holds in this limit. \cite{Belin:2020lsr} studied $\left[\mathcal{W}_{-1}, \mathcal{D}\right]$ in this setting and intriguingly found that the commutation relation is violated at large $N$, but is expected to be  restored non-perturbatively and thereby constrain finite-$N$ effects. If the $\mathcal{W}_m$ commutators were also violated at large $N$, the violation may similarly lead to interesting constraints. 

The connection between one-point functions and universal soft factors is also naively less straightforward when applied to spinning operators as opposed to scalars. Specifically, the building blocks for the one-point functions are conformal three-point functions, which for spinning operators have multiple kinematically-allowed forms and thus depend on more parameters.\footnote{For example, the three-point function involving three stress tensors has three kinematically allowed structures and in four dimensions is parameterized by two constants $a$ and $c$ \cite{Osborn:1993cr}.} It would be interesting, but is beyond the scope of this work to study this connection in detail. 

Beyond the connection to soft theorems in gauge theory and gravity, our one-point function results have important implications for purely conformal/quantum field theoretic realizations of the ${\rm w}_{1+\infty}$ and $S$-algebra. First, although the majority of the generators involve apparently non-local (in the transverse plane) contributions from the stress tensor or conserved current, these bare generators turn out to have simple one-point functions.  In other words, the one-point functions of the local primary descendants turn out to be total derivatives so that the one-point functions of the bare primaries can be readily extracted. This further supports the claim that the maximal-spin ${\rm w}_{1+\infty}$ and $S$-algebra generators (as opposed to their low-spin primary descendants) are indeed the interesting class of operators in conformal/quantum field theory.  

Finally, unlike previously-studied $\mathcal{L}_m$ operators \cite{Besken:2020snx,Belin:2020lsr,Huang:2020ycs,Huang:2021hye}, the one-point functions of $\mathcal{W}_m$ and $\mathcal{S}_m$ generators are manifestly finite for all values of $m$, thus making the ${\rm w}_{1+\infty}$ and $S$-algebra natural candidates for extended symmetry algebras in higher-dimensional conformal field theories.  Mechanically, the accompanying derivatives in the transverse plane help regulate the divergences that would arise from additional powers of $u$.  In particular, we find that the $\mathcal{W}_m$ and $\mathcal{S}_m$ generators themselves (and not only their primary descendants) involve enough transverse derivatives to produce finite one-point functions. 

\subsection{Stress Tensor} \label{subsec:1-pt-W}

The calculation of one-point functions of the $\mathcal{W}_m$ generators begins with the three-point function for two identical scalar fields of conformal dimension $\Delta$ and the stress tensor \cite{Osborn:1993cr}
\begin{equation} \label{eq:3ptTOO}
    \begin{split}
       \langle \mathcal{O}(x_1) T_{\mu\nu}(x_2) \mathcal{O} (x_3)\rangle
         = \frac{a}{x_{12}^4 x_{23}^4 x_{13}^{2\Delta-4}} \left(\frac{X_{13}{}_\mu X_{13}{}_\nu}{X_{13}^2}- \frac{1}{4} \eta_{\mu\nu}\right), 
    \end{split}
\end{equation}
where
\begin{equation} \label{def-capital-X}
    \begin{split}
         X_{13}^\mu \equiv \frac{x_{12}^\mu}{x_{12}^2} + \frac{x_{23}^\mu}{x_{23}^2},
    \end{split}
\end{equation}
and $a$ is a constant that is related by the Ward identity to the normalization of the two-point function.  Choosing this normalization to be unity,
\begin{equation}
    \begin{split}
        \langle \mathcal{O}(x_1) \mathcal{O}(x_2) \rangle = \frac{1}{x_{12}^{2\Delta}},
    \end{split}
\end{equation}
fixes 
\begin{equation}
    \begin{split}
        a = - \frac{2\Delta}{3 \pi^2}. 
    \end{split}
\end{equation}
Note that this input is entirely fixed by conformal symmetry and thus \eqref{eq:3ptTOO} is non-perturbatively exact. In \eqref{eq:3ptTOO}, the operators are ordered as written and this ordering specifies the $i\epsilon$ prescription to be
\begin{equation}
    \begin{split}
        x_{ij}^2= -(t_i-t_j-i\epsilon)^2 + (\vec{x}_{i}-\vec{x}_j)^2 , \quad \quad  i <j. 
    \end{split}
\end{equation}

\subsubsection{Position-space One-point Functions}

We begin by recasting prior calculations of the one-point function of the ANEC operator $\mathcal{W}_{-1}$ in a way that can be readily generalized to the other $\mathcal{W}_m$.  First, we evaluate the $uu$ component of the stress tensor at $\mathscr{J}^+$:
\begin{equation}
    \begin{split}
        \langle \mathcal{O}(x_1) T_{uu}^{(2)}(u, z, \bar{z}) \mathcal{O}(x_3)\rangle
            &=  \frac{a}{4} \frac{1}{x_{13}^{2 \Delta-2}} \lim_{r \to \infty} r^2 \frac{(n \cdot X_{13})^2}{x_{12}^2 x_{23}^2},
    \end{split}
\end{equation}
where $x_2^\mu$ in \eqref{eq:3ptTOO} is re-expressed in terms of $u,r,z,\bar{z}$ according to \eqref{flat-retarded-Bondi-coord}. Using 
\begin{equation}
    \begin{split}
        n \cdot X_{13} = \frac{n \cdot x_1+ r}{x_{12}^2} - \frac{n\cdot x_3 + r}{x_{23}^2}
    \end{split}
\end{equation}
and
\begin{equation}
    \begin{split}
        x_{12}^2 &= - r \left(u+ \hat q \cdot x_1+ i \epsilon\right)- u (n\cdot x_1+i\epsilon) - (t_1-i\epsilon)^2 + \vec{x}_1^2, \\
        x_{23}^2 &= - r \left(u+ \hat q \cdot x_3- i \epsilon\right)- u (n\cdot x_3-i\epsilon) - (t_3+i\epsilon)^2 + \vec{x}_3^2, \\
    \end{split}
\end{equation}
this can be written equivalently as 
\begin{equation} \label{one-point-tuu-pre-int}
    \begin{split}
        \langle&\mathcal{O}(x_1) T_{uu}^{(2)}(u, z, \bar{z}) \mathcal{O}(x_3)\rangle
           \\ &= \frac{a}{4} \frac{1}{x_{13}^{2 \Delta-2}} \left[-\frac{3}{\left(u+ \hat q \cdot x_1+ i \epsilon\right)^2\left(u+ \hat q \cdot x_3- i \epsilon\right)^2}+\frac{1}{2} \partial_u^2 \frac{1}{\left(u+ \hat q \cdot x_1+ i \epsilon\right)\left(u+ \hat q \cdot x_3- i \epsilon\right)}\right].
    \end{split}
\end{equation}
The one-point function of the ANEC operator is then obtained by integrating with respect to $u$
\begin{equation} \label{one-point-anec}
    \begin{split}
         \langle\mathcal{O}(x_1) {\rm \mathcal{W}}_{-1}(z, \bar{z}) \mathcal{O}(x_3)\rangle
         &=  \langle\mathcal{O}(x_1) \int du~T_{uu}^{(2)}(u, z, \bar{z}) \mathcal{O}(x_3)\rangle 
          =  \frac{a}{4} \frac{1}{x_{13}^{2 \Delta-2}} \frac{12 \pi i }{(\hat q \cdot x_{13}+ i\epsilon)^3},
    \end{split}
\end{equation}
where note that the second term in \eqref{one-point-tuu-pre-int} does not contribute upon integration. 

Proceeding similarly for $\mathcal{W}_0$ and $\mathcal{W}_1$, we need the following additional components of the stress tensor evaluated at $\mathscr{J}^+$:
\begin{equation}
    \begin{split}
        \langle \mathcal{O}(x_1) T_{uz}^{(2)}(u, z, \bar{z}) \mathcal{O}(x_3)\rangle  
            &= \frac{a}{4} \frac{1}{x_{13}^{2 \Delta-2}}\lim_{r \to \infty} r^3 \frac{(n \cdot X_{13})(\partial_z\hat q \cdot X_{13})}{x_{12}^2 x_{23}^2},\\
         \langle  \mathcal{O}(x_1) T_{zz}^{(2)}(u, z, \bar{z}) \mathcal{O}(x_3)\rangle 
        &=  \frac{a}{4} \frac{1}{x_{13}^{2 \Delta-2}}\lim_{r \to \infty} r^4 \frac{(\partial_z\hat q \cdot X_{13})^2}{x_{12}^2 x_{23}^2}.
    \end{split}
\end{equation}
As before, it is helpful to separate total derivatives in $u$ that will not contribute upon direct integration with respect to $u$: 
\begin{equation}
    \begin{split}
         \langle & \mathcal{O}(x_1) T_{uz}^{(2)}(u, z, \bar{z}) \mathcal{O}(x_3)\rangle   \\
             &= 
        \frac{a}{4} \frac{1}{x_{13}^{2 \Delta-2}}\left[-\frac{3}{2} \frac{\partial_z  \left(\hat q\cdot x_1 + \hat q \cdot x_3 \right)}{  (u+ \hat q \cdot x_1+ i \epsilon)^2(u+ \hat q \cdot x_3- i \epsilon )^2}+\frac{1}{2}\partial_z \partial_u \frac{1}{  (u+ \hat q \cdot x_1+ i \epsilon)(u+ \hat q \cdot x_3- i \epsilon )}\right],
    \end{split}
\end{equation}
and
\begin{equation}
    \begin{split}
        \langle&  \mathcal{O}(x_1) T_{zz}^{(2)}(u, z, \bar{z}) \mathcal{O}(x_3)\rangle \\ 
             &= \frac{a}{4} \frac{1}{x_{13}^{2 \Delta-2}} \left[-\frac{1}{2}\frac{\left(\partial_z\hat q\cdot x_1 \right)^2+4\left(\partial_z\hat q\cdot x_1 \right)\left(\partial_z\hat q\cdot x_3 \right)+\left(\partial_z\hat q\cdot x_3 \right)^2 }{(u+ \hat q \cdot x_1+ i \epsilon)^2(u+ \hat q \cdot x_3- i \epsilon )^2} 
             \right. \\& \quad \quad \quad \quad\quad~\left.   
             +\frac{1}{2} \partial_z \partial_u \frac{\partial_z \left(\hat q\cdot x_1 + \hat q \cdot x_3 \right)}{(u+ \hat q \cdot x_1+ i \epsilon)(u+ \hat q \cdot x_3- i \epsilon )}-\frac{1}{2} \partial_u^2 \frac{(\partial_z\hat q \cdot x_1)(\partial_z\hat q \cdot x_3)}{(u+ \hat q \cdot x_1+ i \epsilon)(u+ \hat q \cdot x_3- i \epsilon )}\right].
    \end{split}
\end{equation}
Then, up to total derivatives in $u$, the expectation value of the combination of $T_{uu}^{(2)}$ and $T_{uz}^{(2)}$ that appears in $\mathcal{W}_0$ can be expressed in terms of the expectation value of $T_{uu}^{(2)}$: 
\begin{equation}
    \begin{split}
        \langle \mathcal{O}&(x_1) \frac{1}{2} \left(u \partial_z T_{uu}^{(2)}(u, z, \bar{z})- 2 T_{uz}^{(2)}(u, z, \bar{z}) \right) \mathcal{O}(x_3)\rangle \\
        &=- \frac{1}{4} \left[3 \left( \partial_z\hat q \cdot x_1\right)+   \left(\hat q \cdot x_1\right) \partial_z + 3 \left( \partial_z\hat q \cdot x_3\right)+   \left(\hat q \cdot x_3\right) \partial_z \right] \langle \mathcal{O}(x_1)  T_{uu}^{(2)}(u, z, \bar{z}) \mathcal{O}(x_3)\rangle\\
       & \quad- \frac{a}{16} \frac{1}{x_{13}^{2 \Delta-2}} \left[  \partial_u \partial_z \frac{1}{(u+ \hat q \cdot x_1+ i \epsilon)(u+ \hat q \cdot x_3- i \epsilon )} -  \partial_u^2  \frac{\partial_z \left(\hat q \cdot x_1+ \hat q \cdot x_3\right) }{\left(u+ \hat q \cdot x_1+ i \epsilon\right)\left(u+ \hat q \cdot x_3- i \epsilon\right)} \right.
       \\& \quad \quad \quad\quad \quad \quad\quad  \left.
       -\frac{1}{2} \partial_u^2 \partial_z   \left(\frac{1}{u+ \hat q \cdot x_1+ i \epsilon}+ \frac{1}{u+ \hat q \cdot x_3- i \epsilon} \right) \right].
    \end{split}
\end{equation}
As before, the total $u$-derivatives do not contribute upon integrating with respect to $u$ and we thus find 
\begin{equation}
    \begin{split}
          \langle \mathcal{O}&(x_1)\mathcal{W}_{0} (z, \bar{z})  \mathcal{O}(x_3)\rangle\\
         & =
          \langle \mathcal{O}(x_1) \frac{1}{2} \int du\left(u \partial_z T_{uu}^{(2)}(u, z, \bar{z})- 2 T_{uz}^{(2)}(u, z, \bar{z}) \right) \mathcal{O}(x_3)\rangle\\
        &= - \frac{1}{4} \left[3 \left( \partial_z\hat q \cdot x_1\right)+   \left(\hat q \cdot x_1\right) \partial_z + 3 \left( \partial_z\hat q \cdot x_3\right)+   \left(\hat q \cdot x_3\right) \partial_z \right] \langle \mathcal{O}(x_1)\mathcal{W}_{-1} (z, \bar{z})  \mathcal{O}(x_3)\rangle. 
    \end{split}
\end{equation}
Similarly, the expectation value of the combination of $T_{uu}^{(2)}$, $T_{uz}^{(2)}$ and $T_{zz}^{(2)}$ that appears in $\mathcal{W}_1$ can be expressed in terms of the expectation value of $T_{uu}^{(2)}$ up to total derivatives in $u$ that vanish upon integration:
\begin{equation}
    \begin{split}
        \langle \mathcal{O}(x_1) &\frac{1}{8} \left(u^2 \partial_z^2 T_{uu}^{(2)}(u, z, \bar{z})- 4 u \partial_z T_{uz}^{(2)}(u, z, \bar{z})+6T_{zz}^{(2)}(u, z, \bar{z}) \right) \mathcal{O}(x_3)\rangle \\
        &= \frac{1}{32} \prod_{j=3}^{4} \left[  j  \partial_z\hat q \cdot \left(x_1+x_3\right)+  \hat q \cdot \left( x_1+x_3\right) \partial_z \right] \langle \mathcal{O}(x_1)  T_{uu}^{(2)}(u, z, \bar{z}) \mathcal{O}(x_3)\rangle + \partial_u(\cdots),
    \end{split}
\end{equation}
thus yielding
\begin{equation}
    \begin{split}
        \langle \mathcal{O}(x_1)\mathcal{W}_{1} (z, \bar{z})  \mathcal{O}(x_3)\rangle  =\frac{1}{32} \prod_{j=3}^{4}\left[  j  \partial_z\hat q \cdot \left(x_1+x_3\right)+  \hat q \cdot \left( x_1+x_3\right) \partial_z \right]\langle \mathcal{O}(x_1)\mathcal{W}_{-1} (z, \bar{z})  \mathcal{O}(x_3)\rangle.
    \end{split}
\end{equation}

At the next order, we begin to encounter expressions that involve non-local contributions in the transverse plane.  The contributions from $T_{uu}^{(2)}$, $T_{uz}^{(2)}$ and $T_{zz}^{(2)}$ remain local and in particular, at the next order we find 
\begin{equation} \label{pre-int-leading-w2}
    \begin{split}
        \langle \mathcal{O}(x_1) \frac{1}{2^3 3!
        }&\left(u^3 \partial_z^3 T_{uu}^{(2)} - 6u^2 \partial_z^2 T_{uz}^{(2)} + 18 u \partial_z T_{zz}^{(2)}\right)\mathcal{O}(x_3) \rangle \\
        &=-   \frac{1}{2^6 3!}\prod_{j=3}^{5}\left[  j  \partial_z\hat q \cdot \left(x_1+x_3\right)+  \hat q \cdot \left( x_1+x_3\right) \partial_z \right]\langle \mathcal{O}(x_1)  T_{uu}^{(2)}(u, z, \bar{z}) \mathcal{O}(x_3)\rangle   
        \\& \quad \quad -  \frac{3a}{16}\frac{1}{x_{13}^{2 \Delta-2}}   \frac{(\partial_z \hat q \cdot x_1+\partial_z \hat q \cdot x_3)(\partial_z \hat q \cdot x_1)(\partial_z \hat q \cdot x_3)}{\left(u+ \hat q \cdot x_1+ i \epsilon\right)^2\left(u+ \hat q \cdot x_3- i \epsilon\right)^2} + \partial_u \left(\cdots \right),
    \end{split}
\end{equation}
where again the total-$u$ derivatives vanish upon integration with respect to $u$. The primary descendant of $\mathcal{W}_2$ also involves
\begin{equation} \label{one-point-trr5}
    \begin{split}
         \langle &\mathcal{O}(x_1) \frac{1}{2^3}\left(\partial_z^2T_{rr}^{(5)} + 4 \partial_z  T_{rz}^{(4)} + 4 T_{zz}^{(3)}\right)\mathcal{O}(x_3) \rangle\\
            &=\frac{3a}{16} \frac{1}{x_{13}^{2 \Delta-2}} \partial_{\bar{z}} \left(\frac{(\partial_z \hat q \cdot x_1+\partial_z \hat q \cdot x_3)(\partial_z \hat q \cdot x_1)(\partial_z \hat q \cdot x_3)}{\left(u+ \hat q \cdot x_1+ i \epsilon\right)^2\left(u+ \hat q \cdot x_3- i \epsilon\right)^2}\right) \\& \quad - \frac{3a}{16} \frac{1}{x_{13}^{2 \Delta-2}}  \partial_u \left( \left(\partial_z \frac{\partial_z \hat q \cdot x_1}{u+\hat q \cdot x_1+i\epsilon}\right) \left(\partial_{\bar{z}} \frac{\partial_z \hat q \cdot x_3}{u+\hat q \cdot x_3-i\epsilon}\right)+\left(\partial_{\bar{z}} \frac{\partial_z \hat q \cdot x_1}{u+\hat q \cdot x_1+i\epsilon}\right) \left(\partial_{z} \frac{\partial_z \hat q \cdot x_3}{u+\hat q \cdot x_3-i\epsilon}\right)\right).
    \end{split}
\end{equation}
Here, the one-point function of $T_{rr}^{(5)}$ vanishes in agreement with general fall-off conditions in conformal field theory \eqref{cft-fall-off}. Notice also that the first term is a total derivative in $\bar{z}$ and, when forming the $\mathcal{W}_2$ generator, exactly cancels the first term in second line of \eqref{pre-int-leading-w2}.  The second line is not a total derivative in $\bar{z}$, but it is a total derivative in $u$ that vanishes upon integrating with respect to $u$. Importantly, the $u$-integral of the last line  continues to vanish even if the $\bar{z}$ integral is performed first to extract the primary from the primary descendant.  Hence, up to terms that vanish upon integration with respect to $u$, the $u$-integrand of the expectation value of $\mathcal{W}_2$ is a local expression on the transverse plane that can be written in terms of the expectation value of $T_{uu}^{(2)}$. Collecting all of these results, the expectation value of $\mathcal{W}_2$ can be written simply as
\begin{equation}
    \begin{split}
        \langle \mathcal{O}(x_1) \mathcal{W}_2(z, \bar{z})\mathcal{O}(x_3) \rangle
       =-   \frac{1}{2^6 3!}\prod_{j=3}^{5}\left[  j  \partial_z\hat q \cdot \left(x_1+x_3\right)+  \hat q \cdot \left( x_1+x_3\right) \partial_z \right]\langle \mathcal{O}(x_1)\mathcal{W}_{-1} (z, \bar{z})  \mathcal{O}(x_3)\rangle  .  
    \end{split}
\end{equation}

We expect this pattern to hold for all $\mathcal{W}_m$ generators. To  compare directly with the convergence properties of the $\mathcal{L}_m$ operators, we explicitly calculate the one-point functions of $\mathcal{W}_3$ and $\mathcal{W}_4$ since $\mathcal{L}_4$ is the first operator in the $\mathcal{L}_m$ family with a divergent one-point function. We find the $T_{uu}^{(2)}$, $T_{uz}^{(2)}$, and $T_{zz}^{(2)}$ contributions to be given respectively by
\begin{equation}
    \begin{split}
        \langle &\mathcal{O}(x_1) \frac{1}{2^4 4!
        }\left(u^4 \partial_z^4 T_{uu}^{(2)} - 8u^3 \partial_z^3 T_{uz}^{(2)} + 36 u^2 \partial_z^2 T_{zz}^{(2)}\right)\mathcal{O}(x_3) \rangle\\
        &= \frac{1}{2^8 4!}\prod_{j=3}^{6}\left[  j  \partial_z\hat q \cdot \left(x_1+x_3\right)+  \hat q \cdot \left( x_1+x_3\right) \partial_z \right] \langle \mathcal{O}(x_1)  T_{uu}^{(2)}(u, z, \bar{z}) \mathcal{O}(x_3)\rangle\\
        & \quad +\frac{3a}{2^7}\frac{1}{x_{13}^{2 \Delta-2}} \left[ \partial_z \frac{\partial_z( \hat q \cdot x_1+ \hat q \cdot x_3)^2(\partial_z \hat q \cdot x_1)(\partial_z \hat q \cdot x_3)}{\left(u+ \hat q \cdot x_1+ i \epsilon\right)^2\left(u+ \hat q \cdot x_3- i \epsilon\right)^2}
        +2\frac{\partial_z^2( \hat q \cdot x_1+ \hat q \cdot x_3)^2(\partial_z \hat q \cdot x_1)(\partial_z \hat q \cdot x_3)}{\left(u+ \hat q \cdot x_1+ i \epsilon\right)^2\left(u+ \hat q \cdot x_3- i \epsilon\right)^2}\right. \\& \quad \quad \quad   \quad \quad  \quad \quad\left. -6\frac{(\partial_z \hat q \cdot x_1)^2(\partial_z \hat q \cdot x_3)^2}{\left(u+ \hat q \cdot x_1+ i \epsilon\right)^2\left(u+ \hat q \cdot x_3- i \epsilon\right)^2}
        \right]+ \partial_u \left(\cdots\right)
    \end{split}
\end{equation}
and 
\begin{equation}
    \begin{split}
        \langle &\mathcal{O}(x_1) \frac{1}{2^5 5!
        }\left(u^5 \partial_z^5 T_{uu}^{(2)} - 10u^4 \partial_z^4 T_{uz}^{(2)} +  60 u^3 \partial_z^3
        T_{zz}^{(2)}\right)\mathcal{O}(x_3) \rangle\\
        &= -
        \frac{1}{2^{10}5!}\prod_{j=3}^{7} \left[  j  \partial_z\hat q \cdot \left(x_1+x_3\right)+  \hat q \cdot \left( x_1+x_3\right) \partial_z \right]\langle \mathcal{O}(x_1)  T_{uu}^{(2)}(u, z, \bar{z}) \mathcal{O}(x_3)\rangle\\
        &\quad  + \frac{a}{2^9} \frac{1}{x_{13}^{2 \Delta-2}} \left[
            -  \partial_z^2 \frac{\partial_z(\hat q \cdot x_1 +\hat q\cdot x_3)^3 (\partial_z \hat q \cdot x_1)(\partial_z \hat q \cdot x_3)}{\left(u+ \hat q \cdot x_1+ i \epsilon\right)^2\left(u+ \hat q \cdot x_3- i \epsilon\right)^2}
            -2 \partial_z \frac{\partial_z^2(\hat q \cdot x_1 +\hat q\cdot x_3)^3 (\partial_z \hat q \cdot x_1)(\partial_z \hat q \cdot x_3)}{\left(u+ \hat q \cdot x_1+ i \epsilon\right)^2\left(u+ \hat q \cdot x_3- i \epsilon\right)^2}
        \right. \\& \quad \quad\quad \quad \quad \quad \quad\left.   
            -3 \frac{\partial_z^3(\hat q \cdot x_1 +\hat q\cdot x_3)^3 (\partial_z \hat q \cdot x_1)(\partial_z \hat q \cdot x_3)}{\left(u+ \hat q \cdot x_1+ i \epsilon\right)^2\left(u+ \hat q \cdot x_3- i \epsilon\right)^2}
            +18\partial_z \frac{(\hat q \cdot x_1 +\hat q\cdot x_3) (\partial_z \hat q \cdot x_1)^2(\partial_z \hat q \cdot x_3)^2}{\left(u+ \hat q \cdot x_1+ i \epsilon\right)^2\left(u+ \hat q \cdot x_3- i \epsilon\right)^2}
        \right. \\& \quad \quad \quad \quad\quad\quad \quad \left. 
            +36\frac{\partial_z(\hat q \cdot x_1 +\hat q\cdot x_3) (\partial_z \hat q \cdot x_1)^2(\partial_z \hat q \cdot x_3)^2}{\left(u+ \hat q \cdot x_1+ i \epsilon\right)^2\left(u+ \hat q \cdot x_3- i \epsilon\right)^2}
        \right]+\partial_u(\cdots),
    \end{split}
\end{equation}
where as before, the total $u$-derivatives vanish upon integration.   We will also need
\begin{equation} \label{pre-int-trr6}
    \begin{split}
        \langle &\mathcal{O}(x_1) \left(\partial_z^2T_{rr}^{(6)} +6 \partial_z  T_{rz}^{(5)} +10 T_{zz}^{(4)}\right)\mathcal{O}(x_3) \rangle\\
        &= \frac{a}{4} \frac{1}{x_{13}^{2 \Delta-2}} \left[ - 3 \partial_{\bar{z}}^2 \frac{\partial_z^2( \hat q \cdot x_1+ \hat q \cdot x_3)^2(\partial_z \hat q \cdot x_1)(\partial_z \hat q \cdot x_3)}{\left(u+ \hat q \cdot x_1+ i \epsilon\right)^2\left(u+ \hat q \cdot x_3- i \epsilon\right)^2} +9  \partial_{\bar{z}}^2 \frac{(\partial_z \hat q \cdot x_1)^2(\partial_z \hat q \cdot x_3)^2}{\left(u+ \hat q \cdot x_1+ i \epsilon\right)^2\left(u+ \hat q \cdot x_3- i \epsilon\right)^2}  \right. \\& \left.  \quad \quad   \quad  - 6 \partial_z \partial_{\bar{z}}  \left[  \left(\partial_z \frac{\partial_z \hat q \cdot x_1}{u+\hat q \cdot x_1+i\epsilon}\right) \left(\partial_{\bar{z}} \frac{\partial_z \hat q \cdot x_3}{u+\hat q \cdot x_3-i\epsilon}\right)+\left(\partial_{\bar{z}} \frac{\partial_z \hat q \cdot x_1}{u+\hat q \cdot x_1+i\epsilon}\right) \left(\partial_{z} \frac{\partial_z \hat q \cdot x_3}{u+\hat q \cdot x_3-i\epsilon}\right)\right]
    \right. \\& \left.  \quad \quad   \quad 
        +6 \partial_u \partial_z \partial_{\bar{z}}^2 \frac{\partial_z( \hat q \cdot x_1+ \hat q \cdot x_3)(\partial_z \hat q \cdot x_1)(\partial_z \hat q \cdot x_3)}{(u+\hat q \cdot x_1+i\epsilon)(u+\hat q \cdot x_3-i\epsilon)}
        -6 \partial_u^2 \partial_{\bar{z}}^2 \frac{ (\partial_z \hat q \cdot x_1)^2(\partial_z \hat q \cdot x_3)^2}{(u+\hat q \cdot x_1+i\epsilon)(u+\hat q \cdot x_3-i\epsilon)}
    \right. \\& \left.  \quad \quad   \quad 
        -6 \partial_u \partial_z \left[
            \frac{\partial_z \hat q \cdot x_1}{u+\hat q \cdot x_1+i\epsilon} \left(\partial_{\bar{z}}^2 \frac{(\partial_z \hat q \cdot x_3)^2}{u+\hat q \cdot x_3-i\epsilon}\right)+ \left(\partial_{\bar{z}}^2 \frac{(\partial_z \hat q \cdot x_1)^2}{u+\hat q \cdot x_1+i\epsilon}\right)\frac{\partial_z \hat q \cdot x_3}{u+\hat q \cdot x_3-i\epsilon}
        \right]
    \right. \\& \left.  \quad \quad   \quad 
        +15 \partial_u \left[ \left(\partial_u \frac{(\partial_z \hat q \cdot x_1)^2}{u+\hat q \cdot x_1+i\epsilon} \right)\left(\partial_{\bar{z}}^2 \frac{(\partial_z \hat q \cdot x_3)^2}{u+\hat q \cdot x_3-i\epsilon}\right)+ \left(\partial_{\bar{z}}^2 \frac{(\partial_z \hat q \cdot x_1)^2}{u+\hat q \cdot x_1+i\epsilon}\right)\left(\partial_u\frac{(\partial_z \hat q \cdot x_3)^2}{u+\hat q \cdot x_3-i\epsilon}\right)\right] 
    \right]
    \end{split}
\end{equation}
and 
\begin{equation}\label{pre-int-trr7}
    \begin{split}
        \langle &\mathcal{O}(x_1) \left(\partial_z^2T_{rr}^{(7)} +8 \partial_z  T_{rz}^{(6)} +18 T_{zz}^{(5)}\right)\mathcal{O}(x_3) \rangle\\
        &=  \frac{a}{4} \frac{1}{x_{13}^{2 \Delta-2}} \Bigg\{\frac{9}{2} \partial_{\bar{z}}^3   \frac{\partial_z( \hat q \cdot x_1+ \hat q \cdot x_3)(\partial_z \hat q \cdot x_1)^2(\partial_z \hat q \cdot x_3)^2}{\left(u+ \hat q \cdot x_1+ i \epsilon\right)^2\left(u+ \hat q \cdot x_3- i \epsilon\right)^2}  
    \\&   \quad 
        -3 \partial_z^2 \partial_{\bar{z}}\left[
             \frac{\partial_z \hat q \cdot x_1}{u+\hat q \cdot x_1+i\epsilon} \left(\partial_{\bar{z}}^2 \frac{(\partial_z \hat q \cdot x_3)^2}{u+\hat q \cdot x_3-i\epsilon}\right)+  
           \left(\partial_{\bar{z}}^2 \frac{(\partial_z \hat q \cdot x_1)^2}{u+\hat q \cdot x_1+i\epsilon}\right)\frac{\partial_z \hat q \cdot x_3}{u+\hat q \cdot x_3-i\epsilon}
        \right] 
    \\&   \quad +\frac{15}{2}\partial_z \partial_{\bar{z}} \left[ \left(\partial_u \frac{(\partial_z \hat q \cdot x_1)^2}{u+\hat q \cdot x_1+i\epsilon} \right)\left(\partial_{\bar{z}}^2 \frac{(\partial_z \hat q \cdot x_3)^2}{u+\hat q \cdot x_3-i\epsilon}\right)+ \left(\partial_{\bar{z}}^2 \frac{(\partial_z \hat q \cdot x_1)^2}{u+\hat q \cdot x_1+i\epsilon}\right)\left(\partial_u\frac{(\partial_z \hat q \cdot x_3)^2}{u+\hat q \cdot x_3-i\epsilon}\right)
        \right]
     \\&   \quad 
        +3 \partial_u \left[
             \frac{\partial_z \hat q \cdot x_1}{u+\hat q \cdot x_1+i\epsilon} \left(\partial_{\bar{z}}^3 \partial_u\frac{(\partial_z \hat q \cdot x_3)^4}{u+\hat q \cdot x_3-i\epsilon}\right)+  
           \left(\partial_{\bar{z}}^3\partial_u \frac{(\partial_z \hat q \cdot x_1)^4}{u+\hat q \cdot x_1+i\epsilon}\right)\frac{\partial_z \hat q \cdot x_3}{u+\hat q \cdot x_3-i\epsilon}
        \right]
      \\&   \quad 
        -3 \partial_u\left[
             \frac{(\partial_z \hat q \cdot x_1)^2}{u+\hat q \cdot x_1+i\epsilon} \left(\partial_{\bar{z}}^3 \partial_u\frac{(\partial_z \hat q \cdot x_3)^3}{u+\hat q \cdot x_3-i\epsilon}\right)+  
           \left(\partial_{\bar{z}}^3\partial_u \frac{(\partial_z \hat q \cdot x_1)^3}{u+\hat q \cdot x_1+i\epsilon}\right)\frac{(\partial_z \hat q \cdot x_3)^2}{u+\hat q \cdot x_3-i\epsilon}
        \right]
      \\&   \quad 
            +\frac{27}{2} \partial_u \partial_{\bar{z}} 
            \left[
             \frac{(\partial_z \hat q \cdot x_1)^2}{u+\hat q \cdot x_1+i\epsilon} \left(\partial_{\bar{z}}^2
             \partial_u\frac{(\partial_z \hat q \cdot x_3)^3}{u+\hat q \cdot x_3-i\epsilon}\right)+  
           \left(\partial_{\bar{z}}^2\partial_u \frac{(\partial_z \hat q \cdot x_1)^3}{u+\hat q \cdot x_1+i\epsilon}\right)\frac{(\partial_z \hat q \cdot x_3)^2}{u+\hat q \cdot x_3-i\epsilon}
        \right]
      \\&   \quad 
            -\frac{9}{2} \partial_u \partial_{\bar{z}}^3
                \left[
             \frac{(\partial_z \hat q \cdot x_1)^2}{u+\hat q \cdot x_1+i\epsilon} \left(  \partial_u\frac{(\partial_z \hat q \cdot x_3)^3}{u+\hat q \cdot x_3-i\epsilon}\right)+  
           \left( \partial_u \frac{(\partial_z \hat q \cdot x_1)^3}{u+\hat q \cdot x_1+i\epsilon}\right)\frac{(\partial_z \hat q \cdot x_3)^2}{u+\hat q \cdot x_3-i\epsilon}
        \right]
        \\& \quad 
            -6 \partial_u^2\left[
             \frac{(\partial_z \hat q \cdot x_1)^2}{u+\hat q \cdot x_1+i\epsilon} \left(\partial_{\bar{z}}^3  \frac{(\partial_z \hat q \cdot x_3)^3}{u+\hat q \cdot x_3-i\epsilon}\right)+  
           \left(\partial_{\bar{z}}^3  \frac{(\partial_z \hat q \cdot x_1)^3}{u+\hat q \cdot x_1+i\epsilon}\right)\frac{(\partial_z \hat q \cdot x_3)^2}{u+\hat q \cdot x_3-i\epsilon}
        \right] \Bigg\}.
    \end{split}
\end{equation}
Using \eqref{one-point-trr5}, together with \eqref{pre-int-trr6} and \eqref{pre-int-trr7}, the integrands for expectation values of $\mathcal{W}_3$ and $\mathcal{W}_4$ can be written as local expressions in the transverse plane up to total $u$-derivative terms that vanish upon integrating with respect to $u$.  Importantly, the total $u$-derivatives in \eqref{pre-int-trr6} and \eqref{pre-int-trr7} converge sufficiently fast when integrating with respect to $u$ that they continue to give a vanishing contribution even upon first integrating in $\bar{z}$ to extract the primary $\mathcal{W}_m$ from the primary descendant $\partial_{\bar{z}}^{m-1}\mathcal{W}_{m}$. 

Thus, up to and including $m=4$, we obtain the explicit result
\begin{equation} \label{one-point-position-space-all-m}
    \begin{split}
        \langle\mathcal{O}(x_1) &\mathcal{W}_m(z, \bar{z})\mathcal{O}(x_3) \rangle
        \\
       &=\frac{(-1)^{m+1}}{2^{2m+2} (m+1)!}\prod_{j=3}^{m+3}\left[  j  \partial_z\hat q \cdot \left(x_1+x_3\right)+  \hat q \cdot \left( x_1+x_3\right) \partial_z \right]  \langle \mathcal{O}(x_1)\mathcal{W}_{-1} (z, \bar{z})  \mathcal{O}(x_3)\rangle .
    \end{split}
\end{equation}
Moreover, we expect the one-point functions for all of the ${\rm w}_{1+\infty}$ generators to take the above form, where the one-point function of the ANEC operator $\mathcal{W}_{-1}$ is given in \eqref{one-point-anec}. 

\subsubsection{Momentum-space One-point Functions and Soft Factors} 

The direct relation of \eqref{one-point-position-space-all-m} to universal soft factors arising in the soft expansion of a graviton scattering amplitude is made explicit by transforming to momentum space.  The momentum-space representation for the one-point function of the ANEC operator
\begin{equation} \label{one-point-ANEC-momentumspace}
    \begin{split}
         \langle \mathcal{O}(p_1)|\mathcal{W}_{-1}(z, \bar{z})| \mathcal{O}(p_3)\rangle
         &\equiv
        \int d^4 x_1 \int d^4 x_3 ~e^{-ip_1\cdot x_1 +i p_3 \cdot x_3}
        \langle \mathcal{O}(x_1) \int du~ T_{uu}^{(2)}(u, z, \bar{z}) \mathcal{O}(x_3)\rangle 
    \end{split}
\end{equation}
is well-known from the literature \cite{Hofman:2008ar,Belitsky:2013xxa,Belitsky:2013bja,Bautista:2019qxj,Belin:2020lsr} and evaluates to
\begin{equation} \label{momentum-2-point}
    \begin{split}
         \langle \mathcal{O}(p_1)| \mathcal{W}_{-1}(z, \bar{z})| \mathcal{O}(p_3)\rangle
        = \frac{1}{4\pi}   \frac{(-p_1^2)^{2}}{\left(- \frac{1}{2} \hat q \cdot p_1\right)^3} \langle \mathcal{O}(p_1)|\mathcal{O}(p_3)\rangle,
    \end{split}
\end{equation}
where the Fourier transform of the two-point function, repeatedly defined here for convenience, is
\begin{equation} \label{eq:actual-momentum-2-point}
    \begin{split}
       \langle \mathcal{O}(p_1)|\mathcal{O}(p_3)\rangle
       & \equiv\int d^4 x_1 \int d^4 x_3 ~e^{-ip_1\cdot x_1+i p_3 \cdot x_3} \langle \mathcal{O}(x_1) \mathcal{O}(x_3) \rangle\\
       & = (2 \pi)^4 \delta^{(4)}(p_1-p_3)\frac{(2 \pi)^3 \Theta(p_1^0- |\vec{p}_1|)}{ \Gamma(\Delta) \Gamma(\Delta-1) }\frac{(-p_1^2)^{\Delta-2}}{2^{2\Delta-2}}. 
    \end{split}
\end{equation}
We then extend the calculation of \eqref{momentum-2-point} to the one-point functions for general $\mathcal{W}_m$ generators
\begin{equation}
    \begin{split}
         \langle\mathcal{O}(p_1)| \mathcal{W}_m(z, \bar{z})|\mathcal{O}(p_3) \rangle  \equiv
        \int d^4 x_1 \int d^4 x_3 ~e^{-ip_1\cdot x_1+i p_3 \cdot x_3}\langle\mathcal{O}(x_1) \mathcal{W}_m(z, \bar{z})\mathcal{O}(x_3) \rangle 
    \end{split}
\end{equation}
by using our position-space formula \eqref{one-point-position-space-all-m}.  The Fourier transform implements the replacement $x_1 \to i \partial_{p_1}$ and $x_3 \to -i \partial_{p_3}$ and we find
\begin{equation} \label{wm-momentum-int}
    \begin{split}
        \langle\mathcal{O}&(p_1)| \mathcal{W}_m(z, \bar{z})|\mathcal{O}(p_3) \rangle \\
        &=\frac{(-1)^{m+1}}{2^{2m+2} (m+1)!}\prod_{j=3}^{m+3}\left[  j  \partial_z\hat q \cdot \left(i \frac{\partial}{\partial p_1}-i \frac{\partial}{\partial p_3}\right)+  \hat q \cdot \left(i \frac{\partial}{\partial p_1}-i \frac{\partial}{\partial p_3}\right) \partial_z \right] 
         \langle\mathcal{O}(p_1)| \mathcal{W}_{-1}(z, \bar{z})|\mathcal{O}(p_3) \rangle.
    \end{split}
\end{equation}
Then, using \eqref{momentum-2-point}, \eqref{wm-momentum-int} simplifies to the form
\begin{equation} \label{wm-momentum-final}
    \begin{split}
        \langle\mathcal{O}(p_1)| \mathcal{W}_m(z, \bar{z})|\mathcal{O}(p_3) \rangle   
        &=\frac{1}{4\pi}{m+3 \choose 2}    \frac{(-p_1^2)^{2}\left( \partial_z \hat n^\mu \hat n^\nu \mathcal{L}_{1\mu\nu} \right)^{m+1}}{\left(-   \hat n \cdot p_1\right)^{m+4}} 
         \langle \mathcal{O}(p_1)|\mathcal{O}(p_3)\rangle,
    \end{split}
\end{equation}
where $\hat n \equiv \frac{1}{2} \hat q$ is the appropriately normalized ``unit" null vector in the coordinate system \eqref{flat-retarded-Bondi-coord} and $\mathcal{L}_{\mu \nu}$ is the momentum-space representation of the orbital angular momentum
\begin{equation} \label{orbital-angular-momentum}
    \begin{split}
        \mathcal{L}_{k\mu\nu} = -i \left( p_{k\mu} \frac{\partial}{ \partial p^\nu_k}-p_{k\nu} \frac{\partial}{ \partial p^\mu_k} \right).
    \end{split}
\end{equation}
The coefficient of the two-point function on the right-hand side of \eqref{wm-momentum-final} can be precisely identified with (derivatives of) the universal soft factor associated to the sub$^{m+1}$leading soft theorem, as we will now explain.  

It has been recently understood that universal behavior of scattering amplitudes in the low-energy limit of an emitted graviton with momentum $\omega \hat{n}$ in fact extends in a certain sense to all orders in the soft expansion \cite{Li:2018gnc,Hamada:2018vrw}. In particular, expanding a tree-level amplitude to all orders,
\begin{equation}
    \begin{split}
        \mathcal{A}_{n+1} (\omega \hat n; p_1, \cdots ,p_n)
             = \frac{\kappa}{2}\sum_{m = -1}^\infty \omega^m \mathcal{A}_{n+1}^{(m)} ( \hat n; p_1, \cdots ,p_n), \quad \quad \kappa^2 = 32 \pi G,
    \end{split}
\end{equation}
each order in the energy expansion can be written as 
\begin{equation}\label{subnleading}
    \begin{split}
        \mathcal{A}_{n+1}^{(m)}(\hat n; p_1,\cdots,p_n) & = \sum_{k = 1}^n S^{(m)}_k(\hat n)
        \mathcal{A}_{n}( p_1, \cdots ,p_n) +\varepsilon^-_{\mu\nu} \mathcal{B}^{\mu\nu}_{(m)} (\hat n; p_1, \cdots, p_n).
    \end{split}
\end{equation}  
Here $S^{(m)}_k$ is the ``universal'' soft factor associated to the sub$^{m+1}$leading soft theorem and $\mathcal{B}^{\mu\nu}_{(m)}$ is the ``non-universal'' contribution that first appears at $m =1$.\footnote{But is highly constrained for purely massless scattering at $m = 1$ \cite{Elvang:2016qvq}.}  From an asymptotic symmetry perspective, $S_k^{(m)}$ specifies the universal transformation of the $k$-th particle under infinite-dimensional asymptotic symmetries and can be chosen to transform under Lorentz transformations as an ${\rm SL}(2, \mathbb{C})$ primary of weight $(\frac{-m-2}{2},\frac{-m+2}{2})$ \cite{Himwich:2021dau,Himwich:2023njb}.  When the $k$-th particle is a massive scalar, the explicit expression for $S_k^{(m)}$ is somewhat subtle but its primary descendant takes the simple unambiguous form \cite{Himwich:2023njb} 
\begin{equation} \label{pd-1}
    \begin{split}
        \partial_{z}^{m+3} S^{(m)}_k (\hat n) 
        &= i^{m+3} {m+3 \choose 2} \frac{p_k^4}{(-\hat n \cdot  p_k)^{m+4}} \left( \partial_{z}\hat n^\mu \hat n^\nu   \mathcal{L}_k{}_{\mu \nu} \right)^{m+1}.
    \end{split}
\end{equation} 
Up to the factors of $i$ and $4\pi$, this  \emph{precisely} matches the expression in \eqref{wm-momentum-final}. Physically, the exact expression in \eqref{wm-momentum-final} or equivalently $\frac{1}{4 \pi} (-i)^{m+3}\partial_{z}^{m+3} S^{(m)}_k (\hat n)$ are the values of Coulombic modes of the field associated with the $(m+2)$-th multipole moment \cite{Hoskins:2026new}. 

\subsection{Currents} \label{subsec:1-pt-S}

The derivation of one-point functions of the $\mathcal{S}_m$ generators in the background of a scalar field proceeds by direct analogy with the derivation for the $\mathcal{W}_m$ generators.  The input for this calculation is the three-point function between two complex scalars and a ${\rm U}(1)$ current \cite{Osborn:1993cr,Freedman:1998tz},
\begin{equation}
    \begin{split}
        \langle \mathcal{O}^\dagger(x_1) j_{\mu}(x_2) \mathcal{O} (x_3)\rangle
         = \frac{bX_{13}{}_\mu}{x_{12}^2 x_{23}^2 x_{13}^{2\Delta-2}} , 
    \end{split}
\end{equation}
where $X_{13}$ is defined in \eqref{def-capital-X} and $b$ is a constant that is proportional to the ${\rm U}(1)$ charge $Q$ of the scalar field.  To determine the one-point function of $\mathcal{S}_0$, we first evaluate the $u$ component of $j_\mu$ at null infinity:
\begin{equation}
    \begin{split}
        \langle \mathcal{O}^\dagger&(x_1) j_{u}^{(2)}(u,z,\bar{z}) \mathcal{O} (x_3)\rangle\\
        & = \frac{b}{2x_{13}^{2\Delta-2}} \lim_{r \to \infty} r^2 \frac{(n \cdot X_{13})}{x_{12}^2 x_{23}^2}\\& 
        =-\frac{b}{2x_{13}^{2\Delta-2}} \left[\frac{1}{\left(u+ \hat q \cdot x_1+ i \epsilon\right)^2\left(u+ \hat q \cdot x_3- i \epsilon\right)}-\frac{1}{\left(u+ \hat q \cdot x_1+ i \epsilon\right)\left(u+ \hat q \cdot x_3- i \epsilon\right)^2} \right],
    \end{split}
\end{equation}
and then integrate with respect to $u$:
\begin{equation}
    \begin{split}
        \langle \mathcal{O}^\dagger(x_1) \mathcal{S}_0(z, \bar{z})  \mathcal{O}  (x_3)\rangle
        &= \langle \mathcal{O}^\dagger(x_1) \int du~j_{u}^{(2)}(u,z,\bar{z}) \mathcal{O}  (x_3)\rangle
          = \frac{b}{2x_{13}^{2\Delta-2}} \frac{-4 \pi i}{\left(\hat q \cdot x_{13}+i \epsilon\right)^2}.
    \end{split}
\end{equation}
The one-point functions of general $\mathcal{S}_m$ generators can be calculated in a similar way and are ultimately related to the one-point function of $\mathcal{S}_0$:
\begin{equation} \label{s-one-point-position}
    \begin{split}
         \langle \mathcal{O}^\dagger(x_1) \mathcal{S}_m(z, \bar{z})  \mathcal{O}  (x_3)\rangle
         = \frac{(-1)^{m}}{2^{2m}m!} \prod_{j=2}^{m+1} \left[j \partial_z \hat q \cdot (x_1+x_3) + \hat q\cdot (x_1+x_3) \partial_z\right]\langle \mathcal{O}^\dagger(x_1) \mathcal{S}_0(z, \bar{z})  \mathcal{O}  (x_3)\rangle. 
    \end{split}
\end{equation}
We have checked this relation explicitly up to $m=4$ and expect it to hold for general $m$.  Notice that as for the $\mathcal{W}_m$ generators, although the expression for $\mathcal{S}_m$ in terms of null integrals of $j_\mu$ \eqref{def-s-operator} is non-local in the transverse plane for $m>1$, the expression for the one-point function \eqref{s-one-point-position} is local. 

Fourier-transforming to momentum space,
\begin{equation}
    \begin{split}
        \langle \mathcal{O} (p_1) |\mathcal{S}_0(z, \bar{z})  |\mathcal{O}  (p_3)\rangle
        &\equiv \int d^4x_1 \int d^4 x_3~e^{-ip_1 \cdot x_1 +i p_3 \cdot x_3}
        \langle \mathcal{O}^\dagger(x_1) \mathcal{S}_0(z, \bar{z})  \mathcal{O}  (x_3)\rangle,
    \end{split}
\end{equation}
we recover the one-point function of $\mathcal{S}_0$ \cite{Hofman:2008ar,Belitsky:2013xxa,Belitsky:2013bja}: 
\begin{equation} \label{s0-mom-1-point}
    \begin{split}
         \langle \mathcal{O}(p_1) |\mathcal{S}_0(z, \bar{z})  |\mathcal{O}  (p_3)\rangle
         = \frac{Q}{4 \pi} \frac{(- p_1^2)}{(-\frac{1}{2} \hat q \cdot p_1)^2}\langle \mathcal{O}(p_1)  |\mathcal{O}  (p_3)\rangle,
    \end{split}
\end{equation}
where here 
\begin{equation}
    \begin{split}
        \langle \mathcal{O}(p_1)  |\mathcal{O}  (p_3)\rangle
            \equiv \int d^4 x_1 \int d^4 x_3 ~e^{-ip_1\cdot x_1+i p_3 \cdot x_3} \langle \mathcal{O}^\dagger(x_1) \mathcal{O}(x_3) \rangle. 
    \end{split}
\end{equation}
This readily extends to give the following result for general $m$:
\begin{equation}
    \begin{split}
        \langle \mathcal{O}&(p_1) |\mathcal{S}_m(z, \bar{z})  |\mathcal{O}  (p_3)\rangle\\
        &\equiv\int d^4x_1 \int d^4 x_3~e^{-ip_1 \cdot x_1 +i p_3 \cdot x_3}
         \langle \mathcal{O}^\dagger(x_1) \mathcal{S}_m(z, \bar{z})  \mathcal{O} (x_3)\rangle
        \\
         &= \frac{(-1)^{m}}{2^{2m}m!} \prod_{j=2}^{m+1} \left[j \partial_z \hat q \cdot  \left(i \frac{\partial}{\partial p_1} -i \frac{\partial}{\partial p_3}  \right)  + \hat q\cdot  \left(i \frac{\partial}{\partial p_1} -i \frac{\partial}{\partial p_3}  \right) \partial_z\right]
         \langle \mathcal{O}(p_1) |\mathcal{S}_0(z, \bar{z})  |\mathcal{O}  (p_3)\rangle.
    \end{split}
\end{equation}
Using \eqref{s0-mom-1-point}, this simplifies to 
\begin{equation}
    \begin{split}
        \langle \mathcal{O}(p_1) |\mathcal{S}_m(z, \bar{z})  |\mathcal{O}  (p_3)\rangle
         &= (m+1) \frac{Q}{4 \pi} \frac{(- p_1^2)\left(  \partial_z \hat n^\mu  \hat n^\nu \mathcal{L}_{1\mu\nu}\right)^m}{(-\hat n \cdot p_1)^{m+2}}  
         \langle \mathcal{O}(p_1)  |\mathcal{O}  (p_3)\rangle.
    \end{split}
\end{equation}
Here, as above $\hat n = \frac{1}{2} \hat q$ is the properly normalized ``unit" null vector and $\mathcal{L}_{\mu \nu}$ is the orbital angular momentum \eqref{orbital-angular-momentum}.

\section{Free Field Examples} \label{sec:freefields}

In this section, we present several explicit examples of the $\mathcal{W}$ and $\mathcal{S}$ generators and their commutators for (conformally-coupled) free scalars obeying $\square \phi = 0$. The stress tensor for a conformally-coupled real scalar can be written as \cite{Osborn:1993cr}
\begin{equation} \label{eq:Tfree}
T_{\mu\nu} = \nabla_{\mu}\phi \nabla_{\nu}\phi - \frac{1}{4(d-1)}\left((d-2) \nabla_{\mu}\nabla_{\nu} + \eta_{\mu\nu}\nabla_{\rho}\nabla^{\rho}\right) \phi^2, 
\end{equation}
and the ${\rm U}(1)$ conserved current for complex scalar as
\begin{equation} \label{eq:Jfree}
j_{\mu} = i\bar{\phi} \overleftrightarrow{\partial_{\mu}}\phi. 
\end{equation}

In flat retarded coordinates \eqref{flat-retarded-Bondi-coord}, the large-$r$ expansion of the scalar wave equation is 
\begin{equation} \label{eq:waveEQ}
(n-1)\partial_u \phi^{(n)} + \partial_z \partial_{\bar{z}} \phi^{(n-1)} = 0. 
\end{equation}
Up to a normalization $A$, the commutator of free fields at null infinity is 
\begin{equation} \label{eq:freeCom}
\left[\phi^{(1)}(u,z,\bar{z}), \phi^{(1)}(u',z',\bar{z}')\right] = A~ {\rm sgn}(u-u')\delta^{(2)}(z-z'),
\end{equation}
where ${\rm sgn}(u) = 2 \int_{-\infty}^{\infty} \frac{d\omega}{2\pi i} \frac{1}{\omega} e^{i \omega u}$ and $\partial_u{\rm sgn}(u) = 2 \delta(u)$.

\subsection{Stress Tensors} \label{subsec:freefields-W}

In free scalar field theory, explicit expressions for the $\mathcal{W}_m$ generators in terms of the fundamental field $\phi$ can be obtained by substituting the free-field expression for the stress tensor of a conformally-coupled scalar \eqref{eq:Tfree} into the conformal field theory expressions for the $\mathcal{W}_m$ generators \eqref{def-w} with $T^{(3)} = 0$.  Interestingly, the same expressions can equivalently be obtained by substituting the standard quantum field theory expression for the stress tensor of free scalar field $T_{\mu\nu} = \partial_\mu \phi \partial_\nu \phi - \frac{1}{2}\eta_{\mu\nu} \partial^\rho \phi \partial_\rho \phi$ into the appropriate quantum field theory expressions for the $\mathcal{W}_m$ generators \eqref{def-w} with $T^{(3)} \neq 0$. Note that this equivalence does not extend to all the light-ray operators considered in Section \ref{sec:classification}.  For example, quantum and conformal field-theoretic expressions for the $\mathcal{L}_m$ generators differ by terms involving the trace.  

Previously in the literature, free-field expressions for the $\mathcal{W}$ generators for scalars as well as higher spin free fields were presented as the ``quadratic" $W$ operators in \cite{Hu:2022txx,Hu:2023geb}, following \cite{Freidel:2021ytz}.  When acting on free fields, their expressions were shown to reproduce the ${\rm w}_{1+\infty}$ action and algebra derived from celestial OPEs in \cite{Himwich:2021dau,Freidel:2021ytz,Adamo:2021lrv,Adamo:2021zpw,Himwich:2023njb,Kmec:2024nmu}. We have checked that our expressions for free fields match those of \cite{Hu:2022txx,Hu:2023geb} through sub$^{3}$leading order (equivalently $m=2$) and expect that they match to all orders. 

A notable difference between the free field expressions in \cite{Hu:2022txx,Hu:2023geb} and our general expression for the $\mathcal{W}$ generators \eqref{def-w-intro} is  that the expressions of \cite{Hu:2022txx,Hu:2023geb} involve multiple null integrals in $u$, while ours involve nonlocal integrals in the transverse $\bar{z}$ direction. Recall that although the primary descendants of our $\mathcal{W}_{m}$ generators have local expressions \eqref{def-w} in the transverse $(z,\bar{z})$ plane, the $\mathcal{W}_{m}$ generators appear a priori non-local.  However, when evaluating our expressions in free field theory,  we find that additional integrals in $u$ are interchangeable with non-locality in $\bar{z}$. This flexibility is due to the structure of the scalar wave equation \eqref{eq:waveEQ}.

For concreteness, consider the example of $\mathcal{W}_2$:
\begin{equation} \label{eq:W2des}
\begin{aligned}
\partial_{\bar{z}}\mathcal{W}_{2}(z,\bar{z})  = 
  \frac{1}{48}\int du \left( u^3 \partial_z^3 \partial_{\bar{z}} T_{uu}^{(2)} - 6 u^2 \partial_z^2 \partial_{\bar{z}} T_{uz}^{(2)} + 18 u \partial_z \partial_{\bar{z}} T_{zz}^{(2)} + 24\left( T_{zz}^{(3)} + \partial_z T_{rz}^{(4)} \right) \right).  
  \end{aligned}
\end{equation}
We observe that in  free field theory, the apparently non-local contribution to $\mathcal{W}_2$ can be rewritten as
\begin{equation}
  \int du \left(T_{zz}^{(3)} + \partial_z T_{rz}^{(4)}\right) = \frac{1}{2} \int du \int du'~ {\rm sgn}(u-u')~ \partial_{\bar{z}} \left( \partial_z^2\phi^{(1)}(u)\partial_z\phi^{(1)}(u') \right).
\end{equation}
Since this is a total derivative in $\partial_{\bar{z}}$, we can thus obtain a local expression for the ``pre''-primary $\mathcal{W}_2$ at the price of introducing an additional integral in $u$. For free scalars, we expect this pattern to continue at all further subleading orders. It would be interesting to understand the extent to which this phenomenon persists in generic interacting theories, for instance $\mathcal{N}=4$ super-Yang-Mills. 

We now turn to the algebra for free scalar fields. Free scalar field theory violates our assumption in Section \ref{sec:LRAtensor} of a theory with no light neutral scalars of dimension $1 \leq \Delta \leq 2$.  However, using \eqref{eq:freeCom} to compute several explicit commutators at low orders, we find that all of the deviations from the generic result occur entirely outside the wedge and therefore do not affect the ${\rm w}_{1+\infty}$ algebra that arises upon smearing the generators at these low orders. The analysis of \cite{Hu:2022txx,Hu:2023geb} establishes this pattern at all orders.

All of the free scalar commutators $\left[\mathcal{W}_{m},\mathcal{W}_{m'}\right]$ with $m+m'=-2,-1,0$ match the generic result without corrections. We start to see corrections for $m+m'>0$, and have computed the following examples\footnote{Note that since we are not considering vacuum contributions, we do not keep track of contributions from operator normal ordering.} 
\begin{equation}
    \begin{aligned}   \left[\mathcal{W}_0(z,\bar{z}),\mathcal{W}_1(z',\bar{z}') \right] &=  2A\left[3 \partial_z - 2 \partial_{z'}\right]\left(\delta^{(2)}(z-z')  \mathcal{W}_1 \right) \\
        &\qquad -\frac{A}{4}\partial_z^3 \left(\delta^{(2)}(z-z')\left(3\mathcal{L}_1 +\int du~     \phi^{(1)}\phi^{(1)} \right)\right) ,  \\
     [\mathcal{W}_1(z,\bar{z}), \mathcal{W}_1(z',\bar{z}')]
        & = 6A\left[\partial_z - \partial_{z'} \right] \left(\delta^{(2)}(z-z')\mathcal{W}_2 \right) \\
        &\qquad     - \frac{A}{8}    \left[\partial_z^4-\partial_{z'}^4\right] \left(  \delta^{(2)}(z-z')   \left( \mathcal{L}_2 + \int du ~u\phi^{(1)}\phi^{(1)}    \right)\right), \\
        \big[\mathcal{W}_{0}(z,\bar{z}), \mathcal{W}_{2}(z',\bar{z}')\big] &= - 4 A  \left[ \partial_{z'} -  2\partial_z \right]\left(\delta^{(2)}(z-z')\mathcal{W}_2\right)  \\
        &\qquad + A\partial_{z}^3 \left(\delta^{(2)}(z-z')\left(-\frac{5}{12}\partial_{z}\mathcal{L}_2 +  \int du~ u^2 T_{uz}^{(2)} - \frac{1}{2} \partial_{z}\int du~u \phi^{(1)}\phi^{(1)} \right) \right) \\
        &\qquad +  \frac{A}{4} \partial_z^4\left(\delta^{(2)}(z-z')\mathcal{L}_2 \right).  
    \end{aligned}
\end{equation} 
Taking $A = -\frac{i}{4}$, these commutation relations exactly match the general results presented in \eqref{eq:W0W1}, \eqref{eq:W1W1}, \eqref{eq:W0W2}, inside the wedge, with minor deviations outside the wedge.  Specifically, here we observe additional contributions from the composite operator $\phi \phi$ of scaling dimension $\Delta =2$.  In the first two commutation relations above, these extra contributions can be absorbed into $\mathcal{L}_m$ if the generators are evaluated with the quantum (as opposed to conformal) field theory stress tensor.  This demonstrates that the additional operators can be identified as contributions from the trace of the stress tensor in the non-conformal context.  In the last commutation relation, however, the extra contributions cannot be fully absorbed by evaluating the generators with the quantum field theory stress tensor. Thus, regardless of operator differences between quantum and conformal field theory, the presence of the operator $\phi \phi$ with scaling dimension $\Delta =2$ generally leads to deviations in the algebra.

On general grounds, light neutral scalar operators $\mathcal{O}$ with dimensions in the range $1\leq \Delta \leq 2$ are expected to fall off at large $r$ as $\mathcal{O} \sim \frac{1}{r^\Delta}$.  Hence, the heaviest light-ray operator one can form without additionally specifying a regulator, namely $\int du~ \mathcal{O}^{(\Delta)}$, carries scaling dimension $\Delta=-m = -1$.  Thus, modifications to the algebra due to such operators can only appear in commutators $\left[\mathcal{W}_{m},\mathcal{W}_{m'}\right]$ with $m+m'>0$, as seen explicitly above in the free field example.  General light-ray operators formed from such scalars $\int du~ u^p \mathcal{O}^{(n)}$, where $n \geq \Delta$, transform under ${\rm SL}(2, \mathbb{C})$ with weights $h = \bar{h} = \frac{n-p-1}{2}$.  It is therefore a priori less obvious whether deviations to the algebra will always appear outside the wedge, although it may be possible to establish such a result via an induction proof similar to the one in Subsection \ref{sec:wAlg}. We leave this question to a future investigation.

\subsection{Currents}\label{subsec:freefields-S}

For currents, explicit expressions can again be found by substituting the free field expression \eqref{eq:Jfree} for the current into the generators \eqref{def-s-operator}. As in the previous subsection, we find that  non-localities in the transverse plane can be traded for additional integrals in $u$. As an example, consider $\mathcal{S}_2$, which appears non-local: 
\begin{equation} \partial_{\bar{z}}\mathcal{S}_2(z,\bar{z}) = \frac{1}{8}\int du~ \left( u^2 \partial_z^2 \partial_{\bar{z}} j_u^{(2)} - 2 u \partial_z\partial_{\bar{z}} j_{z}^{(2)} - 2 \partial_z j_r^{(4)} - 2 j_z^{(3)} \right). 
\end{equation}
Using the  expression for the current and the wave equation \eqref{eq:waveEQ}, we find
\begin{equation} \label{free-s2}
\int du \left(j_z^{(3)} + \partial_z j_r^{(4)}\right) =  i\int du \int du' ~{\rm sgn}(u-u') ~\partial_{\bar{z}}\left(\partial_z\bar{\phi}(u) \partial_z\phi(u')\right) .
\end{equation}
Here again, the right-hand side is now a total derivative in $\bar{z}$, so the generator $\mathcal{S}_2$ admits a local expression on the transverse plane.  We expect a similar pattern to  extend to all orders. 

The free-field expression \eqref{free-s2} matches expressions for the light-ray operators considered in \cite{Hu:2023geb}.   We do not study the algebra here, but the results in \cite{Hu:2023geb} establish that any deviations will be outside the wedge.  Moreover, since the current takes the same form in conformal and quantum field theory, there is no ambiguity in the expressions for light-ray operators constructed from the current. Thus, any deviations in the algebra cannot be attributed to differences between conformal and quantum field theory.

\section{Discussion} \label{sec:discussion}

The results presented in this work may have a number of potentially far-reaching consequences, ranging from  new constraints on higher-dimensional conformal field theories to novel extensions and applications in generic relativistic quantum field theories, theories of quantum gravity, and holography.

First,  to gain further insight into the role of the ${\rm w}_{1+\infty}$ algebra in conformal field theories and to check the general operator statements in this paper, it would be instructive to verify the ${\rm w}_{1+\infty}$ algebra explicitly using  four-point functions $\langle \mathcal{O} T T \mathcal{O}\rangle$ of a known interacting conformal field theory such as $\mathcal{N}=4$ super-Yang-Mills. We plan to study this in upcoming work. A conceptually related but distinct study of the $S$ algebra in $\mathcal{N}=4$ super-Yang-Mills was recently carried out in \cite{Alday:2026rso}, in which soft gluon theorems explicitly realize the $S$ algebra.  The symmetry action on colored matter in that work is directly related to our construction of the $S$ algebra from conserved currents.\footnote{\cite{Alday:2026rso} argues that since the $S$ algebra itself is undeformed in conformal field theories, this implies nontrivial conditions on loop corrections to the soft theorems. We expect such conditions to manifest as constraints on the running of our $\mathcal{S}$ generators.}  Given the extensive and ongoing work on $\mathcal{N}=4$ super-Yang-Mills, it is of interest to explore the relation of the ${\rm w}_{1+\infty}$ and $S$ algebras to recent developments on the structure of energy correlators in this theory, including \cite{Dempsey:2025yiv} and \cite{Jackson:2026eml}.

In both the specific case of $\mathcal{N}=4$ super-Yang-Mills and in more general $4d$ conformal field theories, there may be useful consequences of this algebra in the form of sum rules (following \cite{Kologlu:2019bco,Caron-Huot:2020adz,Belin:2020lsr,Chang:2023szz,Caron-Huot:2021enk}) or positivity constraints. In holographic conformal field theories, the algebra will have a corresponding geometric realization in AdS, perhaps involving shockwaves as in  \cite{Hofman:2008ar,Belin:2020lsr}, and it would be interesting to understand whether the aforementioned sum rules lead to new instances of the ``stringy equivalence principle"  formulated in \cite{Kologlu:2019bco}. 

It is also of general interest to attain a more precise understanding of the necessity of conformal symmetry (or lack thereof) in the realization of these infinite-dimensional symmetry algebras. To this end, it would be instructive to study the flow of these light-ray operators under renormalization and to understand how the algebras are modified or deformed. As mentioned above, the dimension of the stress tensor is protected, although in principle the null integration kernel may run or there may be additional operator mixing.  Our stress-tensor light-ray operators and their algebra may, for  instance, further resolve the flow of $a$ and $c$ anomalies under renormalization following, for example, the analyses in \cite{Hartman:2023qdn,Hartman:2023ccw,Hartman:2024xkw}. Our light-ray operators may also provide new, infrared-finite collider observables in quantum gravity (generalizing those studied in \cite{Gonzo:2020xza,Herrmann:2024yai,Chicherin:2025keq}) and gauge theory, particularly in QCD. Indeed, in the context of celestial holography, the ${\rm w}_{1+\infty}$ and $S$ algebras are intimately tied to the multipole structure \cite{Kmec:2026dis,Hoskins:2026new}, soft expansion \cite{Himwich:2023njb}, and symmetry structure \cite{Guevara:2021abz,Strominger:2021lvk,Adamo:2021lrv} of the theory and it would be interesting if these interpretations extend to collider observables. It has also already been shown that in QCD, the subleading soft gluon theorem controls certain types of light-ray operator mixing \cite{Chang:2025zib}. This thus raises the question of whether the further subleading soft gauge theorems govern more general light-ray operator mixing.  Analogous statements may also extend to light-ray operator mixing in quantum gravity. 

Our results also lend themselves to  direct applications to the subject of celestial holography.  For example, our analysis suggests that  the ``cubic terms" appearing in the algebra of soft and hard charges in \cite{Freidel:2021ytz,Hu:2022txx,Hu:2023geb,Freidel:2023gue} are simply related to the terms outside the wedge in the local $\mathcal{W}$ and $\mathcal{S}$ algebras.  Our expressions also hint at the possibility of additional, yet-undiscovered asymptotic symmetries whose associated (hard) charges contain our light-ray operators.  Furthermore, since our construction of the chiral ${\rm w}_{1+\infty}$ and $S$ algebras does not rely on  holomorphic collinear limits typically used in celestial holography,  our methods readily apply to the study of mixed $\mathcal{W}$ and $\overline{\mathcal{W}}$ commutators.  As such, this alternate method may supply a natural prescription for the mixed-helicity algebra in celestial conformal field theory. It would be interesting to compare the results of our method with those of the recent study \cite{Pranzetti:2026pdg}.  In addition, it would be interesting to explore the construction of our light-ray operators in twistor space and in particular, whether  purely local expressions emerge naturally in this uplift, as was previously found for asymptotic symmetry charges and transformations \cite{Adamo:2021lrv,Kmec:2026dis}.  It could also be useful to understand how the analysis presented in this work might accommodate the known quantum corrections to the holographic symmetry algebras \cite{Costello:2022upu,Bittleston:2022jeq,Fernandez:2023abp,Zeng:2023qqp,Fernandez:2024qnu,Serrani:2025oaw}.

Finally, it would be interesting to generalize our analysis to all dimensions $d > 2$. In $d=2$, the stress-tensor analysis reproduces a chiral subalgebra of the Virasoso algebra consisting of modes $L_{n}$ with $n \geq -1$, while the conserved current analysis reproduces a chiral subalgebra of the Kac-Moody algebra consisting of modes $J_m$ with $m \geq 0$.

\section*{Acknowledgments}

We are grateful to Jan Albert, David Simmons-Duffin, Simon Heuveline, Murat Kolo\u{g}lu, Gr\'egoire Mathys, Noah Miller, Prahar Mitra, Ian Moult, Ana-Maria Raclariu, Shu-Heng Shao, Ahmed Sheta, Tianli Wang, and Sasha Zhiboedov for useful conversations. We thank Mathew Calkins, Marcus Hoskins, Gr\'egoire Mathys, Noah Miller, Sabrina Pasterski and Ana-Maria Raclariu for comments on a draft.  This work was completed with support from NSF grant 2310633 and the Simons Collaboration on Celestial Holography.  EH is supported by the John Archibald Wheeler Fellowship at the Princeton Center for Theoretical Science.

\appendix

\section{Derivation of Light $\bar{h}$ Primaries} 

\subsection{Stress-tensor Light-ray Primaries}\label{app:primary-derivation}

First, we consider the transformation of $R_{m, \ell,n}$ defined in \eqref{general-combo-fixed-sl2c} under ${\rm SL}(2, \mathbb{C})$.  The left transformation takes the form:  
\begin{equation}
    \begin{split}
        -&\delta_Y R_{m,\ell, n}[a,b,c,d,e,f]
            \\& = \left( Y^{z}  \partial_{z} +\frac{m+2\ell}{2} \partial_z Y^z\right)  R_{m,\ell, n}[a,b,c,d,e,f]
            \\& \quad 
        + \frac{a_n(m-n+\ell+1)(n+\ell-2)(n-2)(n-3)-b_n(n-2) (2m-n+1)+2c_n (m-n+2)}{2(n-2)(n-3)} \\& \quad \quad \times \partial_z^2 Y^z \partial_z^{-1}
            \int du~(u\partial_z \partial_{\bar{z}})^{m-n+1} \partial_z^{\ell}  T_{rr}^{(n+2)} 
    \\& \quad 
        + \frac{ b_n  (n-2)(m-n+\ell)(n+\ell-1) -2c_n m }{2(n-2)}\partial_z^2 Y^z \partial_z^{-1}
            \int du~(u\partial_z\partial_{\bar{z}})^{m-n+1}  \partial_z^{\ell-1} T_{rz}^{(n+1)} 
    \\& \quad
        +\frac{c_n  (m-n+\ell-1)(n+\ell)}{2} \partial_z^2 Y^z \partial_{z}^{-1}
            \int du~(u\partial_z\partial_{\bar{z}})^{m-n+1} \partial_z^{\ell-2} T_{zz}^{(n)}    
    \\ & \quad 
        + \frac{d_n(m-n+\ell+1)(n+\ell-2)(n-2)(n-3)-\left(c_n(n-1)-  \frac{1}{2}b_n(n+1)(n-2) \right) }{2(n-2)(n-3)}
    \\& \quad \quad \times
        \partial_z^2 Y^z \partial_z^{-1}
            \int du~u(u\partial_z\partial_{\bar{z}})^{m-n+1}\partial_z^{\ell} T^{(n+1)} 
    \\&  \quad  
        +\frac{e_n(m-n+\ell+2)(n+\ell-3)(n-2)(n-3)+2(a_n(n-2)(n-3) - b_n(n-2)+c_n )}{2(n-2)(n-3)} 
    \\& \quad \quad \times
        \partial_z^2 Y^z \partial_z^{-1}
            \int du~u(u\partial_z\partial_{\bar{z}})^{m-n+1} \partial_z^{\ell+1} T_{ r\bar{z}}^{(n)}  
    \\& \quad  
        + \frac{f_n(m-n+\ell +2)(n+\ell-3)}{2}\partial_z^2 Y^z  \partial_z^{-1}
        \int du~u^2 (u\partial_z \partial_{\bar{z}})^{m-n} \partial_z^{\ell +2}T_{\bar{z}\bar{z}}^{(n-1)}
    \\& \quad  
         -\frac{1}{2} \partial_z^2 Y^z\partial_z^{-1}  \int du~(u\partial_z \partial_{\bar{z}})^{m-n+2}
    \\& \quad \quad \times
        \left(
            a_n\partial_z^{\ell}T_{rr}^{(n+1)}  
            + \frac{b_n(n-1)(n-2)-2c_n}{(n-2)(n-3)}\partial_z^{\ell-1}  T_{rz}^{(n)} 
            + \frac{c_n}{n-2}n T_{zz}^{(n-1)}   
            +d_n u \partial_z^{\ell}T^{(n)} \right) 
    \\& \quad  
         -\frac{e_n}{2} \partial_z^2 Y^z\partial_z^{-1}  \int du~(u\partial_z \partial_{\bar{z}})^{m-n+2}  u  \partial_z^{\ell+1} T_{r \bar{z}}^{(n-1)}  
    \\& \quad 
       -\frac{f_n-e_n}{2} \partial_z^2 Y^z \partial_z^{-1}\int du~u^2 (u\partial_z\partial_{\bar{z}})^{m-n+1} \partial_z^{\ell+2} T_{\bar{z}\bar{z}}^{(n-2)} .
    \end{split}
\end{equation}
Demanding that the non-primary terms proportional to $\partial_z^2 Y^z$ cancel between $R_{m, \ell, n}$ with $n$ differing by one implies that the coefficients satisfy the following recursion relations:
\begin{equation}
    \begin{split}
       a_{n+1}(n-2)(n-3)  & =a_n (n-2)(n-3)(m-n+\ell+1)(n+\ell-2)-b_n(n-2) (2m-n+1)\\& \quad \quad +2c_n (m-n+2)
        ,\\
       b_{n+1}n(n-1)-2c_{n+1} &=   b_n  (n-1)(n-2)(m-n+\ell)(n+\ell-1) -2c_n m (n-1) 
        ,\\
        (n+1)c_{n+1}& = c_n(n-1) (m-n+\ell-1)(n+\ell)
        ,\\
        d_{n+1}(n-2)(n-3)&= d_n(m-n+\ell+1)(n+\ell-2)(n-2)(n-3)+\frac{1}{2}b_n(n+1)(n-2)\\& \quad \quad - c_n(n-1) 
        ,\\
        e_{n+1}(n-2)(n-3)&=e_n(m-n+\ell+2)(n+\ell-3)(n-2)(n-3)
            \\& \quad \quad +2(a_n(n-2)(n-3) - b_n(n-2)+c_n ) 
        ,\\
        f_{n+1}-e_{n+1}&=f_n(m-n+\ell +2)(n+\ell-3). 
    \end{split}
\end{equation}
The initial value $c_3$ is fixed to cancel the non-primary transformation of $\mathcal{O}_{m, \ell}$ \eqref{sl2c-op} and given by
\begin{equation}
    \begin{split}
        c_3 = \frac{1}{3!} \ell(\ell+1)(\ell+2) (m+\ell-3)(m+\ell-2)(m+\ell-1).
    \end{split}
\end{equation}
The recursion for $c_n$ can then be solved and the solution is
\begin{equation}
    c_n = \frac{(m+\ell-1)!}{ (\ell-1)!}\frac{(n-2)!(n+\ell-1)!}{n!(m-n+\ell-1)!}.
\end{equation}
Using the recursion for $b_n$, this then determines the initial value $b_3$:
\begin{equation}
    \begin{split}
        b_{3}    &=   c_3 =\frac{1}{3!} \ell(\ell+1)(\ell+2) (m+\ell-3)(m+\ell-2)(m+\ell-1),
    \end{split}
\end{equation}
and the $b_n$ recursion relation can be subsequently solved:
\begin{equation}
    \begin{split}
        b_n   = \frac{(m+\ell-1)!}{ (\ell-1)!} \frac{(n-3)!   (n+\ell-2)! }{  n! (m-n+\ell)!}(2 (n-1) ((\ell-1) (m+\ell)-n)-n(\ell-2) (m+\ell+1)).
    \end{split}
\end{equation}
Next, using the boundary condition $a_{m+2} = 0$, the recursion for $a_n$ can be solved and we find
\begin{equation}
    \begin{split}
        a_n = \frac{(m+\ell-1)!}{ (\ell-1)!} \frac{ (n-3)!(n+\ell-3)!}{(n-1)!   (m-n+\ell+1)!} (m-n+2)     (\ell (m+\ell-1)-m-n+1).
    \end{split}
\end{equation}
After that, we solve for $d_n$ using the boundary condition $d_{m+2}=0$:
\begin{equation}
    \begin{split}
         d_n = \frac{(m+\ell-1)!}{ (\ell-1)!} \frac{(2-\ell) (\ell+m+1) (m-n+2)   (n+\ell-3)!}{2 m (n-2) (m-n+\ell+1)!}.
    \end{split}
\end{equation}
Finally, demanding that $e_{m+2} = 0$ and $f_{m+1}=0$, we solve the remaining two recursion relations and find 
\begin{equation}
    \begin{split}
        e_n = \frac{(m+\ell+1)!}{ (\ell-3)!}\frac{(m(m+1)-  (n-2) (n-1))   (n+\ell-4)!}{m(m+1)(n-2) (n-1) (m-n+\ell+2)!}
    \end{split}
\end{equation}
and 
\begin{equation}
    \begin{split}
        f_n = \frac{(m+\ell+1)!}{ (\ell-3)!}\frac{  (n+\ell-4)! (-m (m+1)-(n-1) (n-2(m+1)))}{m (m+1) (n-1)  (m-n+\ell+2)!}.
    \end{split}
\end{equation}
Notice that $e_n$ and $f_n$ vanish for $\ell\leq 2$ and $a_n$, $b_n$, $c_n$, and $d_n$ vanish for $\ell \leq 1$.  Hence there are only nontrivial solutions for $\ell \geq 1$.

The right ${\rm SL}(2, \mathbb{C})$ transformation 
\begin{equation} 
    \begin{split}
          -& \delta_{\bar{Y}}R_{m,\ell, n}[a,b,c,d,e,f]\\
         & = \left(\bar{Y}^{\bar{z}}\partial_{\bar{z}} +\frac{m}{2} \partial_{\bar{z}} \bar{Y}^{\bar{z}}\right)
               R_{m,\ell, n}[a,b,c,d,e,f]  
            + \frac{(m-n+1)(n-2)}{2}\partial_{\bar{z}}^2 \bar{Y}^{\bar{z}}\partial_{\bar{z}}^{-1}
                R_{m,\ell, n}[a,b,c,d,e,f]  
       \\& \quad - \frac{1}{2} \frac{2mf_n}{n-3}\partial_{\bar{z}}^2 \bar{Y}^{\bar{z}}\partial_{\bar{z}}^{-1}
            \int du~u  (u\partial_z\partial_{\bar{z}})^{m-n+1} 
                \partial_{z}^{\ell+1}T_{r \bar{z}}^{(n)} 
    \\& \quad 
        +\frac{f_n(m-2n+2)}{2}\partial_{\bar{z}}^2 \bar{Y}^{\bar{z}}\partial_{\bar{z}}^{-1}
        \int du~u^2(u\partial_z\partial_{\bar{z}})^{m-n}\partial_{z}^{\ell+2}T_{\bar{z}\bar{z}}^{(n-1)}  
    \\& \quad  
         -\frac{1}{2} \partial_{\bar{z}}^2 \bar{Y}^{\bar{z}}\partial_{\bar{z}}^{-1}  \int du~(u\partial_z \partial_{\bar{z}})^{m-n+2} \left([a_n-e_n]\partial_{z}^\ell  T_{rr}^{(n+1)}  
         +[b_n-2 a_n]\partial_{z}^{\ell-1}T_{rz}^{(n)}  
        +[c_n-b_n]\partial_{z}^{\ell-2}T_{zz}^{(n-1)} \right)
    \\& \quad  
         -\frac{1}{2} \partial_{\bar{z}}^2 \bar{Y}^{\bar{z}}\partial_{\bar{z}}^{-1}  \int du~u(u\partial_z \partial_{\bar{z}})^{m-n+2} \left( \left(d_n-\frac{1}{2}e_n \right) \partial_{z}^{\ell }T^{(n)}+e_n \partial_{z}^{\ell+1}T_{r\bar{z}}^{(n-1)}
        +f_n   \partial_{z}^{\ell+2} T_{\bar{z}\bar{z}}^{(n-2)}\right)
    \\&  \quad  
        -\frac{1}{2}\frac{1}{n-4}\frac{1}{n-3} \partial_{\bar{z}}^2 \bar{Y}^{\bar{z}}\partial_{\bar{z}}^{-1} \int du~   (u\partial_z\partial_{\bar{z}})^{m-n+2}\\
        & \quad \times  \left( 
         2\left[e_n(n-3)(m-1)-f_n(m-n+3)\right] \partial_{z}^{\ell}T_{r r}^{(n+1)} 
         +  2  \left[e_n(n-3)-f_n\right] (u \partial_z  \partial_{\bar{z}})\partial_{z}^{\ell-1}T_{ rz}^{(n-1)} 
            \right) 
    \\& \quad 
       -\frac{1}{2}\frac{1}{n-4}\frac{1}{n-3}\partial_{\bar{z}}^2 \bar{Y}^{\bar{z}}\partial_{\bar{z}}^{-1}\int du~  (u\partial_z\partial_{\bar{z}})^{m-n +2}  
    \\& \quad \times
       \left(   
           -   \left[2e_n(n-3)-f_n(n-2) \right] u \partial_{z}^{\ell}T^{(n)}
       +2\left[e_n (n-3)-f_n\right] u \partial_{z}^{\ell+1} T_{r\bar{z}}^{(n-1)}
       +2f_n(n-4)  u  \partial_{z}^{\ell+2} T_{\bar{z}\bar{z}}^{(n-2)}      \right)  ,   
    \end{split}
\end{equation}
imposes further restrictions on the coefficients, specifically implying the additional recursion relations: 
\begin{equation}
    \begin{split}
        (m-n+2)(n-3) a_{n-1} &=a_n -e_n + 2~\frac{e_n(n-3)(m-1)-f_n(m-n+3)}{(n-3)(n-4)} , 
        \\
        (m-n+2)(n-3) b_{n-1}
            &= b_n-2 a_n+2~ \frac{e_{n+1}(n-2)-f_{n+1}}{(n-2)(n-3)} ,
        \\
        (m-n+2)(n-3) c_{n-1} 
            &=c_n-b_n,
        \\
        (m-n+2)(n-3) d_{n-1}
            &=d_n-\frac{1}{2}e_n -\frac{2e_n(n-3)-f_n(n-2)}{(n-3)(n-4)} ,
        \\
        (m-n+2)(n-3) e_{n-1}- \frac{2m f_{n-1}}{n-4}    
            &=e_n + 2~\frac{e_n (n-3)-f_n}{(n-3)(n-4)} ,
        \\
        (m-n+2)(n-3) f_{n-1}+ (m-2n+4)f_{n-1}
            &=f_n+ \frac{2f_n}{n-3}.
    \end{split}
\end{equation} 
We can check that the third recursion involving $c_n$ is always satisfied by our solution, while the first, second and fourth recursion relations are satisfied provided $\ell \leq 2$.  Finally the last two recursion relations are trivially satisfied if $\ell\leq 2$ because this sets $e_n = f_n =0$. Hence, here we find that there are only non-trivial solutions for $\ell \leq 2$.  Combining with the previous result, we thus conclude that there are only non-trivial solutions for $\ell =1$ and $\ell=2$, which are the following: 
\begin{equation} 
    \begin{split}
          \underline{\ell = 1}: 
         &\quad a_n = - \frac{m! (n-3)!}{(m-n+1)!}, 
            \quad \quad b_n = -\frac{m! (n-3)!}{(m-n+1)!}(2n-m-4), 
            \quad  \\
            & \quad c_n = \frac{m! (n-3)!}{(m-n+1)!}(n-2)(m-n+1),\quad\quad d_n =  \frac{m! (n-3)!}{(m-n+1)!}  \frac{m+2}{2m},
            \quad \quad e_n = f_n = 0,\\
            \\
         \underline{\ell = 2}: 
         &\quad a_n = \frac{(m+1)!(n-3)!}{(m-n+1)!}, \quad \quad b_n  = \frac{(m+1)!(n-3)!}{(m-n+1)!} 2(n-1), \\
        &\quad c_n  = \frac{(m+1)!(n-3)!}{(m-n+1)!} (n+1)(n-2),\quad \quad d_n = e_n  = f_n = 0. 
    \end{split} 
\end{equation} 
These are exactly the coefficients that appear in \eqref{def-v} and \eqref{def-w}.

\subsection{Current Light-ray Primaries}\label{app:primary-derivation-current}

First, we observe that  
\begin{equation} \label{sl2c-Mdesc}
    \begin{split}
         -\delta_Y    \partial_{\bar{z}}^{m-1} \mathcal{M}_{m, \ell+m-1}   
        &=  
           \left(Y^{z}\partial_{z} +\frac{ m+2\ell}{2} \partial_z Y^z \right) \partial_{\bar{z}}^{m-1} \mathcal{M}_{m, \ell+m-1}  
        \\&
         \quad  \quad - \frac{\ell (\ell+1)(\ell+m-1)(\ell+m-2)}{2}\partial_z^2 Y^z \partial_z^{-1} \int du~(u \partial_z\partial_{\bar{z}})^{m-1} \partial_z^{\ell-1} j_{z}^{(2)}  , 
    \\
         -\delta_{\bar{Y}}    \partial_{\bar{z}}^{m-1} \mathcal{M}_{m, \ell+m-1}   
        &=  
           \left(\bar{Y}^{\bar{z}}\partial_{\bar{z}} +\frac{ m}{2} \partial_z \bar{Y}^{\bar{z}} \right) \partial_{\bar{z}}^{m-1} \mathcal{M}_{m, \ell+m-1}. 
    \end{split}
\end{equation}
Next, we consider the transformation of the remaining elements in the basis, grouped accordingly: 
\begin{equation}
    \begin{split} 
        -\delta_Y &\int du~(u\partial_z \partial_{\bar{z}})^{m-n+1}  \left (a_n\partial_z^{\ell}j_{r}^{(n+1)}+ b_n\partial_z^{\ell-1}j_{z}^{(n)}
             +c_n u\partial_z^{\ell+1} j_{\bar{z}}^{(n-1)}\right)
        \\&= 
            \left(Y^{z}  \partial_{z} +\frac{m+2\ell}{2} \partial_z Y^z \right)
                \int du~(u\partial_z\partial_{\bar{z}})^{m-n+1} 
                    \left (a_n\partial_z^{\ell}j_{r}^{(n+1)}
                            +b_n \partial_z^{\ell-1}j_{z}^{(n)}
                                +c_n u\partial_z^{\ell+1} j_{\bar{z}}^{(n-1)} \right)
        \\& \quad
            +\frac{a_n(\ell+m-n+1) (n+\ell-2)(n-2) -m b_n}{2(n-2)}\partial_z^2 Y^z \partial_z^{-1}\int du~(u\partial_z\partial_{\bar{z}})^{m-n+1} \partial_z^\ell j_{r}^{(n+1)} 
            \\& \quad  
                +\frac{a_n(n-2)-b_n+ c_n(\ell+m-n+2)(n+\ell-3)(n-2)}{2(n-2)} \partial_z^2 Y^z\partial_z^{-1}\int du~u (u\partial_z\partial_{\bar{z}})^{m-n+1}\partial_z^{\ell+1}j_{\bar{z}}^{(n-1)}, \\
        & \quad  +
            \frac{b_n(\ell+m-n)(n+\ell-1)}{2}\partial_z^2 Y^z \partial_z^{-1}\int du~(u\partial_z\partial_{\bar{z}})^{m-n+1}\partial_z^{\ell-1} j_{z}^{(n)}  
        \\& \quad 
            -\frac{1}{2} \partial_z^2 Y^z \partial_z^{-1}\int du~ (u \partial_z \partial_{\bar{z}})^{m-n+2}  \left(a_n\partial_z^{\ell}j_{r}^{(n)} 
                +b_n \frac{n-1}{n-2}\partial_z^{\ell-1} j_{z}^{(n-1)}
            + c_nu \partial_z^{\ell+1} j_{\bar{z}}^{(n-2)} \right) .
    \end{split}
\end{equation}
The cancellation of all terms involving $\partial_z^2 Y^z$ for any $m$ implies the following recursion relations: 
\begin{equation}
    \begin{split}
         a_n(\ell+m-n+1) (n+\ell-2)(n-2) -m b_n 
            &=a_{n+1} (n-2) 
        ,\\
         a_n(n-2)-b_n+ c_n(\ell+m-n+2)(n+\ell-3)(n-2) 
            &=c_{n+1}(n-2) 
        ,\\
        b_n  (n-1)(\ell+m-n)(n+\ell-1)
            &= b_{n+1}n.
    \end{split}
\end{equation}
\eqref{sl2c-Mdesc} fixes an initial condition for $b_3$
\begin{equation}
    b_3 = - \frac{1}{2}\ell (\ell+1)(\ell+m-1)(\ell+m-2).
\end{equation}
The recursion relations can then be solved, where in addition we set $a_{m+2} = c_{m+2}=0$.  The solution is 
\begin{equation} \label{current-coefficient-solution}
    \begin{split}
    a_n &=
    - \frac{  (m-n+2) (m+\ell-1)! (n+\ell-3)!}{  (n-2)  (\ell-1)!  (m-n+\ell+1)!},\\ 
       b_n&= -\frac{(m+\ell-1)! (n+\ell-2)!}{(n-1) (\ell-1)! (m-n+\ell)!}, \\
        c_n &=-\frac{(m-n+2)(m+\ell)! (n+\ell-4)! }{m(n-2) (\ell-2)! (m-n+\ell+2)!}.
    \end{split}
\end{equation}
We observe that there is no non-trivial solution for $\ell < 1$. 

The transformation under the right-moving part of ${\rm SL}(2, \mathbb{C})$
\begin{equation}
    \begin{split} 
       - &\delta_{\bar{Y}} \int du~(u\partial_z \partial_{\bar{z}})^{m-n+1}  \left (a_n\partial_{z}^{\ell}j_{r}^{(n+1)}+ b_n\partial_{z}^{\ell-1}j_{z}^{(n)}
             +c_n u\partial_{z}^{\ell+1} j_{\bar{z}}^{(n-1)}\right)
        \\&=    \left(\bar{Y}^{\bar{z}}  \partial_{\bar{z}} +\frac{m}{2} \partial_{\bar{z}} \bar{Y}^{\bar{z}}+ \frac{(m-n+1)(n-2)}{2} \partial_{\bar{z}}^2 \bar{Y}^{\bar{z}}\partial_{\bar{z}}^{-1}\right)\\& \quad \quad \times  \int du~(u\partial_z \partial_{\bar{z}})^{m-n+1}  \left   (a_n\partial_{z}^{\ell}j_{r}^{(n+1)}+ b_n\partial_{z}^{\ell-1}j_{z}^{(n)}
             +c_n u\partial_{z}^{\ell+1} j_{\bar{z}}^{(n-1)}\right)
        \\&    -\frac{1}{2}\partial_{\bar{z}}^2 \bar{Y}^{\bar{z}}\partial_{\bar{z}}^{-1}\int du(u\partial_z \partial_{\bar{z}})^{m-n+2}  \left[\left(a_n+ \frac{m c_n}{n-3}\right)\partial_{z}^\ell j_{r}^{(n)} 
        -\left(a_n -b_n\right)\partial_{z}^{\ell-1} j_{z}^{(n-1)}+ c_n\frac{n-2}{n-3} u \partial_{z}^{\ell+1}j_{\bar{z}}^{(n-2)}\right]  
      \\&      -\frac{1}{2}\frac{c_n}{n-3}  \partial_{\bar{z}}^2 \bar{Y}^{\bar{z}}\partial_{\bar{z}}^{-1} \int du ~(u\partial_z \partial_{\bar{z}})^{m-n+3} \partial_{z}^{\ell-1}j_{z}^{(n-2)}  
    \end{split}
\end{equation}
places further constraints on the coefficients, in particular, leading to the additional recursion relations 
\begin{equation} \label{right-recursion-current}
    \begin{split}
        (m-n+2)(n-3) a_{n-1} &=  a_n + \frac{m c_n}{n-3} ,\\
        (m-n+2)(n-3) b_{n-1}&= -a_n +b_n+ \frac{c_{n+1}}{n-2},\\
         (m-n+2)(n-3) c_{n-1}&= c_n \frac{n-2}{n-3}. 
    \end{split}
\end{equation}
The first two are satisfied by the general solution \eqref{current-coefficient-solution}, provided that $\ell \leq 1$. Combining this with the fact that the solution \eqref{current-coefficient-solution} is only non-trivial for $\ell \geq 1$, we conclude there is only a non-trivial solution when $\ell =1$.  In this case, $c_n$ vanishes for all $n$, so the last recursion relation in \eqref{right-recursion-current} is trivially satisfied.  The coefficients for $\ell = 1$ take the explicit form
\begin{equation}
    \begin{split}
        \underline{\ell = 1}: 
         &\quad  a_n = - \frac{m!(n-3)!}{(m-n+1)!},  \quad\quad b_n = -\frac{m!(n-2)!}{(m-n+1)!},  \quad\quad c_n = 0,
    \end{split}
\end{equation}
and are precisely the coefficients appearing in \eqref{def-s-operator}. 

\section{Transformation of Inverse Derivatives} \label{app:inverseDer}  

In this appendix, we demonstrate that the ${\rm SL}(2,\mathbb{C})$ transformations of inverse $z,\bar{z}$ derivatives are given by the analytic continuation of the formula for positive derivatives.

Consider an operator $\mathcal{O}' = \partial_z^{-1}\mathcal{O}$ such that $\partial_z\mathcal{O}' = \mathcal{O}$, where $\mathcal{O}$ is an operator of weight $h$ that transforms as 
\begin{equation}
  -\delta_Y\mathcal{O} = \left(Y^{z}\partial_z + h \partial_zY^{z}\right)\mathcal{O} + \partial_z^2 Y^z \widetilde{\mathcal{O}},
\end{equation}
and $\widetilde{\mathcal{O}}$ has weight $h-1$. Note that here we do not want to assume that $\widetilde{\mathcal{O}} \propto \partial_{z}^{-1} \mathcal{O}$. Generally, $\mathcal{O}'$ will transform as 
\begin{equation}
   -\delta_Y \mathcal{O}' = \left(Y^{z}\partial_z + \alpha \partial_zY^{z}\right)\mathcal{O}' + \partial_z^2 Y^z \widetilde{\mathcal{O}}', 
\end{equation}
where $\widetilde{\mathcal{O}}'$ has weight $\alpha-1$. Taking a derivative, we find that
\begin{equation}
  -\partial_z\delta_Y\left( \mathcal{O}'\right) = -\delta_Y\left(\partial_z \mathcal{O}'\right) = \left(Y^{z}\partial_z + \left(\alpha+1\right) \partial_zY^{z}\right)\partial_z\mathcal{O}' + \partial_z^2 Y^z \left( \alpha \mathcal{O}' + \partial_z\widetilde{\mathcal{O}}' \right). 
\end{equation}
Now, matching this with the transformation of $\mathcal{O}$, we find that $\alpha = h-1$ and $\widetilde{\mathcal{O}} = (h-1) \mathcal{O}' + \partial_z\widetilde{\mathcal{O}}'$. This tells us that
\begin{equation}
  -\delta_Y\mathcal{O}' = -\delta_Y\left(\partial_z^{-1}\mathcal{O}\right) = \left(Y^{z}\partial_z + (h-1) \partial_zY^{z}\right)\mathcal{O}' + \partial_z^2 Y^z \left( \partial_z^{-1} \widetilde{\mathcal{O}} - (h-1) \partial_z^{-1}\mathcal{O}'\right). 
\end{equation}
Note that this correctly reproduces $\widetilde{O} = (h-1)\mathcal{O}' = \partial_{z}^{-1}\mathcal{O}$ if $\mathcal{O}$ is the first descendant of a primary. We check that it is also consistent to treat $\partial_{z}^{-1}$ as an integral, as follows:
\begin{equation}
  \begin{aligned}
    -\delta_Y \partial_z^{-1}\mathcal{O} = -\partial_z^{-1}(\delta_Y \mathcal{O}) &= \int dz \left[ \left(Y^z \partial_z + h \partial_z Y^z\right)\mathcal{O} + \partial_z^2 Y^z \widetilde{\mathcal{O}}\right] \\
    &= \left(Y^{z}\partial_z + (h-1) \partial_zY^{z}\right)\mathcal{O}' + \partial_z^2 Y^z \left( \partial_z^{-1} \widetilde{\mathcal{O}} - (h-1) \partial_z^{-1}\mathcal{O}'\right), \\
  \end{aligned}
\end{equation}
which can be derived by integration by parts. Now we consider multiple derivatives of an ordinary descendant
\begin{equation} \label{eq:transSLdesc}
  -\delta_Y\left(\partial_z^{\ell}\mathcal{O}\right) = \left(Y^{z}\partial_z + (h+\ell) \partial_zY^{z}\right)\partial_z^{\ell}\mathcal{O} + \partial_z^2Y^z\left(\partial_z^{\ell}\widetilde{\mathcal{O}} + \ell \left(h + \frac{\ell-1}{2}\right)\partial_z^{\ell-1}\mathcal{O}\right).  
\end{equation}
Similarly we repeat the derivation above for further inverse derivatives to find 
\begin{equation} \label{eq:SLinverseDer}
  -\delta_Y\left(\partial_z^{-\ell}\mathcal{O}\right) = \left(Y^{z}\partial_z + (h-\ell) \partial_zY^{z}\right)\partial_z^{\ell}\mathcal{O} + \partial_z^2Y^z\left(\partial_z^{-\ell}\widetilde{\mathcal{O}} - \ell \left(h + \frac{-\ell-1}{2}\right)\partial_z^{-\ell-1}\mathcal{O}\right),
\end{equation}
which is exactly the analytic continuation of the usual formula from $\ell \to - \ell$.

\section{Low-order $\mathcal{W}$ Commutators} \label{app:Wderivation}

\subsection{Direct Calculation of Further Commutators} \label{app:WderivationInv}

In this appendix, we present results for low-order $\mathcal{W}$ commutators in which the operators on the right-hand side carry scaling dimension $\Delta \geq -3$. We first elaborate on the general procedure for calculating the stress-tensor light-ray commutators that appear in this paper. The procedure directly generalizes the examples presented in Section \ref{sec:LRAtensor}. 

We always consider the commutator of between two ${\rm SL}(2,\mathbb{C})$  primaries of the general form \eqref{eq:generalCommutator}. As in Section \ref{sec:LRAtensor}, we proceed order by order in increasing $m=-\Delta$. To calculate a commutator that produces an operator with scaling dimension $\Delta =-m$, one uses the minimal basis for operators on the right-hand side,  specifically equations \eqref{anec}, \eqref{subleading-basis}, \eqref{delta-1-basis}, \eqref{basis}, and \eqref{basis-caveate} and distributes transverse derivatives according to \eqref{eq:RHweightsNew}. Then, coefficients are fixed by ${\rm SL}(2,\mathbb{C})$ constraints \eqref{eq:SLrecursion} and consistency with the $P_u$ Jacobi identity, which further relates them to coefficients in previously-determined commutators producing operators of scaling dimension $\Delta =-m+1$. 

More precisely, the Jacobi identity with $P_u$ directly relates the coefficients of all operators that are not in kernel of $P_u$ to coefficients in previously-determined commutation relations, leaving unfixed only the coefficients of operators that are in the kernel of $P_u$. At each order in dimension, there are a small number of operators in the kernel of $P_u$, which are of the form \eqref{eq:KernelOp} and treated in Appendix \ref{app:KerP}.\footnote{In particular, the only way new inverse transverse derivatives can appear at a given order is in the kernel of $P_u$.} These operators mix under ${\rm SL}(2,\mathbb{C})$ with operators that are \textit{not} annihilated by $P_u$ as detailed in Appendix \ref{app:KerP}. Thus, coefficients of operators in the kernel of $P_u$ are related by ${\rm SL}(2,\mathbb{C})$ recursion \eqref{eq:SLrecursion} to coefficients of operators \textit{not} annihilated by $P_u$. Since the latter are determined by consistency with the $P_u$ Jacobi identity, the commutator is thereby fully constrained.
 
\subsection{Summary of Results for Low-order $\left[\mathcal{W}_m,\mathcal{W}_n\right]$ Commutators} \label{app:Wsummary}

Here we present several low-order commutators, which can be fully derived from $\left[\mathcal{W}_{-1},\mathcal{W}_{-1}\right]$, $\left[\mathcal{W}_{-1},\mathcal{W}_0\right]$, and Poincar\'e constraints as discussed in the previous subsection. First, we summarize results presented elsewhere throughout the text for commutators that produce operators of scaling dimension $-\Delta \leq 0$:  
\begin{equation}
\begin{aligned}
\left[\mathcal{W}_{-1}(z,\bar{z}), \mathcal{W}_{-1}(z',\bar{z}')\right]  &= 0, \\
\left[\mathcal{W}_{-1}(z,\bar{z}), \mathcal{W}_{0}(z',\bar{z}')\right] &=   i \left[  \frac{1}{2}\partial_{z'} - \partial_z\right] \left(\delta^{(2)}(z-z') \mathcal{W}_{-1} \right), \\
\left[ \mathcal{W}_0(z,\bar{z}), \mathcal{W}_0(z',\bar{z}') \right] &= i \left[\partial_{z'} - \partial_z \right]\left( \delta^{(2)}(z-z') \mathcal{W}_0 \right), \\
\left[\mathcal{W}_{-1}(z,\bar{z}), \mathcal{W}_{1}(z',\bar{z}')\right] &= i \left[ \frac{1}{2} \partial_{z'} -  \frac{3}{2}\partial_z \right]\left(\delta^{(2)}(z-z')\mathcal{W}_{0}  \right) + i \frac{3}{4} \partial_z^2\left(\delta^{(2)}(z-z')\mathcal{D}  \right).  
\end{aligned}
\end{equation}

Next, at $-\Delta = 1$, there is a one-to-one map between the minimal basis \eqref{basis} and classified primaries, so we find
\begin{subequations}
\begin{align}
\left[\mathcal{W}_0(z,\bar{z}), \mathcal{W}_1(z',\bar{z}')\right] &= i \left[ \partial_{z'} - \frac{3}{2}\partial_z \right]\left(\delta^{(2)}(z-z')\mathcal{W}_1  \right) + i \frac{3}{16} \partial_z^3\left(\delta^{(2)}(z-z')\mathcal{L}_1  \right), \label{eq:W0W1}\\
\left[\mathcal{W}_{-1}(z,\bar{z}), \mathcal{W}_{2}(z',\bar{z}')\right] &= i \left[ \frac{1}{2} \partial_{z'} -  2\partial_z \right]\left(\delta^{(2)}(z-z')\mathcal{W}_{1}  \right) \label{eq:Wm1W2} \\
&\quad + i \partial_z^2\left(\delta^{(2)}(z - z')\left( \mathcal{X}_1 - \frac{9}{16} \partial_{z}\mathcal{L}_1  \right)\right) - i \frac{3}{16}\partial_z^3 \left(\delta^{(2)}(z-z') \mathcal{L}_1 \right). \nonumber 
\end{align}
\end{subequations}
For $-\Delta > 1$, the full minimal basis \eqref{basis} is not in one-to-one-correspondence with the set of classified ${\rm SL}(2,\mathbb{C})$ primaries, so the ${\rm SL}(2,\mathbb{C})$ constraints on commutators are more involved as described in Section \ref{sec:GenBasis}.  At $-\Delta = 2$, we find
\begin{subequations}
\begin{align}
\left[\mathcal{W}_1(z,\bar{z}), \mathcal{W}_1(z',\bar{z}')\right] &= i \left[ \frac{3}{2}\partial_{z'} -  \frac{3}{2}\partial_z \right]\left(\delta^{(2)}(z-z')\mathcal{W}_2  \right) +  \frac{i}{32} \left[\partial_z^4 - \partial_{z'}^4\right]\left(\delta^{(2)}(z-z')\mathcal{L}_2  \right), \label{eq:W1W1}\\
\left[\mathcal{W}_{0}(z,\bar{z}), \mathcal{W}_{2}(z',\bar{z}')\right] &= i \left[ \partial_{z'} -  2\partial_z \right]\left(\delta^{(2)}(z-z')\mathcal{W}_2 \right) \label{eq:W0W2} \\
&\hspace{-2.0cm}  + i \partial_{z}^3 \left(\delta^{(2)}(z-z')\left(\frac{5}{48}\partial_{z}\mathcal{L}_2  - \frac{1}{4}\int du~ u^2 T^{(2)}_{uz}  \right) \right)
-  \frac{i}{16} \partial_z^4\left(\delta^{(2)}(z-z')\mathcal{L}_2  \right), \nonumber \\
\left[\mathcal{W}_{-1}(z,\bar{z}), \mathcal{W}_{3}(z',\bar{z}')\right] &= i \left[ \frac{1}{2} \partial_{z'} -  \frac{5}{2}\partial_z \right]\left(\delta^{(2)}(z-z')\mathcal{W}_{2}  \right) \label{eq:Wm1W3} \\
&\hspace{-2.0cm}  + i \partial_z^2\left(\delta^{(2)}(z - z')\left(\frac{15}{32}\mathcal{Y}_2 - \frac{11}{32}\partial^2_{z}\mathcal{L}_2  + \frac{3}{8} \int du~ u^2 \partial_{z}T_{uz}^{(2)}  \right)\right) \nonumber \\
&\hspace{-2.0cm}  + i \partial_z^3\left(\delta^{(2)}(z-z') \left(-\frac{5}{48}\partial_{z}\mathcal{L}_2  + \frac{1}{4} \int du~ u^2 T_{uz}^{(2)} \right)\right) + i \frac{1}{32}\partial_z^4 \left(\delta^{(2)}(z-z') \mathcal{L}_2 \right). \nonumber
\end{align}
\end{subequations} 
Finally at $-\Delta = 3$, we find
\begin{subequations}
\begin{align}
\left[\mathcal{W}_{1}(z,\bar{z}), \mathcal{W}_{2}(z',\bar{z}')\right] &= i \left[ \frac{3}{2}\partial_{z'} -  2\partial_z \right]\left(\delta^{(2)}(z-z')\mathcal{W}_{3}  \right) \label{eq:W1W2} \\
&\hspace{-2.0cm} + i\partial_z^{4} \left(\delta^{(2)}(z - z') \left( \frac{15}{8\cdot 32} \partial_z \mathcal{L}_3  - \frac{1}{24}\mathcal{X}_{3} \right)\right) - \frac{i}{8\cdot 32} \left[ \partial_{z'}^5 + 3\partial_z^5 \right]\left(\delta^{(2)}(z - z') \mathcal{L}_3  \right), \nonumber \\
\left[\mathcal{W}_{0}(z,\bar{z}), \mathcal{W}_{3}(z',\bar{z}') \right] &= i \left[ \partial_{z'} -  \frac{5}{2}\partial_z \right]\left(\delta^{(2)}(z-z')\mathcal{W}_{3}  \right) \label{eq:W0W3} \\
&\hspace{-2.0cm} + i\partial_z^{3} \left(\delta^{(2)}(z - z')\left(-\frac{13}{32\cdot 3} \partial_{z} \mathcal{X}_3  + \frac{21}{128} \partial_{z}^2 \mathcal{L}_3 + \frac{15}{4\cdot16}\int du~ u^2 T_{zz}^{(2)}  \right)\right) \nonumber \\
&\hspace{-2.0cm} + i\partial_z^{4} \left(\delta^{(2)}(z - z')\left( \frac{1}{12}\mathcal{X}_3  - \frac{15}{128}\partial_{z} \mathcal{L}_3  \right) \right) 
+ i\frac{3}{4\cdot 64}\partial_z^{5} \left(\delta^{(2)}(z - z') \mathcal{L}_3  \right), \nonumber \\
\left[\mathcal{W}_{-1}(z,\bar{z}), \mathcal{W}_{4}(z',\bar{z}')\right] &= i \left[ \frac{1}{2} \partial_{z'} -  3\partial_z \right]\left(\delta^{(2)}(z-z')\mathcal{W}_{3}  \right) \label{eq:Wm1W4} \\ &\hspace{-2.0cm} + i \partial_z^2\left(\delta^{(2)}(z - z')\left(60~ \mathcal{V}_3  - \frac{41}{8\cdot 16} \partial^3_{z}\mathcal{L}_3 + \frac{7}{32} \partial_{z}^2\mathcal{X}_3  -\frac{15}{4\cdot 16} \int du~ u^2 \partial_{z}T_{zz}^{(2)}  \right)\right) \nonumber \\
&\hspace{-2.0cm} + i \partial_z^3\left(\delta^{(2)}(z-z') \left(- \frac{21}{8\cdot16}\partial_{z}^2\mathcal{L}_3  + \frac{13}{6 \cdot 16} \partial_{z} \mathcal{X}_{3}  - \frac{15}{4\cdot 16} \int du~ u^2 T^{(2)}_{zz} \right)\right) \nonumber \\
&\hspace{-2.0cm} + i \partial_z^4\left(\delta^{(2)}(z-z') \left(\frac{15}{16^2}\partial_{z}\mathcal{L}_3  - \frac{1}{24} \mathcal{X}_{3} \right)\right) - i \frac{1}{16^2}\partial_z^5 \left(\delta^{(2)}(z-z') \mathcal{L}_3\right). \nonumber 
\end{align}
\end{subequations}  

\section{Operators in the Kernel of $P_u$} \label{app:KerP}

Here we derive the general ${\rm SL}(2,\mathbb{C})$ transformations of operators in the kernel of $P_u$.

\subsection{Stress Tensors} \label{app:KerP-w}

For light-ray operators constructed from stress tensors, the general form of these operators is \eqref{eq:KernelOp}, repeated here for convenience: 
\begin{equation} \label{eq:KernelOpApp}
\mathcal{N}_{m,j,\bar{j}} = \int du~ \partial_z^{j} \partial_{\bar{z}}^{\bar{j}}\left( \alpha \partial_{\bar{z}}^2 T_{zz}^{(m+1)} + \beta \partial_{\bar{z}}^2\partial_z T_{rz}^{(m+2)} + \gamma \partial_{\bar{z}}^2 \partial_z^2 T_{rr}^{(m+3)}+ \mu \partial_{\bar{z}}\partial_z^2 T_{r\bar{z}}^{(m+2)} + \nu \partial_z^2 T_{\bar{z}\bar{z}}^{(m+1)}\right) .
\end{equation}
We first consider $m>3$, then consider $m = 3,2,1$ sequentially as special cases.

For the purposes of the proof in Section \ref{sec:wAlg}, we are interested in examining the transformation of \eqref{eq:KernelOpApp} under $Q_{\bar{Y}}$:  
\begin{equation} \label{eq:TransKerOpApp}
\begin{aligned}
- \delta_{\bar{Y}} &\mathcal{N}_{m,j,\bar{j}} = \left(\bar{Y}^{\bar{z}} \partial_{\bar{z}} + \frac{m+4+2\bar{j}}{2}  \partial_{\bar{z}} \bar{Y}^{\bar{z}}\right)\mathcal{N}_{m,j,\bar{j}}\\
&+ \partial_{\bar{z}}^2\bar{Y}^{\bar{z}} \Bigg(\frac{\beta - \alpha}{2} \int du~ u \partial_z^{j+1}\partial_{\bar{z}}^{\bar{j}+2}T_{zz}^{(m)} + \frac{\alpha}{2} (2+\bar{j}) (1 + \bar{j} + m)\int du~ \partial_z^{j}\partial_{\bar{z}}^{\bar{j}+1}T_{zz}^{(m+1)} \\
&\qquad \qquad  + \frac{2\nu + (m-1)((m-2)(2\gamma-\beta) - 2  \mu)}{2(m-1)(m-2)} \int du~ u\partial_z^{j+2}\partial_{\bar{z}}^{\bar{j}+2}T_{rz}^{(m+1)}  \\
& \qquad \qquad + \frac{\beta}{2}(2+\bar{j})(1+\bar{j}+m) \int du~ \partial_z^{j+1}\partial_{\bar{z}}^{\bar{j}+1}T_{rz}^{(m+2)} - \frac{\gamma}{2} \int du~ u \partial_z^{j+3}\partial_{\bar{z}}^{\bar{j}+2}T_{rr}^{(m+2)} \\
&\qquad \qquad + \frac{2 \nu + (m-1)(-\mu m  + \gamma(2+\bar{j})(m-2)(1+\bar{j}+m))}{2(m-1)(m-2)} \int du~   \partial_z^{j+2}\partial_{\bar{z}}^{\bar{j}+1}T_{rr}^{(m+3)} \\
&\qquad \qquad +  \frac{2\nu - \mu (m-1)m}{2(m-1)(m-2)}\int du~ u \partial_z^{j+3}\partial_{\bar{z}}^{\bar{j}+1}T_{r\bar{z}}^{(m+1)} \\
&\qquad \qquad + \frac{1}{2}\left(-\frac{2 \nu m}{m-1} + \mu(1+\bar{j})(2+\bar{j}+m)\right) \int du~  \partial_z^{j+2}\partial_{\bar{z}}^{\bar{j}}T_{r\bar{z}}^{(m+2)} \\
&\qquad \qquad - \nu \frac{m+1}{2(m-1)}  \int du~ u  \partial_z^{j+3}\partial_{\bar{z}}^{\bar{j}}T_{\bar{z}\bar{z}}^{(m)} +  \frac{\nu}{2} \bar{j}(3+\bar{j} + m) \int du~  \partial_z^{j+2}\partial_{\bar{z}}^{\bar{j}-1}T_{\bar{z}\bar{z}}^{(m+1)} \Bigg).
\end{aligned}
\end{equation}
The transformation \eqref{eq:TransKerOpApp} holds for $m >2$ but the generic case is $m >3$, when all of the operators are non-vanishing. 

For $m >3$, there is no non-trivial  choice of coefficients $\alpha$, $\cdots$, $\nu$ that eliminates all terms in $\delta_{\bar{Y}}\mathcal{N}_{m,j,\bar{j}}$ that are not in the kernel of $P_u$.  In particular, this would require the following eliminations: the coefficient of $u T_{rr}^{(m+2)}$ sets $\gamma$ to zero. The coefficient of $u T_{\bar{z}\bar{z}}^{(m)}$  sets $\nu$ to zero, which implies $\mu$ is zero from the coefficient of $u T_{r\bar{z}}^{(m+1)}$. Then the coefficient of $u T_{rz}^{(m+1)}$ implies $\beta = 0$ and finally the coefficient of $u T_{zz}^{(m)}$ implies $\alpha = 0$. Thus, all coefficients must be set to zero and only the trivial solution remains. A similar conclusion can be reached for $\delta_{Y}\mathcal{N}_{m,j,\bar{j}}$ by observing that the two transformations exchange $z \leftrightarrow \bar{z}$, $Y^z \leftrightarrow \bar{Y}^{\bar{z}}$, $j \leftrightarrow \bar{j}$, $ \alpha \leftrightarrow \nu$, and $\beta \leftrightarrow \mu$.

For illustration, we also consider examples at $m=3,2,1$, which are relevant for some of the explicit calculations summarized in Appendix \ref{app:Wderivation}. 

Suppose we would like to find an operator in the kernel of $P_u$ for which the transformation under $Q_Y$ and $Q_{\bar{Y}}$ contains only terms in the kernel of $P_u$ at $m=3$. This is relevant for example in $\left[\mathcal{W}^p,\mathcal{W}^q\right]$ commutators with $p+q = \frac{11}{2}$. At $m=3$ there is no longer a constraint from the coefficient of $u T_{rr}^{(5)}$ since $T_{rr}^{(5)}$ vanishes. Still, the coefficient of $u T_{\bar{z}\bar{z}}^{(3)}$ implies $\nu$ is zero and the coefficient of $u T_{r\bar{z}}^{(4)}$ implies that $\mu = 0$. Now the coefficient of $uT_{zz}^{(3)}$ implies that $\alpha = \beta$ and of $u T_{rz}^{(4)}$ implies $\beta = 2\gamma$. For these coefficients, the full ${\rm SL}(2,\mathbb{C})$ transformation of \eqref{eq:KernelOpApp} is
\begin{equation}
\begin{aligned}
- &\left(\delta_Y +\delta_{\bar{Y}}\right)\mathcal{N}_{3,j,\bar{j}} = \left( Y^z \partial_z + \frac{7+2j}{2}  \partial_z Y^z  + \bar{Y}^{\bar{z}} \partial_{\bar{z}} + \frac{7+2\bar{j}}{2}  \partial_{\bar{z}} \bar{Y}^{\bar{z}}\right)\mathcal{N}_{3,j,\bar{j}}\\
&\qquad \qquad +  \partial_z^2 Y^{z} \frac{\gamma}{2} \partial_z^{j-1}\partial_{\bar{z}}^{\bar{j}+2} \Big[ \int du~ 4\left(\partial_z^2 T_{rr}^{(6)} + \partial_zT_{rz}^{(5)}\right) + j(6+j)\left( \partial_z^2 T_{rr}^{(6)}+ 2\partial_zT_{rz}^{(5)} + 2 T_{zz}^{(4)}\right)  \\
&\qquad \qquad \qquad \qquad  \qquad \qquad \qquad \qquad - 4 u \partial_z \partial_{\bar{z}}\left(T_{zz}^{(3)} + \partial_z T_{rz}^{(4)} \right) \Big] \\
&\qquad \qquad + \partial_{\bar{z}}^2 \bar{Y}^{\bar{z}} \frac{\gamma}{2} (2+\bar{j})(4+\bar{j}) \partial_z^j \partial_{\bar{z}}^{\bar{j}+1} \int du\left( \partial_z^2 T_{rr}^{(6)} + 2 \partial_z T_{rz}^{(5)} + 2 T_{zz}^{(4)}\right), 
\end{aligned}
\end{equation}
which from $Q_Y$ always contains terms not annihilated by $P_u$. Note that $\mathcal{N}_{3,j,-4}$ is a right-primary and has the appropriate $\bar{h}$ weight so that its contribution to $\left[\mathcal{W}^p,\mathcal{W}^q\right]$ involves no anti-holomorphic derivatives.

For the case of $m=2$, there is no $\gamma$ term in \eqref{eq:KernelOpApp}, and its full ${\rm SL}(2,\mathbb{C})$ transformation is   
\begin{equation}
\begin{aligned}
- (\delta_Y &+\delta_{\bar{Y}})\mathcal{N}_{2,j,\bar{j}} = \left( Y^z \partial_z + \frac{6+2j}{2}  \partial_z Y^z  + \bar{Y}^{\bar{z}} \partial_{\bar{z}} + \frac{6+2\bar{j}}{2}  \partial_{\bar{z}} \bar{Y}^{\bar{z}}\right)\mathcal{N}_{2,j,\bar{j}}\\
&+ \partial_z^2 Y^z \frac{1}{2}  \partial_z^{j-1} \partial_{\bar{z}}^{\bar{j}} \int du \Big[ \alpha j (5+j) \partial_{\bar{z}}^{2} T_{zz}^{(3)} -3\alpha u \partial_{\bar{z}}^3\partial_z T_{zz}^{(2)} + (-4 \alpha + \beta(1+j)(4+j) \partial_{\bar{z}}^2 \partial_z T_{rz}^{(4)} \\
&\qquad \qquad \qquad \qquad \qquad + 2(\beta-\alpha) u \partial_{\bar{z}}^{2}\partial_z^2 T_{z\bar{z}}^{(2)} + \mu(2+j)(3+j)\partial_{\bar{z}}\partial_z^2 T_{r\bar{z}}^{(4)} \\
&\qquad \qquad \qquad \qquad \qquad + (\mu-\nu) u \partial_{\bar{z}}\partial_z^3 T_{\bar{z}\bar{z}}^{(2)} + \nu(2+j)(3+j)\partial_z^2 T_{\bar{z}\bar{z}}^{(3)}   \Big] \\
&+ \partial_{\bar{z}}^2 \bar{Y}^{\bar{z}} \frac{1}{2}  \partial_z^{j} \partial_{\bar{z}}^{\bar{j}-1} \int du \Big[\nu \bar{j}(5+\bar{j}) \partial_z^2 T_{\bar{z}\bar{z}}^{(3)} - 3 \nu u \partial_z^3 \partial_{\bar{z}}T_{\bar{z}\bar{z}}^{(2)} + (-4\nu + \mu(1+\bar{j})(4+\bar{j}))\partial_{\bar{z}}\partial_z^2 T_{r\bar{z}}^{(4)}   \\
&\qquad \qquad \qquad \qquad \qquad + 2(\mu - \nu) u \partial_{\bar{z}}^2 \partial_z^2 T_{z\bar{z}}^{(2)} + \beta(2+\bar{j})(3+\bar{j}) \partial_{\bar{z}}^2 \partial_z T_{rz}^{(4)} \\
&\qquad \qquad \qquad \qquad \qquad + (\beta-\alpha) u \partial_{\bar{z}}^3 \partial_z T_{zz}^{(2)} + \alpha(2+\bar{j})(3+\bar{j})\partial_{\bar{z}}^2 T_{zz}^{(3)} \Big].\\
\end{aligned}
\end{equation}

Finally, at $m=1$ there is no $\beta$, $\gamma$, or $\mu$ term in \eqref{eq:KernelOpApp} and its full ${\rm SL}(2,\mathbb{C})$ transformation is  
\begin{equation}
\begin{aligned}
- \left(\delta_Y + \delta_{\bar{Y}}\right)&\mathcal{N}_{1,j,\bar{j}} = \left( Y^z \partial_z + \frac{5+2j}{2}  \partial_z Y^z  + \bar{Y}^{\bar{z}} \partial_{\bar{z}} + \frac{5+2\bar{j}}{2}  \partial_{\bar{z}} \bar{Y}^{\bar{z}}\right)\mathcal{N}_{1,j,\bar{j}} \\
&+ \partial_z^2 Y^z \frac{1}{2} \partial_{\bar{z}}^{\bar{j}}\partial_z^{j-1} \int du \left[\nu (2+j)^2  \partial_z^{2}T_{\bar{z}\bar{z}}^{(2)} +\alpha\left( 2 u \partial_{\bar{z}}^{2} \partial_z T_{uz}^{(2)} + j(4+j) \partial_{\bar{z}}^{2}T_{zz}^{(2)}\right)\right] \\
&+ \partial_{\bar{z}}^2 \bar{Y}^{\bar{z}} \frac{1}{2} \partial_{\bar{z}}^{\bar{j}-1}\partial_z^{j} \int du \left[\alpha (2+\bar{j})^2  \partial_{\bar{z}}^{2}T_{zz}^{(2)} +\nu\left( 2 u \partial_{z}^{2} \partial_{\bar{z}} T_{u\bar{z}}^{(2)} + \bar{j}(4+\bar{j}) \partial_{z}^{2}T_{\bar{z}\bar{z}}^{(2)}\right)\right]. 
\end{aligned}
\end{equation}

\subsection{Currents} \label{app:KerP-s}

For currents, the most general operator in the kernel of $P_u$ is \eqref{eq:KerPS}, repeated here:
\begin{equation}
\widetilde{\mathcal{N}}_{m,j,\bar{j}} = \partial_z^{j}\partial_{\bar{z}}^{\bar{j}} \int du~ \left( \alpha \partial_{\bar{z}}j_z^{(m+1)} + \gamma \partial_z\partial_{\bar{z}} j_r^{(m+2)} + \nu \partial_z j_{\bar{z}}^{(m+1)}\right),
\end{equation}
which has weights $\left( \frac{m+2}{2} + j, \frac{m+2}{2}+ \bar{j}\right)$. For the purposes of the proofs in Sections \ref{sec:Salg} and \ref{sec:wSalg}, we study its transformations under $Q_{\bar{Y}}$ for $m>1$, which takes the form: 
\begin{equation}
\begin{aligned}
-\delta_{\bar{Y}}& \widetilde{\mathcal{N}}_{m,j,\bar{j}}\\
&= \left(\bar{Y}^{\bar{z}} \partial_{\bar{z}} + \frac{m+2+2\bar{j}}{2}  \partial_{\bar{z}} \bar{Y}^{\bar{z}}\right)\widetilde{\mathcal{N}}_{m,j,\bar{j}}\\
&+\partial_{\bar{z}}^2\bar{Y}^{\bar{z}} \Bigg( -  \frac{\nu + (\alpha - \gamma)(m-1)}{2(m-1)} \int du~ u \partial_z^{j+1}\partial_{\bar{z}}^{\bar{j}+1}j_{z}^{(m)} + \frac{\alpha}{2} (1+\bar{j}) (\bar{j} + m)\int du~  \partial_z^{j}\partial_{\bar{z}}^{\bar{j}}j_{z}^{(m+1)} \\
&\qquad \qquad \quad - \frac{\gamma}{2} \int du~ u \partial_z^{j+2}\partial_{\bar{z}}^{\bar{j}+1}j_{r}^{(m+1)} + \frac{- m \nu +  \gamma(1+\bar{j})(m-1)(\bar{j}+m))}{2(m-1)} \int du~  \partial_z^{j+1}\partial_{\bar{z}}^{\bar{j}}j_{r}^{(m+2)} \\
&\qquad \qquad \quad - \frac{\nu}{2} \frac{m}{(m-1)}  \int du~ u  \partial_z^{j+2}\partial_{\bar{z}}^{\bar{j}}j_{\bar{z}}^{(m)} + \frac{\nu}{2} \bar{j}(1+\bar{j} + m) \int du~  \partial_z^{j+1}\partial_{\bar{z}}^{\bar{j}-1}j_{\bar{z}}^{(m+1)} \Bigg). 
\end{aligned}
\end{equation}
When $m>2$, every non-trivial $\widetilde{\mathcal{N}}_{m,j,\bar{j}}$ mixes under $Q_{\bar{Y}}$ with operators not annihilated by $P_u$, which can be seen explicitly by solving for coefficients $\alpha$, $\gamma$, and $\nu$ that eliminate the mixing.  Namely, to eliminate this mixing, $\nu$ must be zero from the coefficient of $u j_{\bar{z}}^{(m)}$, $\gamma$ must be zero from the coefficient of $u j_{r}^{(m+1)}$, and then $\alpha$ must be zero from the coefficient of $u j_z^{(m)}$. Thus for $m>2$ all non-trivial operators in the kernel of $P_u$ mix under $Q_{\bar{Y}}$ with operators not annihilated by $P_u$. 

Recalling that the fall-off conditions set $j_r^{(3)}=0$, when $m=2$, there is no constraint from the coefficient of $uj^{(m+1)}$.  Thus operators $\widetilde{\mathcal{N}}_{m=2,j,\bar{j}}$ with $\nu=0$ and $\alpha = \gamma$ do not mix under $Q_{\bar{Y}}$ with operators not annihilated by $P_u$.  However, for this choice of coefficients, $\widetilde{\mathcal{N}}_{m=2,j,\bar{j}}$ does mix under $Q_{Y}$ with operators not annihilated by $P_u$ and ultimately only contributes to commutators in a form compatible with both \eqref{eq:localS} and \eqref{eq:InductiveAssumptionWS}.  

We also note that for $m=1$, $j_r^{(3)}=0$ and the transformation is 
\begin{equation}
\begin{aligned}
-\left(\delta_{Y}+\delta_{\bar{Y}}\right)\widetilde{\mathcal{N}}_{1,j,\bar{j}} &= \left( Y^z \partial_z + \frac{3+2j}{2}  \partial_z Y^z  + \bar{Y}^{\bar{z}} \partial_{\bar{z}} + \frac{3+2\bar{j}}{2}  \partial_{\bar{z}} \bar{Y}^{\bar{z}}\right)\widetilde{\mathcal{N}}_{1,j,\bar{j}}\\
&+\frac{1}{2}\partial_z^2 Y^{z} \partial_z^{j-1}\partial_{\bar{z}}^{\bar{j}} \int du \left[ \nu (1+j)^2 \partial_z j_{\bar{z}}^{(2)} + \alpha j (2+j) \partial_{\bar{z}}j_{z}^{(2)} + \alpha u \partial_z\partial_{\bar{z}} j_u^{(2)} \right] \\
&+\frac{1}{2}\partial_{\bar{z}}^2 \bar{Y}^{\bar{z}} \partial_z^{j}\partial_{\bar{z}}^{\bar{j}-1} \int du \left[ \alpha (1+\bar{j})^2 \partial_{\bar{z}} j_{z}^{(2)} + \nu \bar{j} (2+\bar{j}) \partial_{z}j_{\bar{z}}^{(2)} + \nu u \partial_z\partial_{\bar{z}} j_u^{(2)} \right]. 
\end{aligned}
\end{equation}
This transformation is relevant for instance in the computation of $\left[\mathcal{S}^a_{m}, \mathcal{S}^b_{n}\right]$ and $\left[\mathcal{W}_{m}, \mathcal{S}^a_{n}\right]$ with $m+n = 1$ and $\left[\mathcal{K},\mathcal{S}_0^a\right]$.

\section{Low-Order $\mathcal{S}$ Commutators} \label{app:Sderivation}

In this appendix we present low-order commutators involving $\mathcal{S}$. 

\subsection{$\left[\mathcal{S}_m,\mathcal{S}_n\right]$}

Commutators involving two $\mathcal{S}$ operators can be fully derived from $\left[\mathcal{S}_{0}^{a}, \mathcal{S}_0^b\right]$ and Poincar\'e constraints. In analogy with the previous appendix on $\mathcal{W}$ commutators, these commutators are derived by starting from the basis \eqref{eq:basisCurrents} at a given bulk dimension, decorated with  transverse derivative structures compatible with \eqref{eq:RHweightsNew}, and fixing the undetermined coefficients by enforcing consistency under the $P_u$ Jacobi identity with $\left[\mathcal{S}_{0}^{a}, \mathcal{S}_0^b\right]$. We find  
\begin{subequations}
\begin{align}
\left[\mathcal{S}_0^{a}(z,\bar{z}), \mathcal{S}_1^{b}(z',\bar{z}')\right] &= i f^{abc} \delta^{(2)}(z-z') \mathcal{S}^{c}_1 - i \frac{f^{abc}}{2} \partial_z \left(\delta^{(2)}(z-z') \mathcal{J}_1^c\right), \label{eq:S0S1} \\
\left[\mathcal{S}_1^{a}(z,\bar{z}), \mathcal{S}_1^{b}(z',\bar{z}')\right] &= i f^{abc} \delta^{(2)}(z-z') \mathcal{S}^{c}_2 - i \frac{f^{abc}}{8} \left[\partial_z^2 + \partial_{z'}^2\right]\left(\delta^{(2)}(z-z') \mathcal{J}_2^c\right), \label{eq:S1S1} \\
\left[\mathcal{S}_0^{a}(z,\bar{z}), \mathcal{S}_2^{b}(z',\bar{z}')\right] &= i f^{abc} \delta^{(2)}(z-z') \mathcal{S}^{c}_2 + i \frac{f^{abc}}{4} \partial_z \left(\delta^{(2)}(z-z')  \left( \int du~ j_{z}^{(2),c}-\partial_{z}\mathcal{J}_2^c\right) \right) \label{eq:S0S2}\\
&\quad + i \frac{f^{abc}}{8} \partial_z^2 \left( \delta^{(2)}(z-z')   \mathcal{J}_2^c   \right) .\nonumber
\end{align}
\end{subequations}
Here, the full minimal basis \eqref{basis} at $-\Delta > 1$ is not in  one-to-one correspondence with the classified ${\rm SL}(2,\mathbb{C})$ primaries, so the commutators involve the more complicated ${\rm SL}(2,\mathbb{C})$ constraints described in Section \ref{sec:GenBasis}.

\subsection{Mixed $\left[\mathcal{W}_m,\mathcal{S}_n\right]$} \label{app:MixedSW}

The low-order $\left[\mathcal{W}_m,\mathcal{S}_n\right]$ commutators are derived similarly. At  $\Delta = 0$, ${\rm SL}(2,\mathbb{C})$ constraints lead to the commutators 
\begin{align}
\left[\mathcal{W}_{-1}(z,\bar{z}), \mathcal{S}_1^{a}(z',\bar{z}')\right] &= i\left[\frac{1}{2}\partial_{z'} - \frac{1}{2} \partial_z\right]\left( \delta^{(2)}(z-z')\mathcal{S}_0^a\right),  \label{eq:Wm1S1}\\
\left[\mathcal{W}_0(z,\bar{z}), \mathcal{S}_0^{a}(z',\bar{z}')\right] &= i\partial_{z'}\left( \delta^{(2)}(z-z')\mathcal{S}_0^{a}\right), \label{eq:W0S0}
\end{align}
where the coefficients are fixed to be consistent with the action of global translations and Lorentz transformations. At $-\Delta = 1$, the commutators are derived entirely from Poincar\'e constraints and \eqref{eq:Wm1S1},  \eqref{eq:W0S0}, with the results
\begin{align} 
\left[\mathcal{W}_{-1}(z,\bar{z}), \mathcal{S}_2^{a}(z',\bar{z}')\right] &= i\left[\frac{1}{2}\partial_{z'} -  \partial_z\right]\left( \delta^{(2)}(z-z')\mathcal{S}_1^a\right) + i  \frac{1}{2}\partial_{z}^2\left(\delta^{(2)}(z-z')  \mathcal{J}_{1}^a\right),   \label{eq:Wm1S2}\\
\left[\mathcal{W}_{0}(z,\bar{z}), \mathcal{S}_1^{a}(z',\bar{z}')\right] &= i\left[\partial_{z'} - \frac{1}{2} \partial_z\right]\left( \delta^{(2)}(z-z')\mathcal{S}_1^a\right),  \label{eq:W0S1}\\
\left[\mathcal{W}_1(z,\bar{z}), \mathcal{S}^a_0(z',\bar{z}') \right]  &= i \frac{3}{2} \partial_{z'}\left(\delta^{(2)}(z-z') \mathcal{S}_1^a\right) - i\partial_{z'}\left(\delta^{(2)}(z-z')\left[ \partial_{z} \mathcal{J}_{1}^a - \frac{3}{2} \mathcal{S}_1^a \right]\right) \label{eq:W1S0} \\
&\qquad \qquad - \frac{i}{2} \partial_{z'}^2 \left(\delta^{(2)}(z-z')  \mathcal{J}_{1}^a\right) \nonumber. 
\end{align}
Note that $\partial_{z'}$ is outside the wedge of $\mathcal{S}_{0}$ in \eqref{eq:W1S0}.

\bibliography{w-infinity-in-CFT-extended}
\bibliographystyle{utphys}

\end{document}